\documentclass[aps,showkeys,superscriptaddress,nofootinbib]{revtex4}

\usepackage{bm}

\usepackage{amssymb}

\usepackage{graphicx}
\usepackage{dcolumn}

\input epsf

\usepackage{color}

\usepackage{hyperref}

\begin{document}

\title{
Electronic properties of mesoscopic graphene structures: charge
confinement and control of spin and charge transport
}

\author{A.V. Rozhkov}
\affiliation{Advanced Science Institute, RIKEN, Wako-shi, Saitama,
351-0198, Japan}
\affiliation{Institute for Theoretical and Applied
Electrodynamics, Russian Academy of Sciences, 125412, Moscow,
Russia}

\author{G. Giavaras}
\affiliation{Advanced Science Institute, RIKEN, Wako-shi, Saitama,
351-0198, Japan}

\author{Yury P. Bliokh}
\affiliation{Advanced Science Institute, RIKEN, Wako-shi, Saitama,
351-0198, Japan}
\affiliation{Department of Physics, Technion-Israel Institute of
Technology, Haifa 32000, Israel}

\author{Valentin Freilikher}
\affiliation{Advanced Science Institute, RIKEN, Wako-shi, Saitama,
351-0198, Japan}
\affiliation{Department of Physics, Bar-Ilan University, Ramat-Gan 52900,
Israel}

\author{Franco Nori}
\affiliation{Advanced Science Institute, RIKEN, Wako-shi, Saitama,
351-0198, Japan} \affiliation{Department of Physics, The
University of Michigan, Ann Arbor, MI 48109-1040, USA}

\begin{abstract}
This brief review discusses electronic properties of mesoscopic
graphene-based structures. These allow controlling the confinement and
transport of charge and spin; thus, they are of interest not only for
fundamental research, but also for applications. The graphene-related
topics covered here are: edges, nanoribbons, quantum dots,
$pn$-junctions,
$pnp$-structures,
and quantum barriers and waveguides. This review is partly intended as a
short introduction to graphene mesoscopics.
\end{abstract}


\date{\today}

\maketitle
\hfill

\tableofcontents

\section{Introduction}
\begin{figure}[btp]
\centering \leavevmode \epsfxsize=12.5cm
\epsfbox{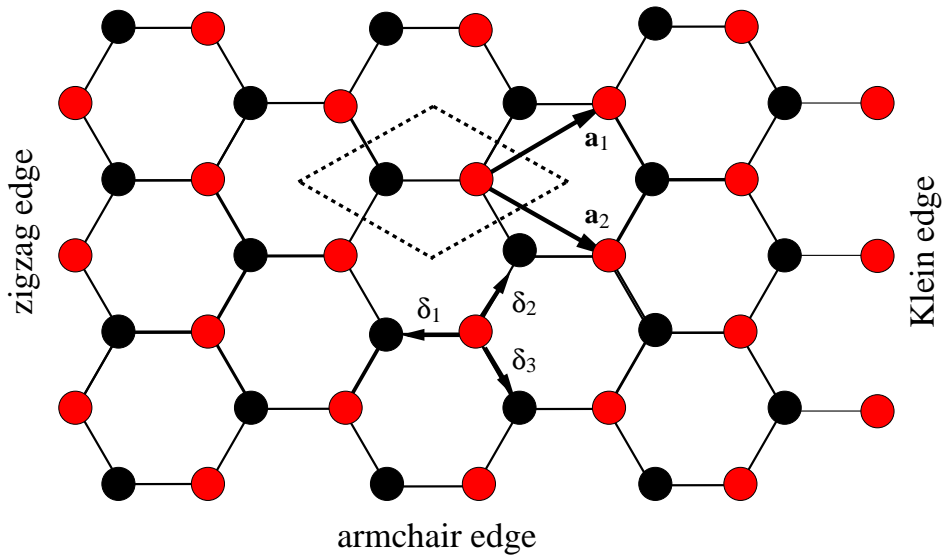} \caption[] {\label{graphene_lattice}
(Color online) Geometry of the graphene lattice showing: primitive lattice
vectors ${\bf a}_{1,2}$, diatomic lattice unit cell (dotted-line
rhombus), and vectors ${\bm \delta}_{1,2,3}$, connecting the
nearest neighbors. Red (black) circles correspond to the
${\cal A}$ (${\cal B}$) sublattice. Three different types of edges
(zigzag, armchair, and Klein edge) are shown. The Klein and zigzag
edge violate the symmetry between the sublattices (the atoms at
the edge sites belong exclusively to sublattice ${\cal A}$: they are all
red), while the armchair edge does not (it has both black and red atoms). }
\end{figure}

Graphene is a two-dimensional (2D) layer of carbon atoms ordered
into a honeycomb lattice as shown in Fig.~\ref{graphene_lattice}.
It is a material with a host of unusual properties
\cite{Geim2007,neto_etal,ISI:000224756700045,chakraborty_review,
chem_review_graphene,review_mag_yazaev,
cresti_disord_review,beenakker_colloq,sarma_review,mucciolo_review_j_phys}
including (among others): Dirac
spectrum of low-lying quasiparticles \cite{neto_etal}, large
mean-free-path \cite{ISI:000224756700045}, and high electron
mobility
\cite{high-mobility,du_high-mobility}.

Besides its purely fundamental importance, researchers view
graphene as a promising new material for electronic
\cite{electronic_device}, chemical \cite{chem_sensing}, or
electromechanical \cite{electromechanical} applications, where
graphene's unique properties may be of substantial benefit. Unlike
3D matter, whose bulk is hidden from direct observation and
influence, graphene's ``bulk'', its 2D surface, is always exposed,
and its structure may be inspected or modified with greater ease.
Furthermore, the Dirac energy dispersion in 2D implies that
graphene is a gapless semiconductor, whose density of states
vanishes linearly when approaching the Fermi energy. As such, it
is ``a bridge material'' separating the worlds of semiconductors
(with an energy gap between the valence and conducting bands) and
metals, with a finite density of electronic states at the Fermi
energy. Depending on the operating regime, graphene can be pushed
in either direction. For example, it is possible to open a gap in
a sample with the help of chemical modifications
\cite{graphane_2003,sofo_graphane},
or lateral confinement
\cite{lattice_distortion,han_experiment_gap,chen_experiment_gap}.
Alternatively, one can make graphene metallic, e.g., by chemical
doping \cite{chem_doping}. Some graphene samples have
spatially-varying electronic properties, due to local
modifications on the sample. The long electronic mean-free-path, which can
be of the order of micrometer, implies that electronic signals can
travel unimpeded large distances through a device. These features
might be very useful in applications.

The unusual properties of graphene motivated significant research
efforts. The field grows very fast: the ISI web site reports that by
October 2010 there were more than 5,000 publications with the word
``graphene'' in their titles. Clearly, this is an enormous volume of
scientific work, of which our brief review covers only a very small
fraction. Its scope is very limited in several respects. As it is obvious
from the title, we direct our attention to mesoscopic graphene systems, a
topic at the boundary between fundamental and applied research.
Furthermore, we mainly discuss the electronic aspects of graphene
mesoscopic systems, especially those which may be relevant for possible
electronic or spintronic applications, for example, charge/spin transport
and confinement, and control over them. Lattice properties are dealt with
only when the lattice affects the electrons significantly. Several topics
are deliberately omitted due to space constraints; these include: quantum
Hall effect, thermal transport phenomena, phonons, and mechanical
properties of graphene.

The review is organized as follows. In Sec.~\ref{graphene} we
discuss the most basic electron properties of an infinite graphene
sheet. The physics of graphene edges is reviewed in
Sec.~\ref{edge}. Sections~\ref{nanoribbon}, \ref{qdot}, and
\ref{pnj} focus on nanoribbons, quantum dots, as well as {\it
pn}-junctions and {\it pnp}-structures, respectively.
Sec.~\ref{barrier} discusses the barriers created by the combined
application of magnetic and/or electric fields. Conclusions are
presented in Sec.~\ref{conclusions}. The main part of the review
is kept non-technical for it to be accessible by a general reader.
More involved discussions are relegated to Appendices.


\section{Basic physics of a graphene sheet}
\label{graphene}

For completeness, in this section we quickly remind the reader the basic
single-electron properties of a graphene sheet. A more detailed
presentation can be found in
Appendix~\ref{appendix::basic}.
It is common to describe a graphene sample in terms of a tight-binding
model on the honeycomb lattice. Lattice parameters for graphene, as well as
some other microscopic characteristics, are given in
Table~\ref{glance}.
\begin{table}
\begin{tabular}{||c|c||}
\hline \hline
\quad    Graphene parameters \quad & Value \\
\hline\hline
    C-C bond length, $a_0$ &  1.4 \AA  \\
\hline
    Lattice constant & 2.46 \AA \\
\hline
    Hopping amplitudes: &   \\
    nearest neighbor, $t$   & 2.8 eV    \\
    next-nearest, $t'$ & 0.1 eV \\
    third-nearest, $t''$ & 0.07 eV  \\
\hline
    Fermi velocity, $v_{\rm F}$ & \quad $1.1 \times 10^6 \textrm{m/s}$
\quad
    \\
\hline \hline
\end{tabular}
\caption{Graphene parameters at a glance.}
\label{glance}
\end{table}
Honeycomb lattice can be split into two sublattices, denoted by
${\cal A}$
and
${\cal B}$.
The Hamiltonian of an electron hopping on a graphene sheet is given by
\begin{eqnarray}
H =
-t
\sum_{{\bf R} \in {\cal A}}
\sum_{i=1,2,3}
c^\dagger_{\bf R}
c^{\vphantom{\dagger}}_{{\bf R} + {\bm \delta}_i}
+
{\rm H.c.},
\label{H}
\end{eqnarray}
where
${\bf R}$
runs over sublattice
${\cal A}$,
and $t = 2.8$ eV is the nearest-neighbor hopping amplitude. The vectors
${\bm \delta}_i$
($i=1,2,3$)
connect the nearest neighbors (see
Fig.~\ref{graphene_lattice}
showing the geometry of the graphene lattice). When necessary, $H$ can be
augmented by interaction or longer-range hopping terms (see
Table~\ref{glance}
for values of the longer-range hopping amplitudes).

\begin{figure}[btp]
\centering \leavevmode \epsfxsize=7.5cm
\epsfbox{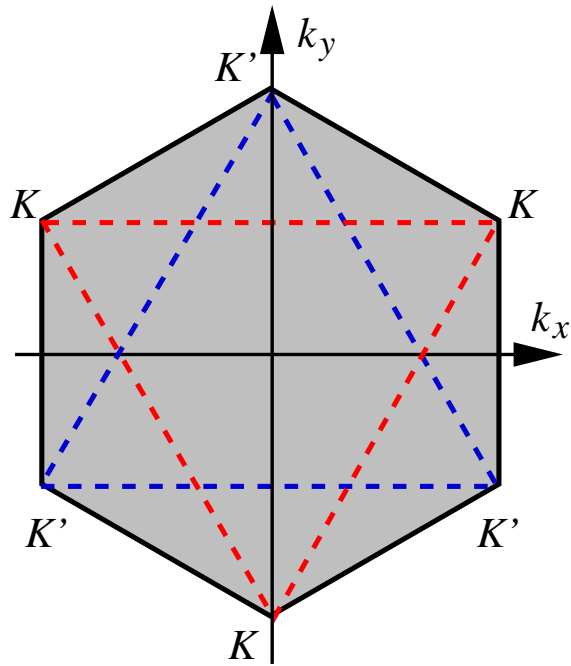}
\caption[]
{\label{bz}
(Color online) The Brillouin zone of graphene is a perfect hexagon. The
Dirac cones are located at the corners of the Brillouin zone. The six cones
can be split into two equivalence classes (cones within the same class are
connected by dashed lines). These classes are commonly referred to as $K$
and $K'$.
}
\end{figure}

Since there are two atoms in graphene's unit cell, it is convenient to
describe the single-electron wave function of graphene as a two-component
spinor $\Psi$. This introduces an isospin quantum number. For every
momentum ${\bf k}$ lying within the Brillouin zone,
Fig.~\ref{bz},
the Hamiltonian $H$ has two eigenvalues
$\varepsilon_{{\bf k} \pm}$,
which have the same magnitude and opposite signs. The eigenvalue
$\varepsilon_{{\bf k} +} > 0 $
($\varepsilon_{{\bf k} -} < 0$) corresponds to the conduction
(valence) band of graphene.

The functions
$\varepsilon_{{\bf k} \pm}$
vanish at the six corners of the Brillouin zone:
${\bf K}_{1,2} = (0, \pm 4\pi/(3\sqrt{3}a_0))$
and
${\bf K}_{3,4,5,6}
=
(\pm 2\pi / (3 a_0), \pm 2\pi/(3\sqrt{3}a_0))$.
Here the symbol $a_0$ denotes the carbon-carbon bond length. Near
${\bf K}_i$,
$i={1,\ldots, 6}$,
the dispersion surface can be approximated by two cones with a common apex
\begin{eqnarray}
\varepsilon_{{\bf k} \pm} \propto \pm |{\bf k} - {\bf K}_i|.
\label{wd_disp}
\end{eqnarray}
The conduction and
valence bands touch each other at the cones' apex.

Of the six cones only two can be chosen to be independent: the remaining
four are connected to these two by a reciprocal lattice vector. Thus, the
cones
${\bf K}_{1, \ldots, 6}$
can be split into two equivalence classes. These classes are commonly
denoted by $K$ and $K'$, and referred to as `valleys'.

When graphene is not doped, its Fermi level passes through the cone
apexes. In such a situation, if one is interested in the low-energy
description, only the states near the cones must be accounted. For states
with energies near the cone apexes, it is possible to use the following
Weyl-Dirac equations
\begin{eqnarray}
\label{dirac}
E\Psi_{1,2}
&=&
H \Psi_{1,2},
\\
H
&=&
- i \hbar v_{\rm F}
(\sigma_y \partial_x
\pm
\sigma_x \partial_y)
=
\hbar v_{\rm F}
\left(
\matrix{
    0&  -\partial_x \pm i\partial_y \cr
    \partial_x \pm i\partial_y &    0
    }
\right),
\end{eqnarray}
which have dispersion as in
Eq.~(\ref{wd_disp}).
These equations become invalid away from the cones. The spinor wave
function
$\Psi_1$ ($\Psi_2$)
corresponds to the electron states near the cone $K$ ($K'$). The plus
(minus) sign in Eq.~(\ref{dirac}) corresponds to $K$ ($K'$). The low-energy
physics of electrons in graphene is equivalent to four species of
two-dimensional massless Dirac electrons: two different spin directions and
two cones, $K$ and $K'$, giving overall fourfold degeneracy.

Pristine undoped graphene is a gapless semiconductor. This means
that its density of states does not have a gap, but vanishes
linearly when the energy approaches the apexes. Sometimes it is
desirable to open a gap in the graphene spectrum. As shown in
Table~\ref{energy_gap} this can be achieved by employing various
mechanical, electronic, and/or chemical methods. In particular, in
monolayer graphene the gap can be induced by substrate or strain
engineering~\cite{Ni2008,low_gap_strain}, as well as by deposition
or adsorption of molecules on the graphene layer, such as, for
example, water and ammonia~\cite{Ribeiro2008}. Based on numerical
studies, the value of the energy gap can range from a few meV to
hundreds of meV. Most importantly it can be larger than room
temperature as required for graphene-based transistors. In bilayer
graphene the gap can be induced and continuously tuned, for
instance, chemically through selective doping~\cite{Ohta2006}, or
even electrically by applying gate voltages~\cite{Zhang2009}. The
fact that graphene's band structure can be controlled externally
and with rather simple processes is a nontrivial result which
reveals the potential of graphene for nanotechnology.

\begin{table}
  \centering
\begin{tabular}{||c|c|c||}
    \hline
    \hline
    \multicolumn{3}{||c||}{\textbf{Inducing an energy gap in graphene}}\\
     \hline
     \hline
  Method & \quad Gap in monolayer (meV) \quad & \quad Gap in bilayer (meV) \quad \\
    \hline
    \hline
 \quad \quad Nanoribbons$^{*}$ (width $\sim$ 15 nm) ~\cite{han_experiment_gap} \quad \quad & 200 & \\
 BN-\textit{h} / Cu(111) substrate ~\cite{Giovannetti2007} &   53 / 11 &\\
 SiC substrate$^{*}$ ~\cite{Zhou2007} & 260  & \\
 External square superlattice ~\cite{Tiwari2009}&  65 & \\
 Strain engineering ~\cite{Ni2008} &  300  & \\
 Adsorption of molecules~\cite{Berashevich2009} & 2$\times 10^{3}$ &  \\
 Graphene covered by H$_{2}$O / NH$_{3}$ ~\cite{Ribeiro2008} & 18 / 11  & 30 / 42 \\
 Nanoribbons$^{*}$ (width $\sim$ 30 nm) ~\cite{Szafranek2010}&  & 50 \\
 Electrical gates$^{*}$ ~\cite{Zhang2009}  & & 250 \\
 Selective doping$^{*}$ (potassium) ~\cite{Ohta2006} & & 100 \\
 Electric field effect$^{*}$ ~\cite{Castro2007} &  &  150  \\
\hline\hline
\end{tabular}
\caption{Brief summary of possible methods to induce an energy gap
in monolayer and bilayer graphene. Asterisks ($^*$) indicate
experimental demonstrations; otherwise the value of the gap is a
theoretical prediction. In some cases, the energy gap is tunable
and its exact value critically depends on the details of the
specific method. Here, `BN-\textit{h}' denotes boron nitride in
the hexagonal configuration.}\label{energy_gap}
\end{table}

\section{Edges of graphene samples}
\label{edge}

The characteristics of a mesoscopic device depend substantially on its
edges. Therefore, it is important to study the electron behavior near the
graphene edge.

\subsection{Edge-stability issues}

Two kinds of edges are often discussed in the literature: zigzag and
armchair. A form of the zigzag is the Klein edge
\cite{klein_edge}.
All three types are shown in
Fig.~\ref{graphene_lattice}.
They are the most symmetric variants of edges in graphene. More complicated
edges were also studied
\cite{complicated_edges,nakada-fujita_nribb_edge_st,
dirac_boundary_cond_graphene,tkachev_zigzag_states_finite_length,
gan_edge_stability}.

\begin{figure}[btp]
\centering \leavevmode \epsfxsize=10.5cm
\epsfbox{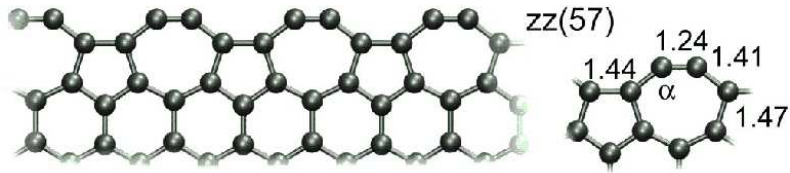}
\caption[]
{\label{reczag}
Reczag, or ZZ 57, edge of a graphene sheet,
from \cite{koskinen_reczag2}.
In the right panel the edge unit cell is shown. It consists of a pentagon
and a heptagon. The latter polygons are the reason why this edge type is
called `57'. Numbers in the right panel are the bond lengths in \AA.
Reprinted figure with permission from 
P. Koskinen, S. Malola, and H. Hakkinen, Phys. Rev. Lett. {\bf 101},
115502 (2008).
Copyright (2008) by the American Physical Society.
}
\end{figure}

Of course, in a laboratory sample some of these edge types may be unstable
chemically or undergo reconstruction. The possibility of the edge
reconstruction has been addressed in several publications. Most
importantly, it appears that the pristine zigzag edge is unstable:
recently, it was predicted on the basis of density-functional theory (DFT)
calculations
\cite{koskinen_reczag2}
that it might undergo reconstruction at room temperature, and become a {\it
reczag}
(short for `reconstructed zigzag'
\cite{koskinen_reczag},
see Fig.~\ref{reczag}).
This kind of edge is often called `ZZ 57'; namely, it is a version of zigzag
(thus ZZ) edge, in which the edge hexagons are replaced by pentagons and
heptagons (hence the 5 and 7). Experimental data supporting the existence
of the reczag edge were presented in
\cite{koskinen_reczag}.
A similar conclusion was reached in
\cite{gan_edge_stability}:
the energy of the zigzag edge is substantially higher than the energy of
the reczag.
In Refs.~\cite{wassmann_edge_stab,wassmann_phys_stat_sol}
the non-hydrogenated zigzag edge was not listed among stable configurations.

The conclusions of Ref.~\cite{koskinen_reczag2}, regarding the
relative stability of zigzag and reczag, were challenged in
Ref.~\cite{magnet_p-orbit}, where the experiment of
Ref.~\cite{ritter_lyding}, proving the existence of the zigzag
edge in a laboratory sample, was quoted. The experimental
demonstration of the zigzag edge stability was also reported in
Ref.~\cite{girit_zigzag_stab}. Results of Ref.~\cite{mol_dyn} are
also in disagreement with Ref.~\cite{koskinen_reczag2}. However,
the authors of Ref.~\cite{mol_dyn} were unsure whether their
molecular dynamics simulations can provide a reliable answer to
the question of the edge stability.

\begin{figure*}
\centering
\leavevmode
\epsfxsize=17cm
\epsfbox{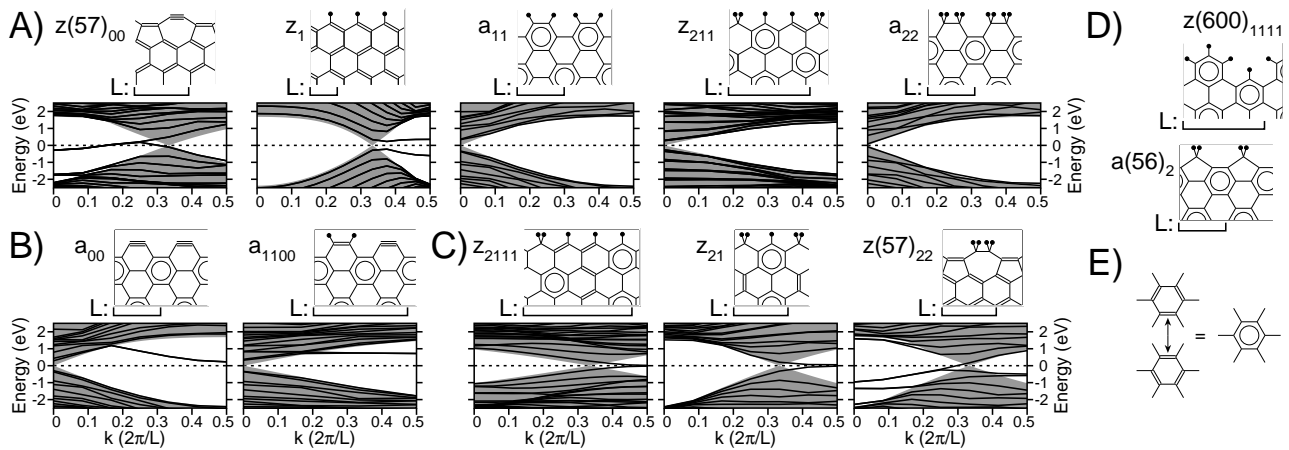}
\caption[]
{\label{wass_table}
Stable types of graphene edges, from
Ref.~\cite{wassmann_edge_stab}.
Row A of the figure shows the five most stable configurations of the
graphene edge with and without hydrogen attached to the unsaturated bonds.
There, hydrogen is represented by small black circles. Rows B and C show other
stable armchair and zigzag terminations. The DFT calculations reported in
Ref.~\cite{wassmann_edge_stab}
predict that monohydrogenated zigzag denoted as z$_1$ in the figure and
reczag denoted as z(57)$_{00}$ are stable only at extremely low hydrogen
concentrations. At standard atmospheric conditions, a$_{11}$, z$_{211}$,
and a$_{22}$ are the most stable types of edges. Note that the pristine
zigzag edge (studied in numerous papers) is {\it not} listed as a stable
configuration. More complicated types of the graphene terminations are
presented in panel~D. In panel~E the representation of the benzenoid
aromatic carbon ring as a superposition of two degenerate Kekule
configurations is shown.
Reprinted figure with permission from 
T. Wassmann, A. P. Seitsonen, A. M. Saitta, M. Lazzeri, and F. Mauri,
Phys. Rev. Lett. {\bf 101}, 096402 (2008).
Copyright (2008) by the American Physical Society.
}
\end{figure*}

The chemical stability was also investigated. It was pointed out in
Refs.~\cite{wassmann_edge_stab,wassmann_phys_stat_sol}
on the basis of DFT calculations that the reczag and armchair are stable
only when the concentration of hydrogen in the surrounding media is very
small. If this is not the case, other types of edges, with hydrogen atoms
attached, are stabilized (see
Fig.~\ref{wass_table}).
The results of DFT are consistent with Clar's theory of the aromatic sextet
\cite{clar_poly,clar_sextet}.

The DFT calculations of
Ref.~\cite{gan_edge_stability}
demonstrated that the energy of zigzag, armchair, reczag, and more
complicated regular edges always decreases upon monohydrogenation. This
agrees with
Ref.~\cite{wassmann_edge_stab}:
when enough hydrogen is present in the surrounding media, hydrogenation of
the edge occurs.

These results suggest that the edge stability is a complicated problem in
graphene. The edge stability depends on the orientation of the edge and is
affected by the chemical environment.

\begin{table}
\begin{tabular}{||c|c|c|c|c||}
\hline\hline
\multicolumn{5}{||c||}{\textbf{Graphene edges}} \cr
\hline\hline
          &  Zigzag &  Armchair &\quad Klein \quad \quad &  Reczag \cr
\hline\hline
\quad Stability \cite{koskinen_reczag2} \quad \quad
         & Unstable & Stable    &        & Stable  \cr
          &  (to reczag)&       &        &         \cr
\hline
Edge states & Yes       & No    & Yes    &         \cr
\hline
\quad Magnetism \cite{wassmann_edge_stab} \quad \quad & Ferromagnetic &
                                                    No   &        & No \cr
\hline
Stress \cite{edge_stress}
         &  \quad Compression \quad \quad& \quad Compression \quad \quad &   & \quad
Tension (weak)\quad \quad \cr
\hline\hline
\end{tabular}
\caption{ Different properties of graphene edges. In addition to the three
types presented in
Fig.~\ref{graphene_lattice}
(zigzag, armchair, and Klein edges), a {\it reconstructed zigzag} (reczag)
edge
\cite{koskinen_reczag}
is now included in this comparison.
}
\label{compare_edges}
\end{table}

\subsection{Electrons near edges}

The simplest way to describe an electron near the edge is to
resort to the Weyl-Dirac equation (\ref{dirac}) with appropriate
boundary conditions. One has to keep in mind that the realistic
boundary condition depends on a variety of factors: the
orientation of the edge, deformation of the chemical bonds near
the edge, edge reconstruction, and possible chemical
functionalization of the unsaturated bonds. Theoretical studies of
these conditions were performed in several papers
\cite{dirac_boundary_cond_nanotube,dirac_boundary_cond_graphene,
complicated_edges,volkov_zagorodnev_bc,volkov_zagorodnev_bc2}.

The physics of electrons near the armchair edge is simple: the edge
always acts as a reflector of the incident electron current. The scattering
is affected by details of the edge structure, such as C-C bond lengths near
the edge and non-carbon radicals attached to the edge. Some additional
details are provided in
Appendix~\ref{appendix::armchair}.

\begin{figure}[btp]
\centering
\leavevmode
\epsfxsize=8.5cm
\epsfbox{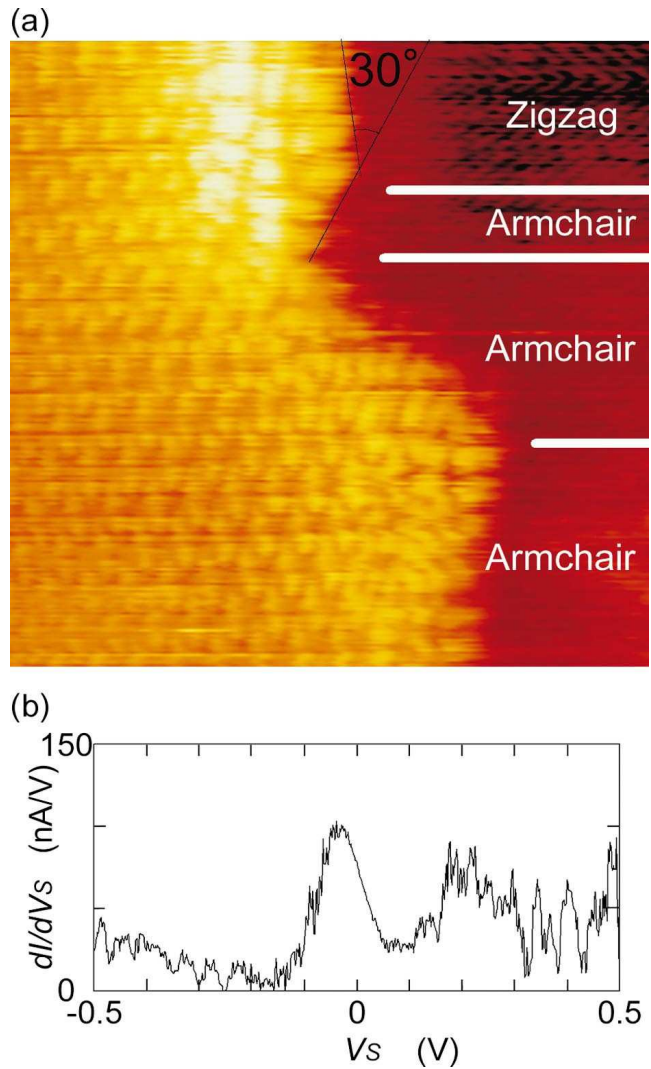}
\caption[]
{\label{zigzag_dos}
(Color online) Scanning tunneling microscope image of different graphene
terminations, from
Ref.~\cite{kobayashi_ldos}.
(a) The termination of a graphene sample is investigated using a scanning
tunneling microscope. Fragments of both zigzag and armchair types are
identified. The edge states near the zigzag edge are clearly visible as
stripes of bright spots stretching along the zigzag edge. No edge states are
present near the armchair termination. (b) A typical dependence of the
differential conductance near the zigzag edge is plotted. The peak near
the zero voltage
($V_{\rm s} = 0$)
corresponds to the edge states.
Reprinted figure with permission from 
Y. Kobayashi, K.-i. Fukui, T. Enoki, K. Kusakabe, and Y. Kaburagi,
Phys. Rev. B {\bf 71}, 193406 (2005).
Copyright (2005) by the American Physical Society.
}
\end{figure}

The physics of Klein and zigzag edges, however, is quite different. These
edges bind electrons. When the nearest-neighbor hopping Hamiltonian is used
to describe graphene, the bound eigenstates (edge states) form a
dispersionless band at the zero of energy (see
Appendix~\ref{appendix::zigzag}).
These edge states can be observed experimentally as a peak in the local
density of states
\cite{niimi_ldos,kobayashi_ldos,ritter_lyding}.
For example,
Fig.~\ref{zigzag_dos}
shows scanning tunneling microscopy data from
Ref.~\cite{kobayashi_ldos}.
There, the edge states are seen near the zigzag edge as a stripe of bright
spots extending along the edge.

The dispersionless band of the bound states is unstable with respect to
different perturbations of $H$. For example, the inclusion of longer-range
hopping makes the band disperse
(Refs.~\cite{el_prop_disordered_graphene,sasaki_nnn,sasaki_gague_nnn,
decomposition,volkov_zagorodnev_bc,volkov_zagorodnev_bc2}
and Appendix~\ref{appendix::zigzag}).

The most interesting way of lifting the degeneracy of edge states is by
adding electron-electron interactions to $H$. It was predicted quite some
time ago
\cite{kusakabe_magnetizm_edge_state}
that magnetic correlations develop at a zigzag edge as a result of the
interaction. This effect was investigated in several papers
\cite{yazayev_magnetic_edge_states,
kumazaki_magnetic_edge_states,
sasaki_magnetic_edge_states,
wunsch_zigzag_mag}
%

For example, a detailed DFT study was reported in
\cite{yazayev_magnetic_edge_states}.
It predicts that an isolated graphene zigzag edge is a ferromagnet with
magnetic moment $m$ of 0.3 of the Bohr magneton per unit cell of the zigzag
edge.  This may be understood qualitatively as follows. The spin coupling
between nearest-neighbor carbon atoms is antiferromagnetic. Thus, different
sublattices have opposite magnetic momenta. In the bulk, this would lead to
cancellation of the total moment. Near the edge, however, the electron
density at the exposed row is higher that at rows located deeper into the
bulk. Most importantly, all sites at the zigzag edge belong to the same
sublattice; therefore, they have the same magnetic moment. This leads to a
local imbalance of the total magnetic moment, which is seen as edge
ferromagnetism.

\begin{table}
\begin{tabular}{||c|c|c|c||}
\hline\hline
\multicolumn{4}{||c||}{\textbf{Zigzag edge magnetism}}\cr
\hline\hline
$M$,        &  $\xi(T)$         &  $\xi(T)$        & \quad Anisotropy\quad \quad \cr
\quad per unit cell\quad \quad
        &\quad $T = 300$ K \quad\quad
                &\quad $T < 10$ K\quad \quad &           \cr
\hline\hline
$0.3\ \mu_{\rm B}$
            & $\sim 1$ nm          &  $\sim 1$ $\mu$m     & Ising, $10^{-4}$ \cr
\hline\hline
\end{tabular}
\caption{Summary of the magnetic properties of the graphene zigzag edge, as
reported
in Ref.~\cite{yazayev_magnetic_edge_states}.
When the electron-electron interaction is taken into consideration, the
edge-state degeneracy is lifted through the magnetization of the
electrons near the edge. This creates a one-dimensional magnetic
system. The magnetic momentum $M$ of such system is 0.3 of the Bohr's
magneton per zigzag edge unit cell. The correlation length $\xi$ at room
temperature is rather short, suggesting that it would be difficult
to utilize the pristine zigzag edge in a spintronic device
operating at room temperature. However, below $T_x = 10$\ K, a crossover to
Ising-like magnetic correlations occurs, and the correlation length
increases exponentially upon approaching $T=0$. It was proposed
\cite{yazayev_magnetic_edge_states}
that $\xi$ could be as large as a micrometer.
}
\label{edge_magnetism}
\end{table}

Reference~\cite{yazayev_magnetic_edge_states}
reports many properties of the edge ferromagnetism
(see also Table~\ref{edge_magnetism}):
spin-wave dispersion
($E = \kappa q^2$, where $\kappa = 320 \,{\rm meV \AA^2}$),
stiffness
($D = 2\kappa/m = 2100 \,{\rm meV \AA^2}$),
magnetic anisotropy
($\sim 10^{-4}$),
and crossover temperature between the Heisenberg and the Ising regimes
($T_x \sim 10$\,K).
The spin correlation length at room temperature is estimated to be of the
order of one nanometer. At temperatures below $T_x$ it increases
exponentially as the temperature decreases. The effects of edge disorder on
the magnetic properties of the zigzag edge were also investigated.

In Ref.~\cite{kumazaki_magnetic_edge_states}
magnetic properties of small fragments of zigzag edge were studied using
the Hubbard model. Such model is relevant for systems with rough edges,
consisting of alternating fragments of different terminations. It is
demonstrated that a very short, of the order of three lattice constants,
zigzag sequence is sufficient to generate a local magnetic moment.

Some other interacting effects have also been studied. For example, the
influence of the long-range Coulomb interaction and doping on the edge
magnetism is discussed in
Ref.~\cite{wunsch_zigzag_mag}
using the Hubbard model with Coulomb interactions.
The interaction of edge states with phonons was studied in
Ref.~\cite{sasaki_phonons}.

\subsection{Graphene/graphane interface}
\label{interface}

\begin{figure}[btp]
\centering
\leavevmode
\epsfxsize=8.5cm
\epsfbox{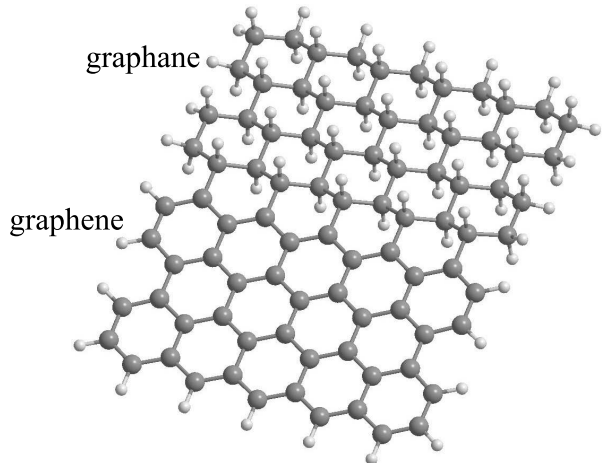} \caption[] {\label{gg_interface}
Graphene/graphane interface, as studied in
Ref.~\cite{openov_gr_interface}.
Larger dark balls correspond to carbon atoms, while the smaller
light balls correspond to hydrogen. The lower part of the sample
that is shown is graphene, while the higher part is graphane. In
bulk graphane every carbon atom has a hydrogen atom attached to
it. }
\end{figure}

Graphane
\cite{sofo_graphane,elias_graphane}
is a hydrogenated sheet of graphene. Unlike graphene, graphane is a
semiconductor with a gap of the order of few eVs. The graphene/graphane
interface, shown in
Fig.~\ref{gg_interface},
can be viewed as a type of graphene edge: low-lying electron states in
graphene decay exponentially inside the gapped media of graphane.
As demonstrated by molecular dynamics simulations
\cite{openov_gr_interface},
the interface remains almost atomically sharp even at sufficiently high
temperatures, which is an extremely attractive feature since it reduces
scattering and simplifies the theoretical description.

Depending on the orientation of the interface relative to the
crystallographic axis of graphene, one can distinguish a zigzag-type
interface (as in
Fig.~\ref{gg_interface}),
or an armchair-type (at the right angle to the zigzag). The zigzag
interface supports edge states whose electronic and magnetic properties
were investigated in
Ref.~\cite{schmidt_gg_interface}.

Besides hydrogenation, graphene may be subjected to fluorination
in order to produce fluoridated graphene
\cite{cheng_flouridation_2010,nair_fluorographene,robinson_fluoridation,
withers_fluoridation,xiang_hydro_fluo}. Like graphene, the latter
is a semiconductor with a gap of the order of a few eV. A similar
conversion occurs upon functionalization of graphene by
nitrophenyl \cite{nitrophenyl}. The properties of the interface
between pure graphene and the functionalized material must be
similar to the properties of the graphene/graphane interface.

\subsection{Fabrication of high-quality edges}

Most of the theoretical work so far has assumed that the edges of the
nanostructures are atomically perfect. Needless to say, this is not easy to
realize experimentally. However, recently, substantial progress in the area
of high-quality edge fabrication has been achieved
(e.g., \cite{sharp_edges,current_edge_rect}).

In Ref.~\cite{sharp_edges}
a chemical method of deriving narrow graphene stripes with sharp edges was
reported. A graphene sample was placed in a solvent and subjected to
sonification. Strips with sharp edges and widths varying from 50~nm to
sub-10~nm were extracted from the solution. The strips produced were
used to fabricate a field-effect transistor-like device.

In Ref.~\cite{current_edge_rect}
it was experimentally demonstrated that during Joule heating of the
graphene sample with disordered edges carbon atoms at the edge were
vaporized, and sharp edges were stabilized. Model calculations shown that
the edge defects were healed through point defect annealing and edge
reconstruction. This process was modelled in
Ref.~\cite{engelund_reconstruction}.
These findings suggest that many theoretical predictions dependent on the
edge quality could be tested experimentally.

\section{Graphene nanoribbons}
\label{nanoribbon}

Nanoribbons, which are strips of graphene, are among the most studied
mesoscopic graphene structures. There are several reasons for this. First,
they demonstrate unusual physical properties, for example, edge states,
which might be used in future spintronics applications. Second, nanoribbons
are easy to produce and demonstrate excellent transport properties.

Third, they have an energy gap in their single-electron spectrum. This gap
is a consequence of the electron confinement, and it is inversely
proportional to the width of a nanoribbon
\cite{han_experiment_gap,chen_experiment_gap}.
This suggests that, at least in principle, a nanoribbon with a desired
value of the gap may be fabricated. Finite gap and high mobility are both
very useful for the design of field-effect transistors (FET): they allow
for large on/off ratios, small losses, and high operating frequencies. For
example, a nanoribbon-based FET realized in
Ref.~\cite{wang_room_t_fet}
demonstrated an on/off ratio of about 10$^6$ at room temperature. Other
nanoribbon-FET devices were described
in Ref.~\cite{Liao2010,fet_nribbon}.
(Carbon nanotubes also have a gapped spectrum. However, their fabrication
process is much more involved).

Nanoribbons are usually classified by their type of edge; for instance,
there are zigzag and armchair nanoribbons. Nanoribbons may also have
disordered
\cite{cresti_disord_review}
or
more complicated regular types of edges
\cite{barone_semicond_nanoribbon}.

\subsection{Zigzag nanoribbons}
\label{zigzag_nanoribbon}

\begin{figure}[btp]
\centering
\leavevmode
\epsfxsize=8.5cm
\epsfbox{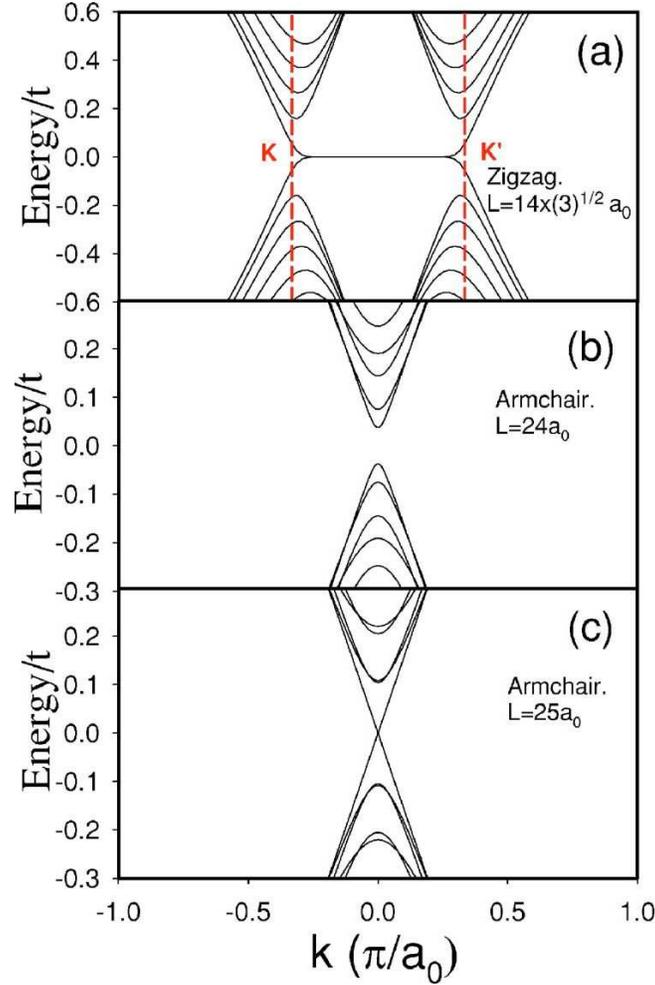}
\caption[]
{\label{nanoribbon_spectr}
(Color online) Single-electron energy spectrum of a nanoribbon calculated
with the help of the Weyl-Dirac
equation~(\ref{dirac}),
from
Ref.~\cite{brey_fertig_dirac_eq_nb}.
Due to the finite width of a nanoribbon, the transverse momentum is
quantized. Therefore, the nanoribbon's spectrum consists of a number of
branches corresponding to different values of the quantized momentum. Since
the graphene lattice is anisotropic, the spectrum of a zigzag nanoribbon
[panel (a)] differs in several respects from the spectrum of an armchair
nanoribbon [panels (b) and (c)]. For the zigzag nanoribbon the remnants of
two Dirac cones, $K$ and $K'$, are visible. The almost flat branch
connecting $K$ and $K'$ corresponds to the edge states. It acquires a very
weak dispersion due to interference of the edge states (exponentially)
localized at the opposite edges of the nanoribbon. For the armchair
nanoribbon both $K$ and $K'$ `coalesce' together. According to the
calculations of
Ref.~\cite{brey_fertig_dirac_eq_nb},
the armchair nanoribbon may be either a semiconductor with a small gap
[panel (b)], or a gapless metal [panel (c)]. The size of the gap depends on
the nanoribbon's width. However, more elaborate treatments accounting for
the electron-electron interaction
\cite{sandler_coulomb_gap,sols_coulomb_blockade},
electron-lattice interaction
\cite{lattice_distortion,son_gap,our_nanoribbon_paper_2009},
or longer-range hopping
\cite{white_gap_longer_range,gunlycke_nanoribbon_gap}
proved that the gap is always non-zero.
Reprinted figure with permission from 
L.~Brey and H.~Fertig, Phys. Rev. B {\bf 73}, 235411 (2006).
Copyright (2006) by the American Physical Society.
}
\end{figure}

Many interesting properties of zigzag nanoribbons are related to the
presence of edge states in the nanoribbon electron spectrum. These states
may be derived, together with other low-lying states, with the help of
Eq.~(\ref{dirac})
supplemented by a boundary condition suitable for the zigzag edge
\cite{brey_fertig_dirac_eq_nb}.
The resultant spectrum is shown in
Fig.~\ref{nanoribbon_spectr}a.
Two almost dispersionless branches connecting the Dirac cones, $K$ and
$K'$, correspond to the edge states. They are analogous to the edge states
discussed in
Sec.~\ref{edge}.

Similar conclusions about edge states may be reached using first-principle
calculations. These reveal that the zigzag nanoribbon is a
semiconductor with a width-dependent gap
\cite{son_half_metal,pisani_monohydrogen,yang_armchair_gw}.
The lowest energy states are edge states.

\subsubsection{Edge magnetism}

Edge states are responsible for magnetism in zigzag nanoribbons. Edge
magnetism is an interesting feature with potential spintronic applications;
since the edge-state branch is both magnetized and able to carry current,
it can be used to couple spin magnetization and current. This property may
be used to control the magnetization with current or vice versa.

The magnetism of a nanoribbon with pristine zigzag edges was
investigated theoretically in 1996
\cite{fujita_nanoribbon_hubbard}.
It is quite similar to the magnetism of an isolated zigzag edge: each edge
of the nanoribbon has a finite magnetization, which is induced due to the
instability of a nearly-flat edge-state band. There is a non-zero coupling
between the magnetizations of the two edges.
In Ref.~\cite{fujita_nanoribbon_hubbard}
it was shown that this coupling is antiferromagnetic, i.e., the
magnetization vectors at the opposite edges are antiparallel. Such result
is easy to understand. Consider the following two statements: $(i)$
repulsive interactions between electrons on a half-filled bipartite lattice
induce an antiferromagnetic correlation between sublattices; $(ii)$ in case
of the zigzag nanoribbon, one of its edges always terminates in atoms of
${\cal A}$
sublattice, the opposite edge terminates in atoms of
${\cal B}$.
As a consequence of $(i)$ and $(ii)$, the local antiferromagnetic tendency
is translated into weak inter-edge antiferromagnetic interactions.

Since the pristine zigzag edge is likely to be chemically unstable
\cite{wassmann_edge_stab,wassmann_phys_stat_sol},
it is therefore important to study nanoribbons with non-carbon atoms or
functional groups attached to the edges. The case of a zigzag nanoribbon
with monohydrogenated edges was discussed in
Ref.~\cite{pisani_monohydrogen}.
Such nanoribbons support edge states and have an edge ferromagnetic moment
with antiferromagnetic coupling between the edges. This is consistent with
the results of
Ref.~\cite{wassmann_edge_stab},
where an extensive list of various edge types and their properties was
presented. However, not all versions of functionalized or reconstructed
zigzag edge support magnetism.

\begin{figure}[btp]
\centering
\leavevmode
\epsfxsize=8.5cm
\epsfbox{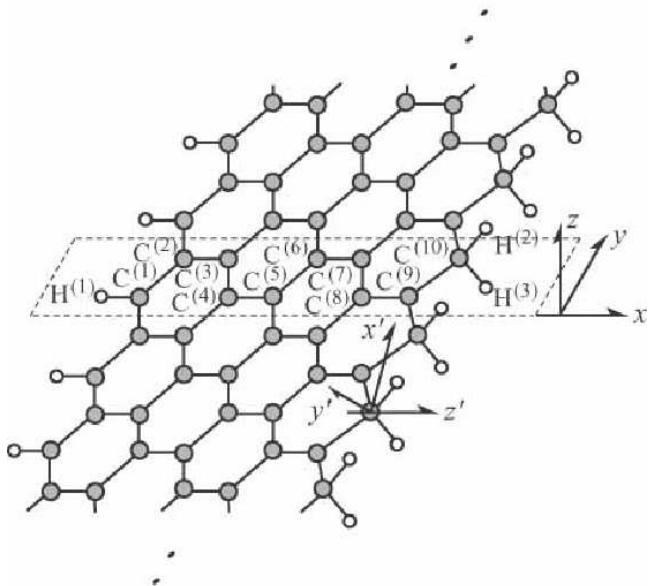}
\caption[]
{\label{kusakabe_ferro}
Zigzag nanoribbon with disparity between edges, from
Ref.~\cite{kusakabe_ferromag}.
The nanoribbon edges are parallel to $y$-axis. The left edge is
monohydrogenated, while the right edge is dihydrogenated. Here, filled
circles are carbon atoms, empty circles are hydrogen atoms. Since the
opposite edges of a pristine zigzag nanoribbon have opposite magnetic
momenta, a disparity between the two edges may induce a non-zero
magnetization of the nanoribbon. Indeed, a local-spin-density approximation
study~\cite{kusakabe_ferromag}
revealed that the nanoribbon in the figure possesses a finite magnetic
moment. This result can be easily understood qualitatively. The
monohydrogenated edge is ferromagnetic, while the dihydrogenated is
non-magnetic. Thus, the whole system is ferromagnetic.
Reprinted figure with permission from 
K. Kusakabe and M. Maruyama, Phys. Rev. B {\bf 67}, 092406 (2003).
Copyright (2003) by the American Physical Society.
}
\end{figure}

A chemical way to produce a nanoribbon with finite magnetic moment was
proposed in
Ref.~\cite{kusakabe_ferromag}:
since the opposite edges of the zigzag nanoribbon have opposite magnetic
momenta, a disparity (e.g., a non-equivalent chemical functionalization)
between the two edges may create a nanoribbon with non-zero magnetization.
Local-spin-density calculations reported
in Ref.~\cite{kusakabe_ferromag}
proved that fact for the nanoribbon whose one edge is monohydrogenated,
while the other is the dihydrogenated (see
Fig.~\ref{kusakabe_ferro}).
This result can be easily understood qualitatively. The monohydrogenated
edge is ferromagnetic, while the dihydrogenated is non-magnetic
\cite{wassmann_edge_stab}.
Thus, the whole system is ferromagnetic.

Reference~\cite{cervantes_func_mag}
presented a very detailed DFT study of the effect the edge
functionalization exerts on the zigzag nanoribbon's magnetism. The main
focus there was the monohydrogenated zigzag nanoribbon, where some of the
hydrogen atoms were replaced by other radicals. When the edges have
non-identical chemical structure (i.e., one edge is purely hydrogenated,
the other has some of its hydrogens substituted), it was determined that a
finite magnetization may be generated. The effect is particularly strong
for the oxygen substitution. This result is consistent with
Ref.~\cite{kusakabe_ferromag}.
It was also reported that such nanoribbon has different band gaps for
different spin orientations. This state can be described as a
spin-selective semiconductor.

In addition, it was shown in
Ref.~\cite{cervantes_func_mag}
that not only the magnetic properties but the band gap of a zigzag
nanoribbon is sensitive to the chemical functionalization of the edges as
well. For example, doping with oxygen may close the gap, provided that its
concentration is sufficiently high. A variety of other effects dependent on
the edge chemistry was also discussed.

\subsubsection{Half-metallicity}

\begin{figure}[btp] \centering \leavevmode \epsfxsize=12.5cm
\epsfbox{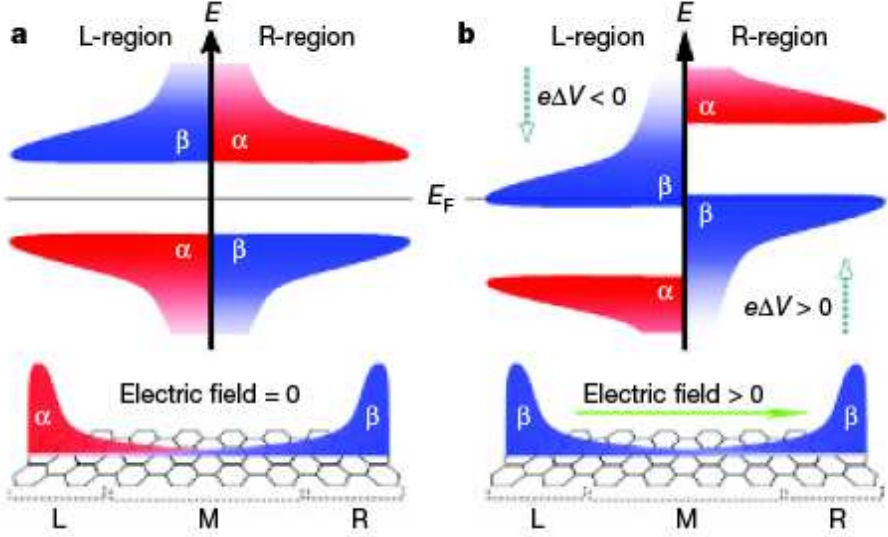} \caption[] {\label{half_metal} (Color online)
Zigzag nanoribbon without (a) and with (b) a transverse electric field,
from~Ref.\cite{son_half_metal}.
Symbols $\alpha$ ($\beta$), in red (blue), represent the
spin-up (spin-down) orientations of the edge band electrons. In the lower
part of panel (a) both orientations are present: spin-up is on the left
edge (L) of the nanoribbon, and spin-down is on the right edge (R). The
letter `M' stands for `middle' of the nanoribbon. In the upper part of (a)
the energies of different edge states are plotted. On the left edge, the
spin-up states are filled ($E<0$), and the spin-down are empty. On the
right edge, the situation is reversed. Panel (b) shows what happens when a
transverse electric field is applied. The electrostatic potential pushes
down the states on the left edge, and pushes up the states on the right
edge. As a result, the density of states at the Fermi energy is zero
(finite) for spin-up (spin-down) electrons. The system becomes half-metal:
it is a conductor (metal), since there is a finite density of
current-carrying states at the Fermi energy, yet, these states correspond
only to one spin polarization (spin-down).
Reprinted by permission from Macmillan Publishers Ltd: 
\href{http://www.nature.com/nature/index.html}{Nature},
Y.-W. Son, M. L. Cohen, and S. G. Louie, Nature {\bf 444}, 347 (2006),
copyright 2006.
}
\end{figure}

A half-metal is a conductor whose charge carriers are fully spin-polarized.
This is a very desirable property with potential applications to
spintronics because it can be used to create a fully spin-polarized
current.
Reference~\cite{son_half_metal}
suggested that the application of a transverse electric field to a zigzag
nanoribbon closes the gap for one spin orientation. The gap for another
spin orientation is increased even more (see Fig.~\ref{half_metal}).
A similar conclusion was reached in
Ref.~\cite{Hod2008}.
Note that this idea is also based on inducing a disparity between the two
edges: this time the transverse field is the agent producing the
disparity.

The proposal
\cite{son_half_metal}
summarized in the previous paragraph was disputed in
Ref.~\cite{rudberg_half_semicond},
where it was argued that, when the transverse field is applied, two spin
polarizations have different gap values. However, both gaps are finite for
any value of the electric field, producing a spin-selective semiconductor,
and not half-metal.
Reference~\cite{rudberg_half_semicond}
attributed the discrepancy to the artifacts of the computational technique
of
Ref.~\cite{son_half_metal}.

A suitable functionalization can enhance the half-metallic features of
the zigzag nanoribbon
\cite{hod_half_metal_func}.
Also,
Ref.~\cite{dutta_half_metal_chemical_mod} explored a wider range
of chemical modifications of zigzag nanoribbons in search for
robust half-metallicity. The results in
Ref.~\cite{hod_half_metal_func,dutta_half_metal_chemical_mod}
suggested that the chemical modifications may by a powerful tool for the
control of the zigzag nanoribbon's half-metallic properties.

\subsection{Armchair nanoribbons}
\label{armchair nanoribbon}

\begin{figure}[btp]
\centering
\leavevmode
\epsfxsize=8.5cm
\epsfbox{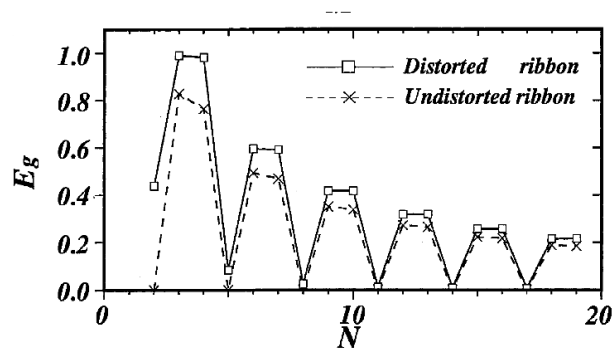}
\caption[]
{\label{armchair_gap}
Armchair nanoribbon's gap as a function of the nanoribbon's width, from
Ref.~\cite{lattice_distortion}.
The tight-binding calculation (dashed line) predicts that the gap of the
armchair nanoribbon vanishes periodically as a function of the nanoribbon's
width. However, the numerical calculations within a more elaborate model,
which allows for deformation of the carbon-carbon bonds, prove that the gap
is non-zero (albeit very close to it) for any width (solid line), although,
the dependence on the width remains oscillatory. An analytical demonstration
of the instability of the zero-gap state is given in
Ref.~\cite{our_nanoribbon_paper_2009}.
This instability occurs because the increase of the elastic energy due to
the lattice deformation is smaller than the decrease of the electron
kinetic energy due to the gap opening.
Figure is reprinted from:
M. Fujita, M. Igami, and K. Nakada, J. Phys. Soc. Jpn. {\bf 66}, 1864 (1997).
}
\end{figure}

The electron properties of the armchair nanoribbon are simpler than those of the
zigzag nanoribbon. Both tight-binding model and Weyl-Dirac equation
calculations show that graphene armchair nanoribbons with pristine edges
may be either in a semiconducting (finite gap,
Fig.~\ref{nanoribbon_spectr}b),
or metallic (zero gap,
Fig.~\ref{nanoribbon_spectr}c),
state with the gap oscillating as a function of the nanoribbon's width. A
more accurate numerical study
\cite{lattice_distortion}
from 1997, which allowed for deformation of the carbon-carbon bonds
dangling at the edges, proved that the metallic state is unstable: the
dangling bonds deform, inducing a finite gap in the electron spectrum.
Thus, according to
Ref.~\cite{lattice_distortion},
the armchair nanoribbon is always a semiconductor (see
Fig.~\ref{armchair_gap}).

The above line of reasoning was generalized in
Ref.~\cite{our_nanoribbon_paper_2009}, where it was shown
analytically that the metallic state of an armchair nanoribbon is
generically unstable: the edge bond instability is only one
possibility. It was also discussed
\cite{our_nanoribbon_paper_2009}
how the electron gap of a {\it finite-length} armchair nanoribbon can be
effectively closed with the help of chemical modifications of the
nanoribbon's edge.

A gap may also be generated by electron-electron interactions
\cite{sandler_coulomb_gap},
or longer-range hopping
\cite{white_gap_longer_range,gunlycke_nanoribbon_gap}.
In several papers, the electronic gap was determined with the help
of first-principles techniques.
Refs.~\cite{barone_semicond_nanoribbon,son_gap} reported DFT
calculations of the gap for the armchair nanoribbon with different
widths, with both pristine and monohydrogenated armchair edges. The
results of Ref.~\cite{barone_semicond_nanoribbon} for the gap are
summarized in Table~\ref{armchair_edge}.
References~\cite{han_experiment_gap,chen_experiment_gap}
reported the experimental measurement of the gap. The experimental value
for the gap was found to be consistent with the results of DFT
calculations. However,
Ref.~\cite{yang_armchair_gw}
claimed that DFT underestimates the gap, and the use of the so-called
{\it GW} approximation \cite{mahan} is more appropriate. {\it GW} values of
the gap are significantly higher than DFT values.

The effect of the edge functionalization on the spectral gap was studied in
Ref.~\cite{cervantes_func_mag}.
The gap was found to be robust against functionalization. This is different
from the case of zigzag nanoribbons whose gap is very sensitive to the
chemical structure of the edges (see
subsection~\ref{zigzag_nanoribbon}).

Thus, it is expected on the basis of theoretical studies that an armchair
nanoribbon is a semiconductor with a width-dependent gap, whose value is
rather insensitive to the edge chemical structure.

\begin{table}
\begin{tabular}{||c|c|c|c||}
\hline\hline \multicolumn{4}{||c||}
{\textbf{Armchair nanoribbon energy gap}} \cr
\hline\hline 
\quad Energy gap [eV]   \quad\quad   
	&  \quad Bulk semiconductors with similar values of the gap \quad \quad
		& \quad Width [nm]  \quad\quad  
			& \quad Width [$\sqrt{3}a_0$] \quad \quad\cr
\hline\hline
0.7         & Ge, InN &  2--3          &  8--12               \cr
\hline
from 1.1 to 1.4     &Si, InP, GaAs &  1--2          &  4--8                \cr
\hline\hline
\end{tabular}
\caption{Gap values for the armchair nanoribbons of different widths found
using the density functional theory,
from Ref.~\cite{barone_semicond_nanoribbon}.
The quantity
$\sqrt{3} a_0$
represents the lattice constant. The gap of the armchair nanoribbon
oscillates with its width; however, the gap is never zero. The details of
the oscillations depend on the edge functionalization (for example, in
Ref.~\cite{barone_semicond_nanoribbon},
both pristine and monohydrogenated armchair nanoribbons are discussed). The
largest values of the gap are quite insensitive to functionalization. In
order to have a gap value similar to known bulk semiconductors, a very
narrow armchair nanoribbon must be used. These results were reconsidered
in Ref.~\cite{yang_armchair_gw},
where it was claimed that density functional theory underestimates the gap,
and that the use of the so-called {\it GW} approximation
\cite{mahan}
is more appropriate. Within the {\it GW} framework, the values of the gap
substantially increase.}
\label{armchair_edge}
\end{table}

\subsection{Nanoroads}
\label{nanoroads}

\begin{figure}[btp]
\centering
\leavevmode
\epsfxsize=8.5cm
\epsfbox{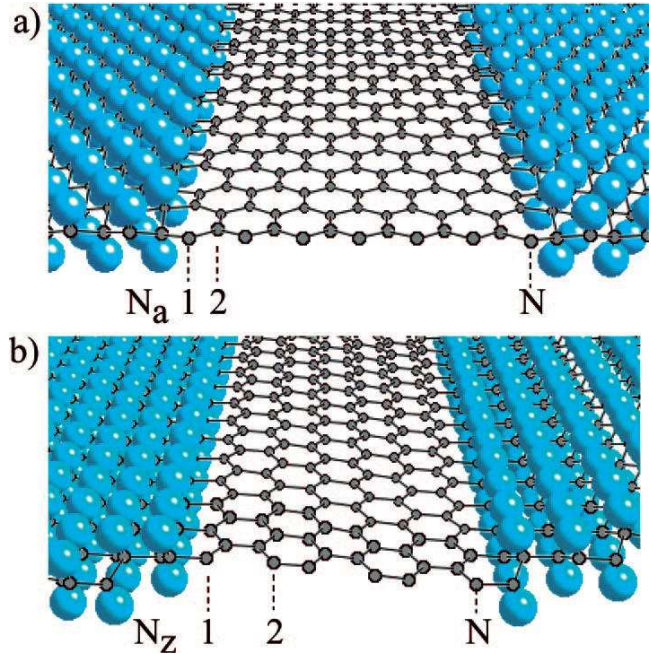}
\caption[]
{\label{nanoroad}
(Color online)
Nanoroads (i.e., graphene nanoribbons bounded by two graphene/graphane
interfaces), from
Ref.~\cite{singh_nanoroad}.
Small dark balls are carbon atoms, while the large blue balls are hydrogen
atoms. The armchair nanoroad is on panel~(a), while the zigzag nanoroad is
on panel~(b). The quantities
$N_a = 1, \ldots, N$
and
$N_z = 1, \ldots, N$
characterize the width of the nanoroads. For the nanoroad on panel~(a)
$N=13$,
while
$N=6$
for the nanoroad on panel~(b).
Reprinted with permission from:
K. Singh and B. I. Yakobson, Nano Lett. {\bf 9}, 1540 (2009).
Copyright 2009 American Chemical Society.
}
\end{figure}

So far we have assumed that graphene nanoribbons are formed by cutting a
piece of graphene into a narrow strip. Another way to define a nanoribbon
was proposed in
Ref.~\cite{singh_nanoroad}:
to sculpture a graphene nanoribbon by removing hydrogen atoms along a
narrow strip inside a wider graphane sample, as shown in
Fig.~\ref{nanoroad}.
In such a case, a nanoribbon, called nanoroad in
Ref.~\cite{singh_nanoroad},
is bound by two graphene/graphane interfaces, which are discussed in
subsection~\ref{interface}.

As it was established in
Ref.~\cite{openov_gr_interface},
the graphene/graphane interface remains almost atomically sharp even at
high temperatures. This makes nanoroads a promising candidate to observe
ballistic transport.

The magnetic and electronic properties of graphene nanoroads, including the
effects of spin-orbit coupling, were discussed in
Refs.~\cite{singh_nanoroad,tozzini_nanoroad,Xiang2009,schmidt_nanoroad}.
In
Ref.~\cite{singh_nanoroad}
the electronic structure of nanoroads was studied using DFT. It was found
that the armchair-type nanoroads are semiconducting. As for zigzag
nanoribbons, when they are wide enough they demonstrate edge magnetism.
Zigzag nanoribbons are semiconducting in the antiferromagnetic state and
metallic in the ferromagnetic state.

Several other works also used DFT to analyze nanoroad properties.  In
Ref.~\cite{tozzini_nanoroad},
the zigzag nanoroad stability and electronic structure were investigated,
and it was established that even extremely narrow zigzag nanoroads are
stable. Narrow nanoroads are always semiconducting due to Peierls
instability,
which opens a gap. A similar mechanism is responsible for the gap in
polyacetylene.
Reference~\cite{Xiang2009}
studied the adsorption of hydrogen on a graphene nanoribbon, also
formulating the rules governing such adsorption. It was proposed to use
such process in order to create narrow nanoroads.

Reference~\cite{schmidt_nanoroad} showed that, due to enhanced
spin-orbit coupling at the interface, a nanoroad might be used to
convert spin polarization into valley polarization and vice versa.
Such a device can operate at temperatures of about 1\,K.

The theoretical research summarized in this subsection indicates that
nanoroads may be an attractive alternative to usual nanoribbons, able to
sustain ballistic propagation of electrons, and also exhibit unusual spin
features.

\subsection{Transport properties of nanoribbons}
\label{nribb_transport}

For applications, such as FET
\cite{Liao2010,wang_room_t_fet},
the transport properties of graphene nanoribbons must be investigated.
A study of the conductance through a pristine nanoribbon within the
framework of the Landauer formalism was presented in
Ref.~\cite{peres_landauer}
(see also the review
\cite{peres_review}).
In such a case, the nanoribbon's conductance is quantized: it changes in
discrete steps when the gate voltage is varied. For zigzag nanoribbons it
was found that, when the gate potential is tuned to the charge neutrality
point (i.e., the Fermi energy is at the apex of the Dirac cone), the
conductance is finite due to the edge states. These are the only
current-carrying modes under such conditions.
Reference~\cite{peres_landauer}
studied the transport through metallic (zero-gap) armchair nanoribbons.
Since it is understood now that, strictly speaking, all armchair
nanoribbons are semiconducting (see
subsection~\ref{armchair nanoribbon}),
the results obtained for such objects are valid as long as the gap may be
neglected, e.g., when the temperature exceeds the gap. Another study of
electron transport through a disorder-free short-and-wide nanoribbon was
presented in
Ref.~\cite{tworzydlo_landauer}.
Its findings were compared well with the experiments in
Ref.~\cite{miao_nribb_transp_exp}.
It was concluded in
Ref.~\cite{miao_nribb_transp_exp}
that the electron propagation in graphene is ballistic up to lengths of the
order of 1\,$\mu$m. 

However, in a typical experimental situation ballistic propagation can be
spoiled both by edge disorder
\cite{areshkin_transport,evaldsson_edge_disorder_nribb,gunlycke_edge_disorder,
mucciolo_transport}
and bulk disorder
\cite{areshkin_transport,mucciolo_transport}.
For bulk disorder, it was found that electron transport is rather
insensitive to long-range disorder. Yet, when the disorder becomes
short-range, it leads to Anderson localization and to destruction of the
ballistic propagation
\cite{areshkin_transport,mucciolo_transport}.

If a nanoribbon is sufficiently narrow, the edge disorder is
important
\cite{areshkin_transport,evaldsson_edge_disorder_nribb,gunlycke_edge_disorder,
mucciolo_transport}. The armchair nanoribbons are more sensitive
to edge disorder than the zigzag nanoribbons
\cite{areshkin_transport,mucciolo_transport}.
Reference~\cite{evaldsson_edge_disorder_nribb} reported that when
edge disorder is present, the difference between transport
properties of the armchair and zigzag nanoribbons disappears.
These results suggest that to experimentally produce a ballistic
nanoribbon one has to overcome very stringent limitations on the
edge purity \cite{mucciolo_transport}. However, for a
semiconducting armchair nanoribbon of finite length one can
optimize the width so that the localization effects are, to some
extent, masked \cite{gunlycke_edge_disorder}.

Doping \cite{biel_transport} and edge functionalization
\cite{gunlycke_edge_funct,our_nanoribbon_paper_2009} affect the
transport as well. For example, if a finite-length armchair
nanoribbon has a gap induced by the edge-bond deformations (see
subsection~\ref{armchair nanoribbon}), then a suitably chosen
chemical disorder at the edges may actually increase the
conductivity by effectively closing the gap
\cite{our_nanoribbon_paper_2009}. Namely, if the edges are
functionalized with two different kinds of radicals randomly
distributed along the length of the ribbon, then the term of the
Hamiltonian responsible for the opening of the gap becomes
disordered, and the gap closes when this term vanishes on average.
Clearly, one has to counteract the Anderson localization in such a
situation. Fortunately, this phenomenon is not important for a
nanoribbon of sufficiently short length. Transport through this
nanoribbon is effectively metallic
\cite{our_nanoribbon_paper_2009}.

Reference~\cite{sols_coulomb_blockade} pointed out that for nanoribbons
with very corrugated edges the interaction effects can seriously affect the
charge transport. According to
Ref.~\cite{sols_coulomb_blockade}, a nanoribbon with corrugated
edges can be viewed as a series of weakly coupled quantum dots
defined by the random geometry of the edges. Electron transport
through such system is limited by the Coulomb blockade effect in
these dots.

\begin{figure}[btp]
\centering
\leavevmode
\epsfxsize=8.5cm
\epsfbox{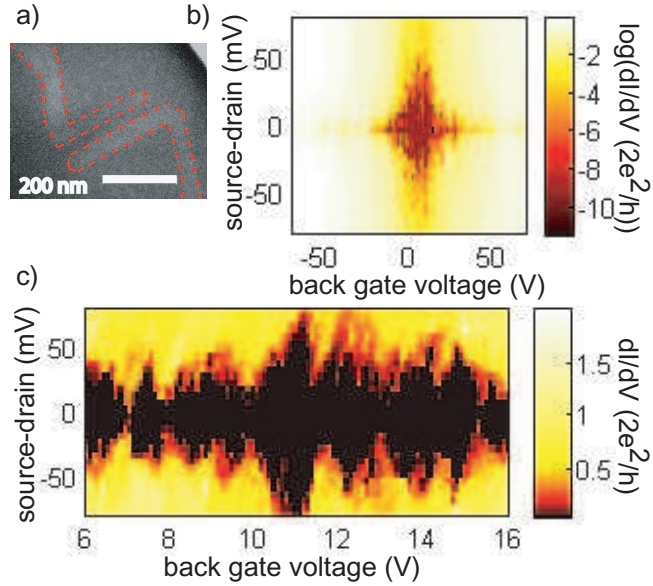}
\caption[]
{\label{tr_gap}
(Color online)
Graphene nanoribbon (a) and the nanoribbon's electrical conductance data
(b,c), from
Ref.~\cite{Todd2009}.
Panel~(a) shows a nanoribbon defined by etching a graphene sheet. The
darker area is graphene, while the lighter areas inside the red dashed
lines are graphene-free. In panel~(b) the differential conductivity is
plotted as a function of the back-gate voltage and source-drain voltage.
Note that for small source-drain voltage the conductivity remains
suppressed (dark area) for small back-gate voltages, and grows when the
latter exceeds a certain value. This is a manifestation of the transport
gap. In panel~(c) the same data is plotted in a smaller back-gate voltage
window. The plot has the characteristic shape of overlapping noisy
``Coulomb diamonds", suggesting that the transport occurs through several
``charge puddles" acting as quantum dots.
Reprinted with permission from:
K. Todd, H. T. Chou, S. Amasha, and D. Goldhaber-Gordon, Nano Lett. {\bf 9},
416 (2009).
Copyright 2009 American Chemical Society.
}
\end{figure}

Recent experiments
\cite{molitor_transport_gap,stampfer_transport_exp,Todd2009,
gallagher_transport_gap} on gated nanoribbons discovered that
there is a Fermi energy interval where the conductance is
suppressed, see 
Fig.~\ref{tr_gap}.
This phenomenon is called the transport gap. It was pointed out in
Ref.~\cite{stampfer_transport_exp}
that the size of the experimentally observed transport gap is too big to be
consistent with conclusions of simple one-particle nanoribbon models, where
the gap is generated due to the transverse quantization. To develop a more
realistic description, the effect of both disorder and interactions must be
accounted, in addition to the transverse quantization. It was proposed that
the transport occurs through tunneling between consecutive ``charge
puddles", which are the areas with non-zero charge induced by external
the disorder potential.

Our discussion shows that transport properties of the nanoribbons are
affected most prominently by bulk and edge disorder, transverse
quantization, edge type, and interactions.

\section{Quantum dots}\label{qdot}

Quantum dots formed in semiconductor heterostructures have been
studied extensively because they are considered promising
candidates for applications in optoelectronics on the nanometer
scale~\cite{Wiel2003,Reimann2002,Michler2009,Buluta2009,Buluta2010}.
For instance, dots might be used in detectors, diodes, memory and
laser devices. Furthermore, single-electron transport devices
which make use of quantum dots could be employed as transistors,
and spin-based dot devices might be useful for quantum logic
gates. Electrons confined in usual semiconductor dots, with a
typical size of a few hundreds of nanometers, are described by the
Schr\"odinger equation and most of their electronic properties are
now well-understood and have been experimentally studied by many
research groups.

The physics of graphene quantum dots is very different from that
in usual semiconductor dots. The reason is twofold: $(i)$ charge
carriers in graphene are massless and obey the relativistic 2D
Weyl-Dirac equation~(\ref{dirac}), and $(ii)$ the different
configurations of the carbon atoms at the boundaries of the dot
affect significantly the dot properties. There are two basic
methods of defining a graphene quantum dot. In the first method,
the dots are defined by the actual geometry of the graphene layer
and they are usually referred to as graphene islands. In the
second method, the dots are defined through the application of
electric and magnetic fields. Of course quantum dots can also be
defined by combining these two methods and recently some other
ideas have been put forward for dot formation, which for example
include the application of strain to the graphene sheet, a
spectral gap opening, and even chemical techniques.

\subsection{Geometry-induced dots and graphene islands}

It is now possible to mechanically cut (i.e., etch) a graphene
flake into various shapes of a few tens of nanometers, which can
confine electrons and thus act as quantum dots. These
geometry-induced dots or graphene islands have well-defined
discrete energy levels whose spectrum depends on the size, shape
and the edge type of the dot. Further, disorder and interaction
effects are also important for the electronic properties of any
realistic graphene system.

\begin{figure}
\begin{center}
\epsfxsize=12.cm \epsfbox{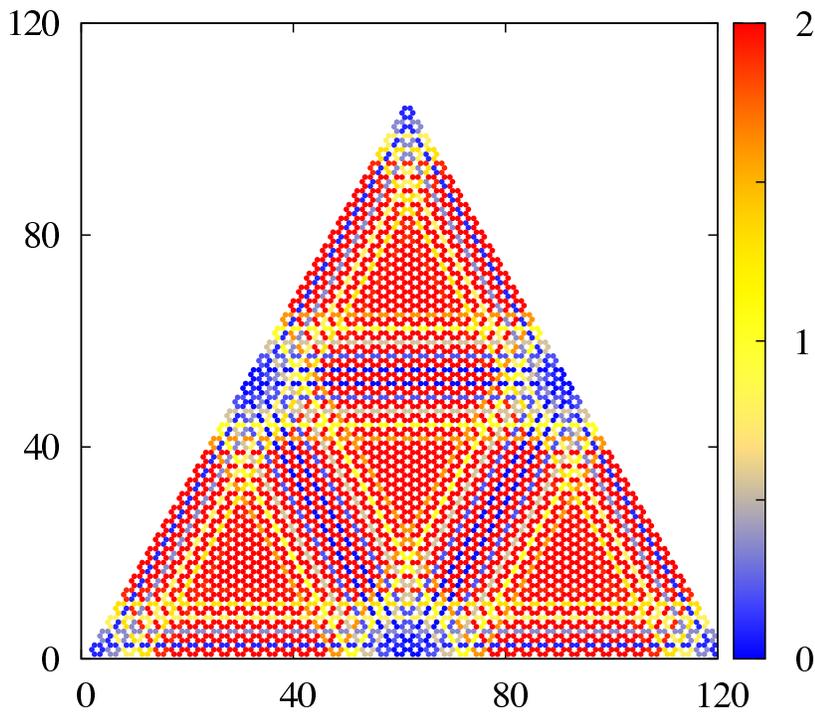}\caption{(Color
online) The Schr\"odinger equation for the tight-binding
Hamiltonian on a triangular armchair quantum dot can be solved
exactly~\cite{Rozhkov2010}. Moreover, the algebraic structure of
the wave functions found is sufficiently simple to allow for
analytical expressions for some matrix elements. In this figure
the exact probability density for certain electron eigenstate on a
triangular quantum dot is plotted.}\label{exactstate}
\end{center}
\end{figure}

A range of typical dot geometries including triangular, hexagonal,
rectangular and circular have been studied numerically, mainly
within the tight-binding and DFT
models~\cite{Ezawa2007,Viana-Gomes2009,Fernandez-Rossier2007,Tang2008,
Bhowmick2008,akola_trigonal,heiskanen_trigonal,Kim2010}. In some
cases exact analytical solutions are also
possible~\cite{potasz_exact_trigonal,ezawa_exact_trigonal,Rozhkov2010}.
For example, for a triangular armchair dot, exact tight-binding
eigenfunctions and eigenenergies were obtained, and a technique
for matrix element calculation was developed~\cite{Rozhkov2010}
(see Fig.~\ref{exactstate}). Quantum dots with arbitrary shapes
have also been examined~\cite{Wimmer2010}.

An important feature of the nano-islands is the appearance of
degenerate zero-energy states that are mostly localized at the
edges, as predicted for triangular and circular zigzag dots, as
well as rectangular
dots~\cite{Ezawa2007,Viana-Gomes2009,Fernandez-Rossier2007,Tang2008,Bhowmick2008,Kim2010}.
For triangular dots there is a sublattice imbalance, i.e.,
$N_{Z}$=$N_{A}-N_{B}$$\neq$0, where $N_{A}$ $(N_{B})$ is the
number of carbon atoms of sublattice $\mathcal{A}$
$(\mathcal{B})$, and this condition is sufficient in order to have
$N_{Z}$ zero-energy states~\cite{Fernandez-Rossier2007}. The
number of these states is proportional to the size of the edges
which, in principle, can be made quite large.

Nanostructures with degenerate zero-energy states are useful for
applications, since the electrons inside such structures may order
magnetically. Magnetism is a consequence of the Coulomb
interaction and Hund's rule. For example, the ground state of
rectangular dots can support antiferromagnetic ordering whereby
the magnetic moments are localised at the zigzag edges with
opposite orientation (for edge magnetism, see also the previous
section). As shown in Ref.~\cite{Tang2008} for rectangular dots
there is a critical minimum width between the zigzag edges that
gives rise to magnetic ordering. If the width is smaller, then the
state is nonmagnetic. On the other hand, triangular zigzag dots
favor ferromagnetic ordering (see Table~\ref{island1}).

Interestingly, external uniaxial strain on square dots enhances
the magnetization and leads to a spatial displacement (drift) of
the magnetization from the zigzag to the armchair
edges~\cite{Viana-Gomes2009}. A magnetization enhancement of 100\%
was predicted for a strain on the order of 20\%, which might be
possible to induce by mechanical methods~\cite{mohiuddin2009}.

\begin{figure}
\begin{center}
\includegraphics[height=7.50cm,angle=0]{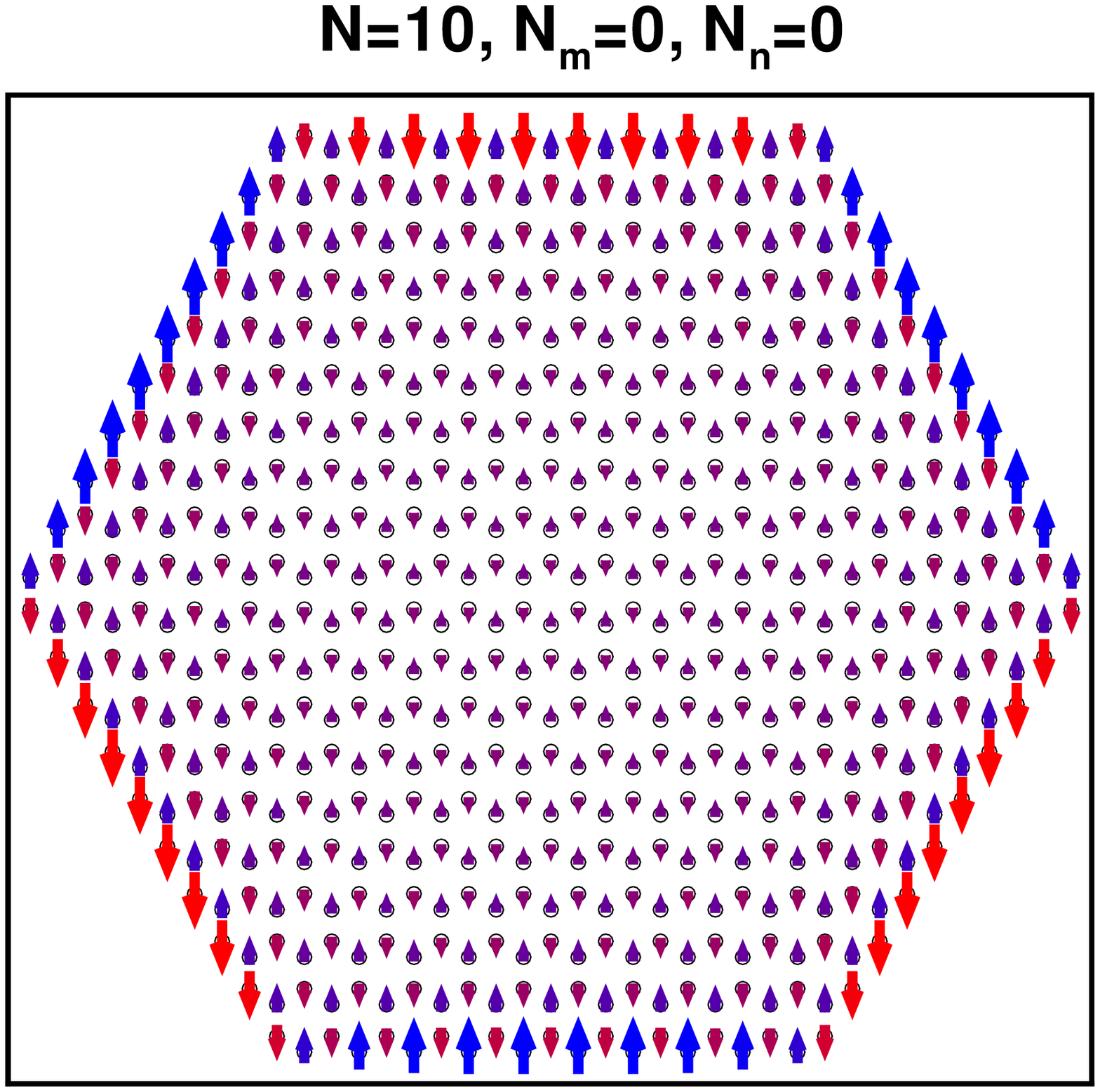}
\includegraphics[height=7.50cm,angle=0]{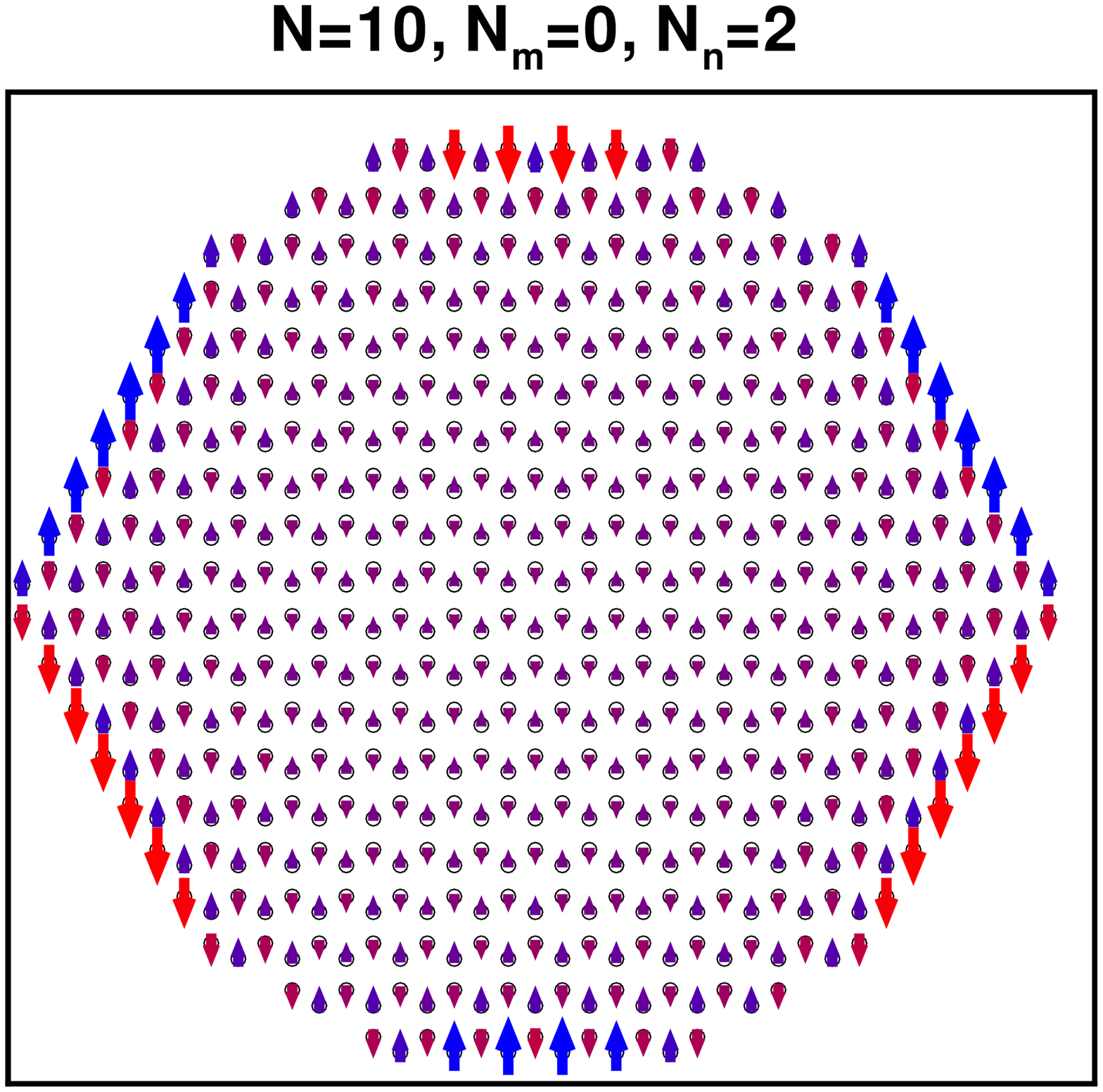}
\includegraphics[height=7.50cm,angle=0]{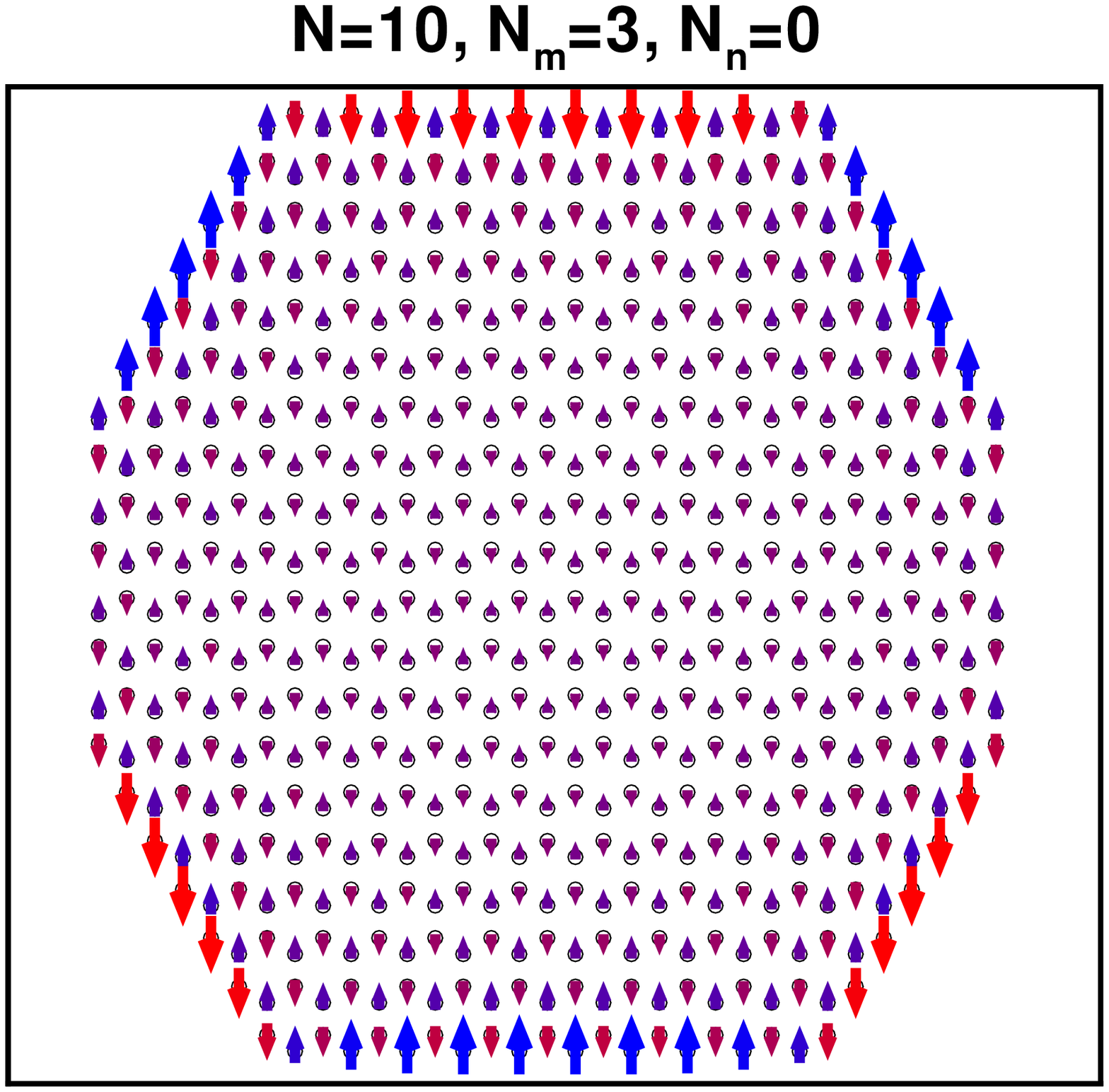}
\includegraphics[height=7.50cm,angle=0]{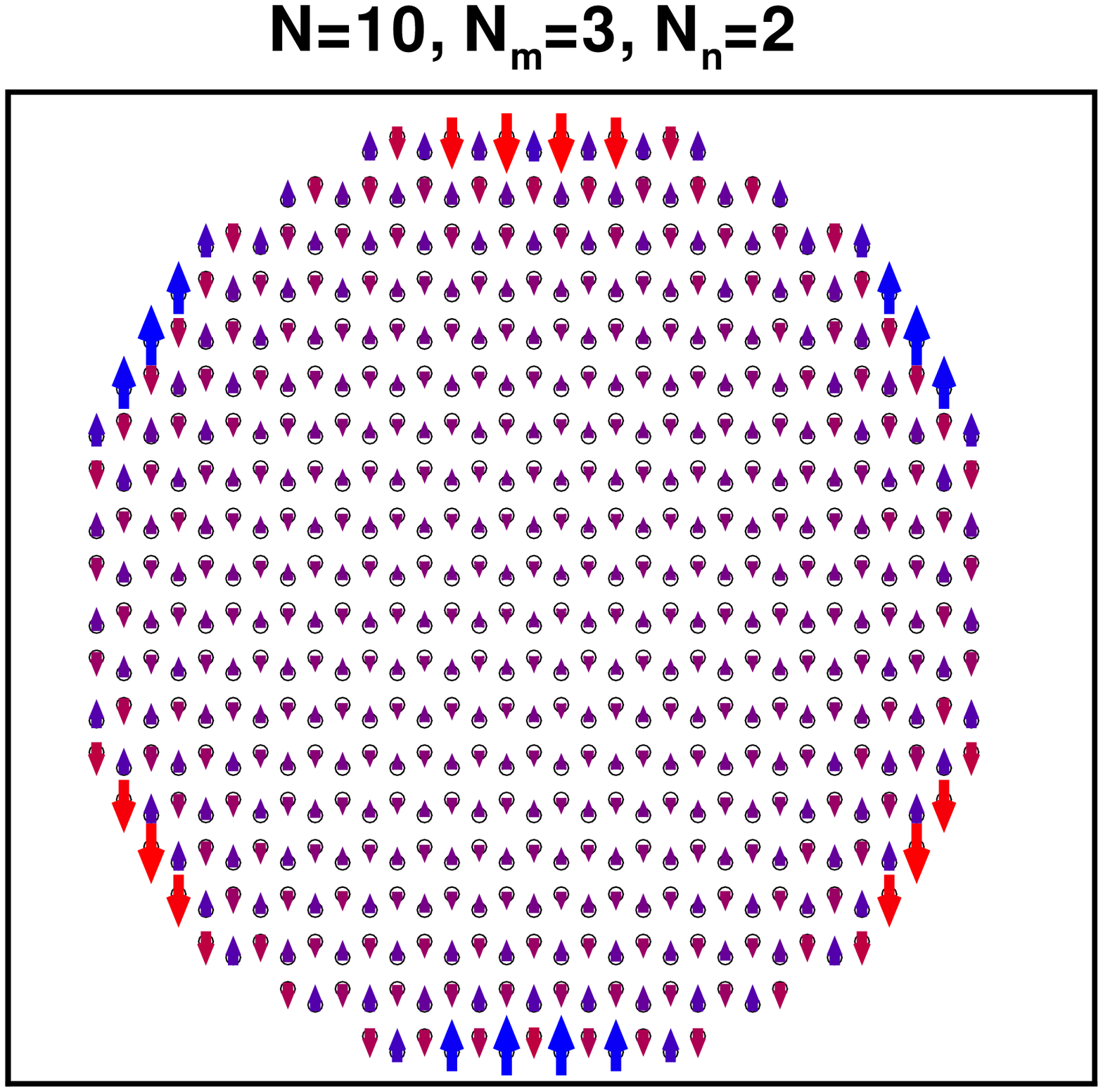}
\caption{(Color online) Magnetic properties of graphene dots were
investigated numerically in Ref.~\cite{Bhowmick2008} within the
framework of a mean field theory of the Hubbard model. Expectation
value of the spin magnetization $S^{z}$ in four different dots.
The size of the arrows is proportional to the magnitude of the
magnetic moment. In this figure, $N$, $N_{m}$ and $N_{n}$ denote
the size of the dot in the perfect hexagon configuration, the
number of armchair edges per vertical side, and the number of
armchair edges per slanted edge, respectively. The magnetic
moments are much larger at the zigzag edges than at the armchair
edges and the internal sites. Reprinted with permission from S.
Bhowmick and V. B. Shenoy, Journal of Chemical Physics {\bf128},
244717 (2008). Copyright 2008 American Institute of
Physics.}\label{nanodot}
\end{center}
\end{figure}

Moreover, it was theoretically predicted that the edge-state
magnetism is robust to impurities and edge-defects, and survives
even to irregular structures as long as there are three to four
repeat units of zigzag edges~\cite{Bhowmick2008}. A similar
conclusion was reached in
Ref.~\cite{kumazaki_magnetic_edge_states}. Some examples of
magnetic structures for different dots are presented in
Fig.~\ref{nanodot}. The results are derived from the mean field
analysis of the Hubbard model~\cite{Bhowmick2008}.

Another promising property, which may be used in memory devices,
is that the spin relaxation time can be long enough, as shown for
triangular dots. In general, the spin relaxation time increases
with the interaction strength and system size, but remains long
even for a relatively small system~\cite{Ezawa2007}.

\begin{table}
  \centering
\begin{tabular}{||c|c|c||}
\hline\hline \multicolumn{3}{||c||}{\textbf{Graphene dots or
islands (theoretical studies)}}\cr \hline\hline
   Type & \quad Zero-energy edge states \quad \quad & Magnetic ordering  \cr
    \hline\hline
  Triangular   &  Yes & Ferromagnetic \cr
  Hexagonal  &  No  &  No \cr
\quad \quad Parallelogram \quad \quad &  No & No \cr
 Rectangular &  Yes & \quad \quad Antiferromagnetic \quad \quad \cr
   \hline
   \hline
\end{tabular}
\caption{Graphene dots have attracted considerable theoretical
interest. Various geometries were examined and degenerate
zero-energy edge states were predicted for triangular zigzag and
rectangular dots (the rectangular dot is defined by two zigzag and
two armchair edges). These states are mainly localized at the
zigzag edges having only a small amplitude at the centre of the
dot. Edge states are absent at arcmhair edges. The existence of
such states leads to magnetic ordering that critically depends on
the specific geometry. For rectangular dots there exists a minimum
width between the two zigzag edges for stable antiferromagnetic
ordering~\cite{Tang2008}. The magnetic properties of the
geometry-induced dots are robust to defects and impurities of the
edges, can survive to irregular structures~\cite{Bhowmick2008},
and can be tuned by the application of an external
strain~\cite{Viana-Gomes2009}.}\label{island1}
\end{table}

\begin{figure}
\begin{center}
\includegraphics[height=10.0cm,angle=0]{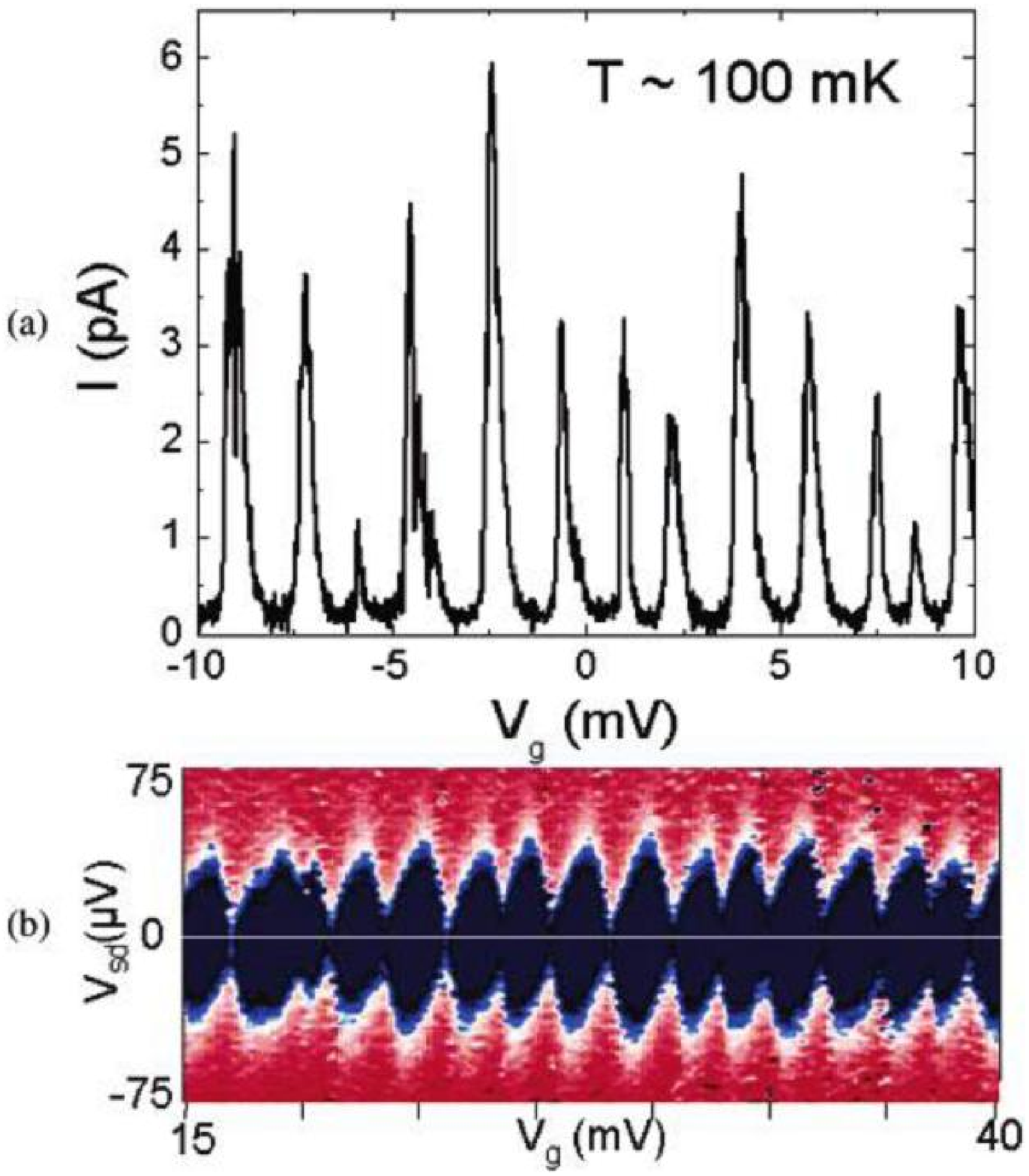}
\caption{(Color online) Gated graphite quantum dots were
fabricated and low-temperature electrical transport measurements
were performed~\cite{Bunch2005}. (a) Current versus gate voltage
$V_{g}$ with source-drain bias $V_{sd}=10$ $\mu$V at temperature
$T\sim 100$ mK. Coulomb oscillations are observed with a period in
gate voltage of $\Delta V_{g}=1.5$ mV. (b) The differential
conductance ($dI/dV_{sd}$) is plotted as a color scale versus gate
voltage ($V_{g}$) and source-drain bias $V_{sd}$. Blue (red)
signifies low (high) conductance. The charging energy of the dot
is equal to the maximum height of the diamonds: $\Delta
V_{sd}=0.06$ mV. Reprinted with permission from J. S. Bunch, Y.
Yaish, M. Brink, K. Bolotin, and P. L. McEuen, Nano Letters
{\bf5}, 287 (2005). Copyright 2005 American Chemical
Society.}\label{graphitedot}
\end{center}
\end{figure}

Reference~\cite{Bunch2005} reported low-temperature electrical
transport measurements on gated quasi-2D graphite quantum dots.
These were the first measurements on mesoscopic samples of
graphite which consists of many stacked layers of graphene held
together by weak van der Waals forces. Coulomb charging phenomena
were demonstrated with the help of data in Fig.~\ref{graphitedot},
where the electrical current through the dot as a function of gate
voltage and source-drain bias is plotted. More recent experiments
probed the energy spectrum of quantum dots formed in a single
layer of
graphene~\cite{Ponomarenko2008,Stampfer2008a,Stampfer2008}. An
all-graphene single-electron transistor, exhibiting
Coulomb-blockade behaviour, was operational well-above
liquid-helium temperatures~\cite{Ponomarenko2008}. The
Coulomb-blockade peaks are (nearly) periodic as a function of gate
voltage for large islands ($>100$ nm), and nonperiodic for small
ones ($<100$ nm). The distance between the peaks is proportional
to the sum of charging and confinement energies. The former, being
typically constant for a specific dot geometry, dominates for
large islands~\cite{Ponomarenko2008}. For small islands the size
quantization becomes important, and the confinement energy
prevails, leading to nonperiodic peaks (see Table~\ref{island2}).
The energy-level statistics of graphene islands was also probed,
and it was shown to agree well with the theory of chaotic Dirac
billiards~\cite{Ponomarenko2008}.

Coulomb-blockade measurements on a graphene island ($\sim 200$ nm)
with an integrated charge detector were also
reported~\cite{Guttinger2008}. A nanoribbon placed 60 nm from the
island acts as a detector, which enhanced the resolution of single
charging events on the island. In addition, tunable double quantum
dots were fabricated whereby the coupling to the leads and the
interdot coupling were tuned by graphene in-plane
gates~\cite{Molitor2009}. Spin spectroscopy has also been
investigated in graphene dots~\cite{Guttinger2010}.

\begin{figure}
\begin{center}
\includegraphics[height=9.50cm,angle=0]{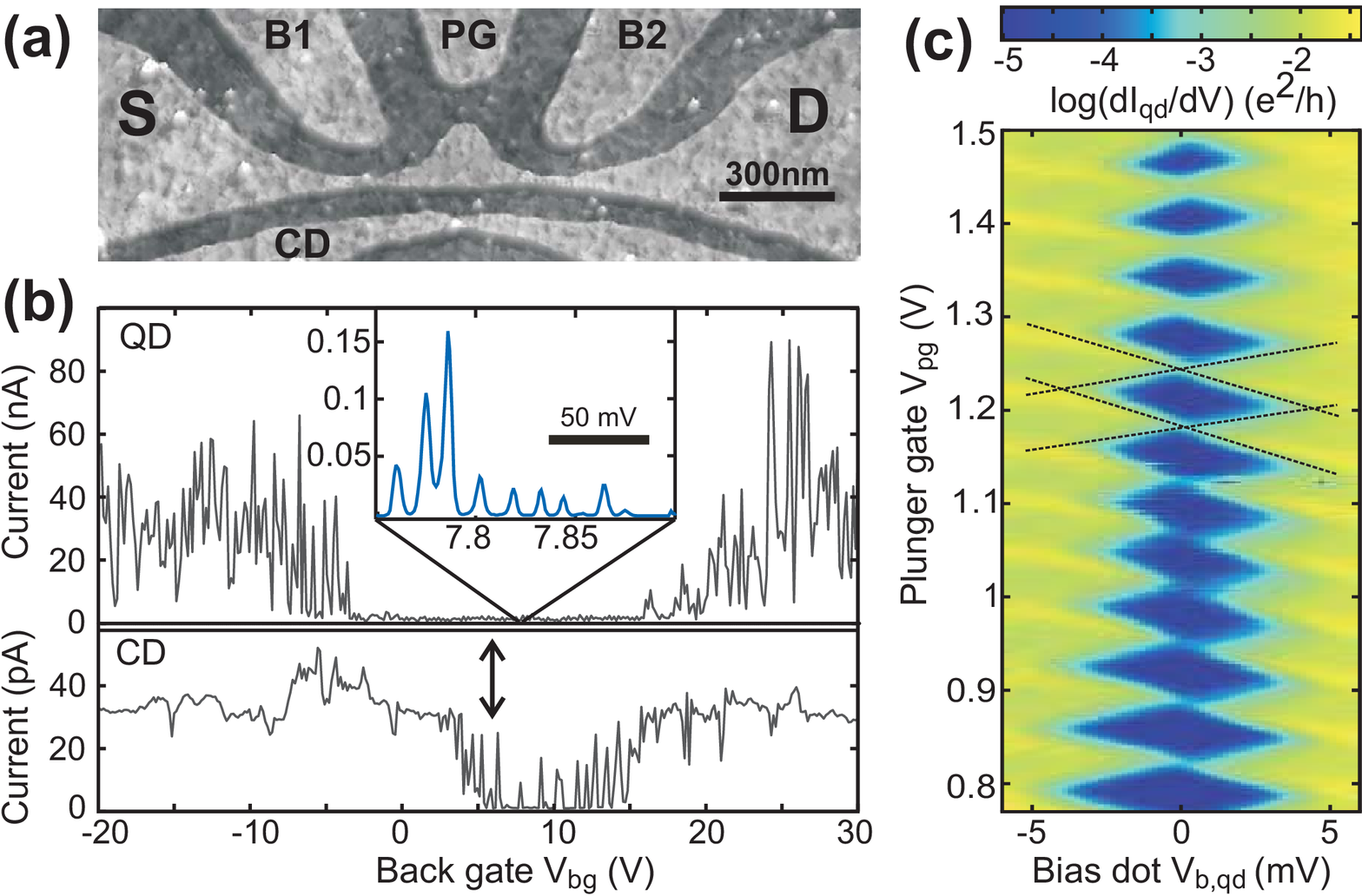}
\caption{(Color online) Graphene quantum dot device with charge
detector and transport measurement data~\cite{Guttinger2008}. (a)
Scanning force micrograph of the measured device. The central
island, that acts as the quantum dot (QD), is connected to source
(S) and drain (D) contacts via two narrow constrictions. The
diameter of the dot is 200 nm and the constrictions are 35 nm
wide. The charge detector (CD) is a graphene nanoribbon and the
lateral gates B1, B2 and PG are used to tune the device. (b)
Current as a function of back-gate voltage of the QD (upper panel)
and CD (lower panel). The source-drain voltage is 500 $\mu$V and
the temperature is 1.7 K. The inset shows typical Coulomb-blockade
features as expected in a dot device. (c) Differential conductance
is plotted as a color scale versus source-drain bias and PG gate
voltage, for a back-gate voltage of 2 V. The charging energy was
estimated to be about 4.3 meV. Reprinted with permission from J.
Guttinger, C. Stampfer, S. Hellmuller, F. Molitor, T. Ihn, and K.
Ensslin, Applied Physics Letters {\bf93}, 212102 (2008). Copyright
2008 American Institute of Physics.}\label{chargedetector}
\end{center}
\end{figure}


\begin{table}
  \centering
\begin{tabular}{||c|c|c||}
\hline\hline \multicolumn{3}{||c||}{\textbf{Graphene dots or
islands (experimental studies)}}\cr \hline\hline
 \quad  Size of island ($D$) \quad & \quad Coulomb peaks \quad \quad & Energy scale  \cr
    \hline\hline
  $<$ 100 nm   &  Nonperiodic & \quad Confinement $\sim v_{F}h/2D\approx41$ meV ($D=$ 50 nm) \quad \cr
  $>$ 100 nm  &  Periodic  & Charging $\approx$ 3 meV ($D\approx$ 250 nm) \cr
   \hline
   \hline
\end{tabular}
\caption{Graphene islands were investigated experimentally via
electrical transport measurements, and single-electron transport
was demonstrated~\cite{Ponomarenko2008}. When the diameter of the
island is large ($D>100$ nm) the Coulomb peaks in the conductance,
as a function of back-gate voltage, are periodic and their
position is determined mainly by the characteristic charging
energy. For small-diameter islands ($D<100$ nm), the position of
the peaks is nonperiodic and the dominant energy scale is the
confinement energy on the order of $E_{D}\sim v_{F}h/2D$ ($h$ is
Planck's constant). This is much larger than the corresponding
energy $E_{S}\sim h^{2}/8 mD^2$ of Schr\"odinger electrons with
effective mass $m$; $E_{D}/E_{S}\sim40$ for $D=100$ nm, and $m$ is
the effective mass for GaAs. Stable, robust and conductive dot
islands as small as 15 nm were fabricated, showing the potential
of graphene for nanoelectronics.}\label{island2}
\end{table}

\subsection{Field-induced dots}

Charge confinement within the ``bulk'' graphene sheet is tricky
due to the Klein tunneling effect~\cite{Katsnelson2006}. In case
of normal incidence, this allows for perfect transmission of
massless relativistic particles through high and wide potential
barriers. A key point is that in the barrier region the states of
massless Weyl-Dirac particles have an oscillatory character, even
at energies lower than the potential height, as happens exactly
outside the barrier. This is completely different from
Schr\"odinger particles with non-zero mass, for which the states
in the barrier region decay exponentially and therefore perfect
transmission is not feasible. Experimentally, the Klein tunneling
was demonstrated in graphene through electrical transport
measurements in steep potential barriers generated by metallic
gates~\cite{Stander2009}.

\begin{figure}
\begin{center}
\includegraphics[height=9.50cm,angle=0]{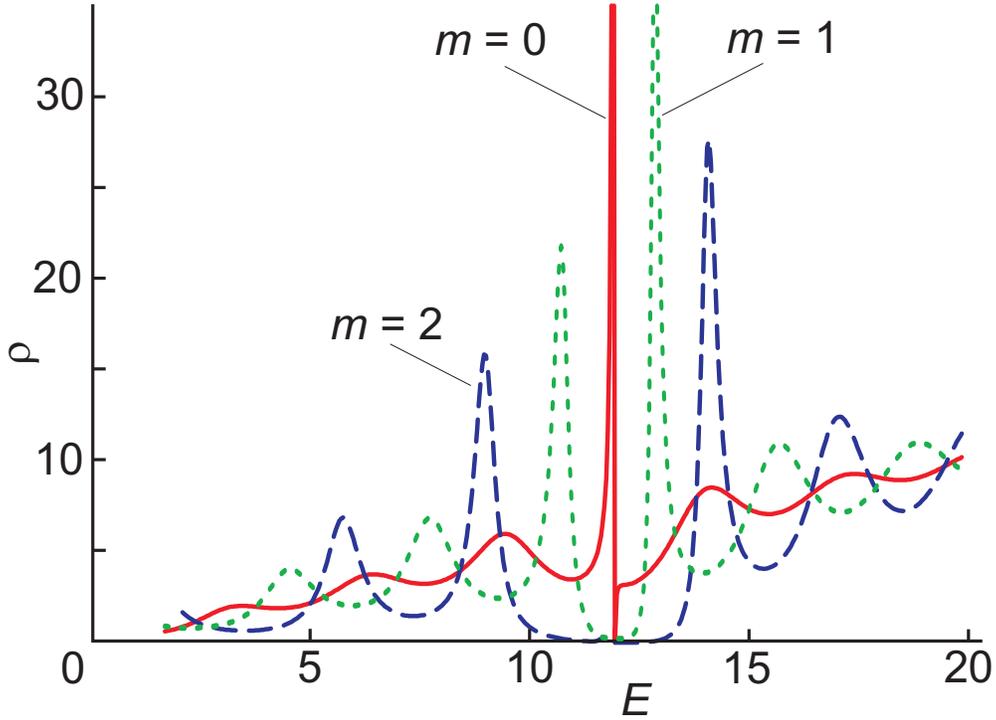}
\caption{(Color online) The physics of quasi-bound states in
circular graphene dots was examined theoretically by solving the
Weyl-Dirac equation~\cite{Matulis2008}. The figure shows the local
density of states as a function of energy in the case of a barrier
height of $V=12$ (dimensionless units) for three angular momentum
numbers: $m=0$ (solid curve), $m=1$ (dotted curve), and $m=2$
(dashed curve). The peaks become narrower as the momentum
increases, within a specific energy range, thus the lifetime of
the corresponding states becomes longer. Notice the very narrow
peak when the energy is close to the barrier height (for $m=0$),
as a consequence of the total reflection of the wavefunction at
the dot edge. Reprinted figure with permission from A. Matulis and
F. M. Peeters, Physical Review B {\bf77}, 115423 (2008). Copyright
(2008) by the American Physical Society.}\label{densitymatulis}
\end{center}
\end{figure}

Because of the Klein tunneling, an electrostatic potential minimum
in graphene leads to quasi-bound states, i.e., resonant states
(see Table~\ref{states} for a classification of the dot states)
and therefore it is inadequate to confine
electrons~\cite{Chen2007,Matulis2008,Hewageegana2008}.
Nevertheless, the finite lifetime of the states, characterizing
the trapping time of an electron in the dot region, can be
relatively long. It depends on the potential profile and
eigenenergy. A smooth potential and a large angular momentum
enhance the lifetime of the quasi-bound
states~\cite{Chen2007,Hewageegana2008}, as happens also with
eigenenergies close to the maximum of the potential
barrier~\cite{Matulis2008}. Figure~\ref{densitymatulis} shows the
effect of the quasi-bound states on the local density of states of
a circular dot.

The electrostatic confinement of electrons in graphene dots was
also examined through the dependence of the conductance on the dot
area, which is tunable with a metal gate~\cite{Bardarson2009}.
Both disc-shaped dots, in which the classical dynamics is regular,
and stadium-shaped dots where the classical dynamics is chaotic
were studied. Confinement can be achieved only in the former when
the corresponding Weyl-Dirac equation is separable.

\begin{figure}
\begin{center}
\includegraphics[height=10.0cm,angle=0]{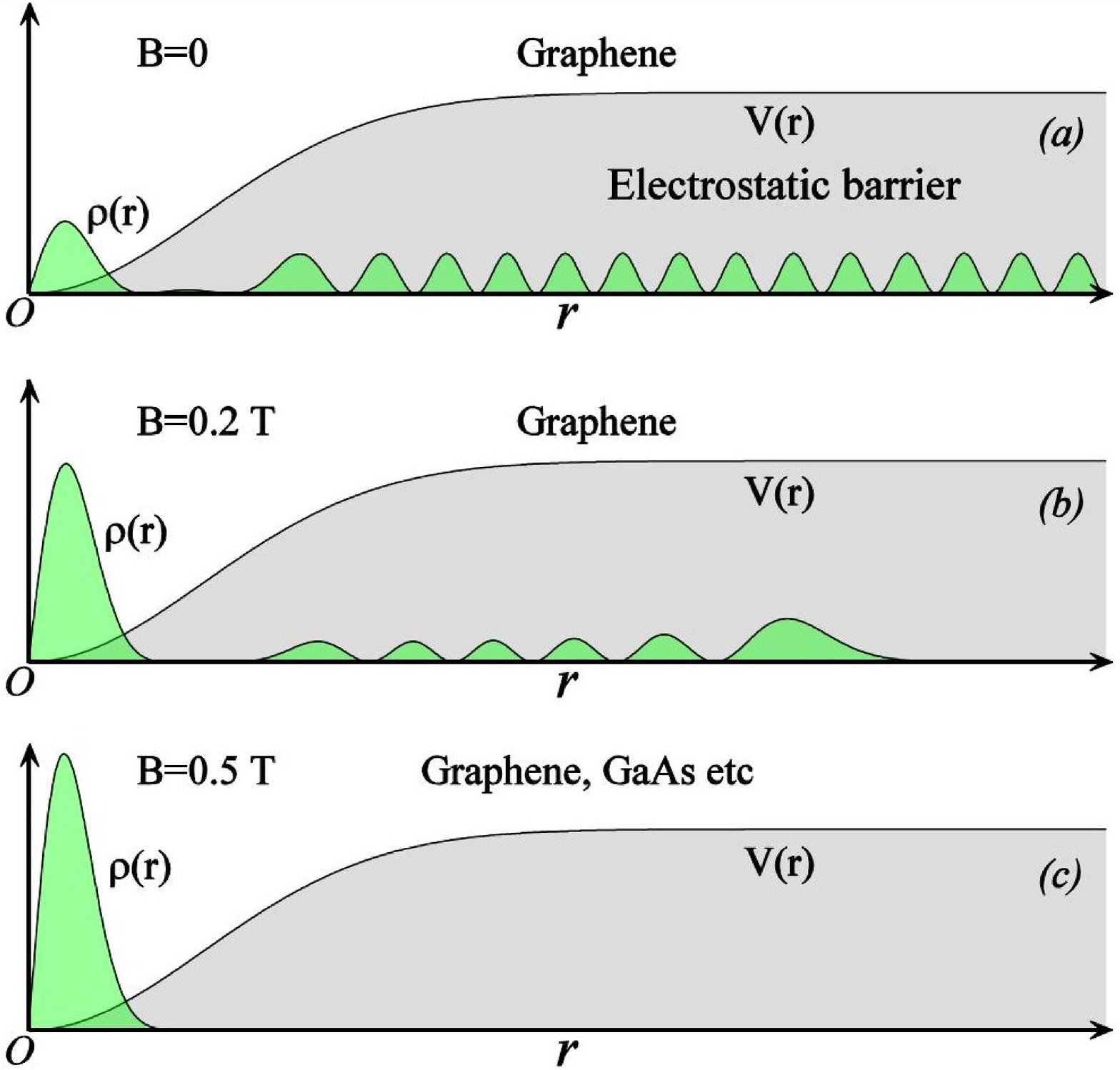}
\caption{(Color online) Klein tunneling in a circular graphene
quantum dot in a uniform magnetic field. The dot is defined by the
electrostatic potential $V(r)$, which vanishes at the centre
(\textit{O}) of the dot and asymptotically rises to a constant
value, while the magnetic field $B$ is perpendicular to the
graphene sheet. (a) For $B=0$ and small angular momentum, the
radial probability distribution $\rho(r)$, for one of the spinor
components, has a large amplitude near the centre of the dot and
oscillates inside the barrier region because of the Klein
tunneling. In this case, which is unique to graphene, the quantum
state is quasi-bound and has an oscillatory asymptotic character.
(b) When the magnetic field is nonzero, the Klein tunneling is
partially suppressed. At large $r$ the oscillatory behavior is
replaced by exponential decay, indicating that the state is bound.
As in (a), this case is also unique to graphene. (c) With
increasing magnetic field, the Klein tunneling is completely
suppressed and the probability distribution decays exponentially
inside the electrostatic barrier. The state is now bound near the
centre of the dot and such a state can be seen in both graphene
dots and usual semiconductor dots, e.g., GaAs. For the latter,
this state can be seen even when $B=0$. A magnetic-field-induced
confinement-deconfinement transition in a graphene dot due to the
Klein tunneling was theoretically examined in
Ref.~\cite{Giavaras2009}.}\label{dotklein}
\end{center}
\end{figure}

\begin{figure}
\begin{center}
\includegraphics[height=7.50cm,angle=0]{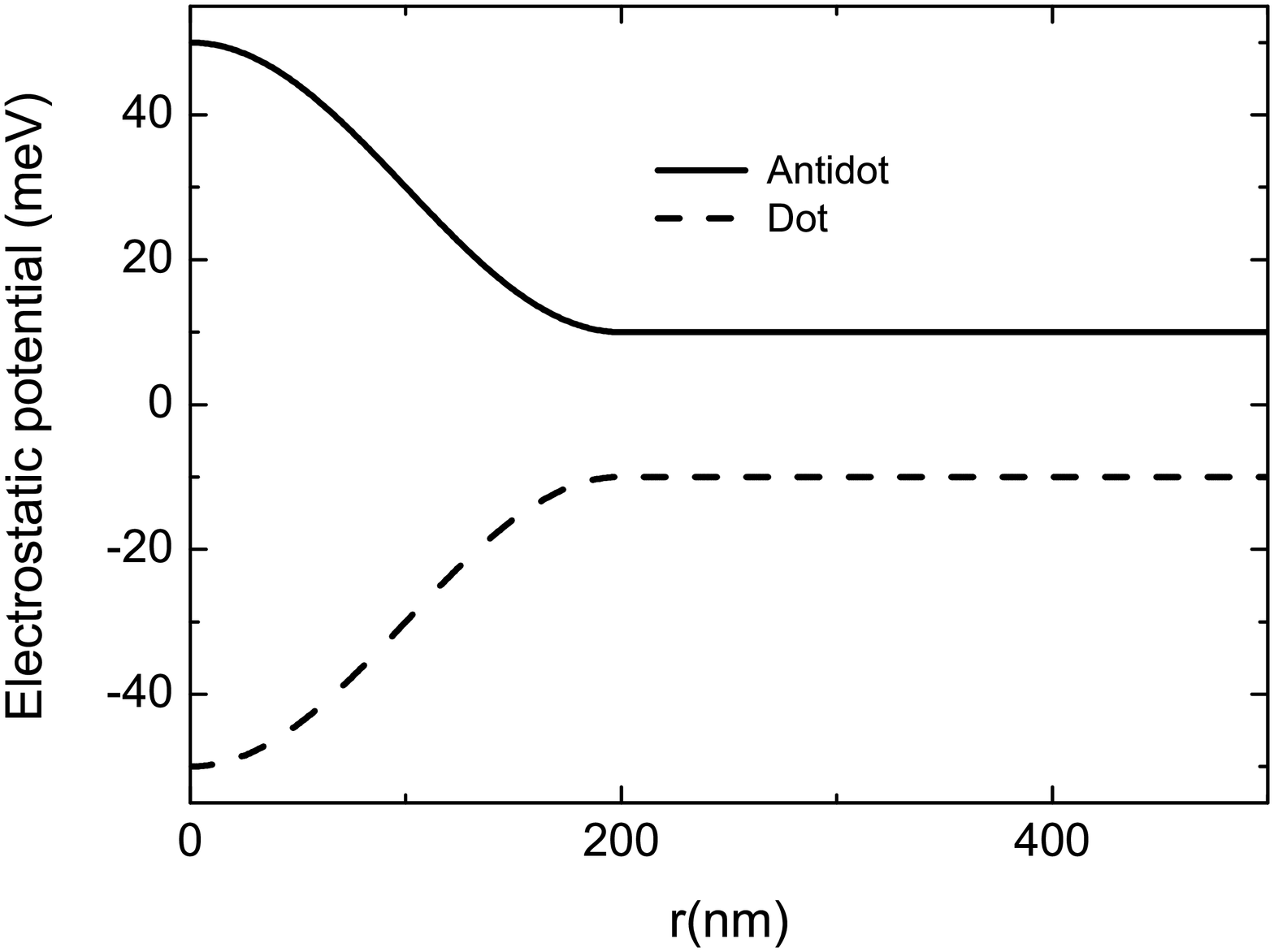}
\includegraphics[height=8.0cm,angle=270]{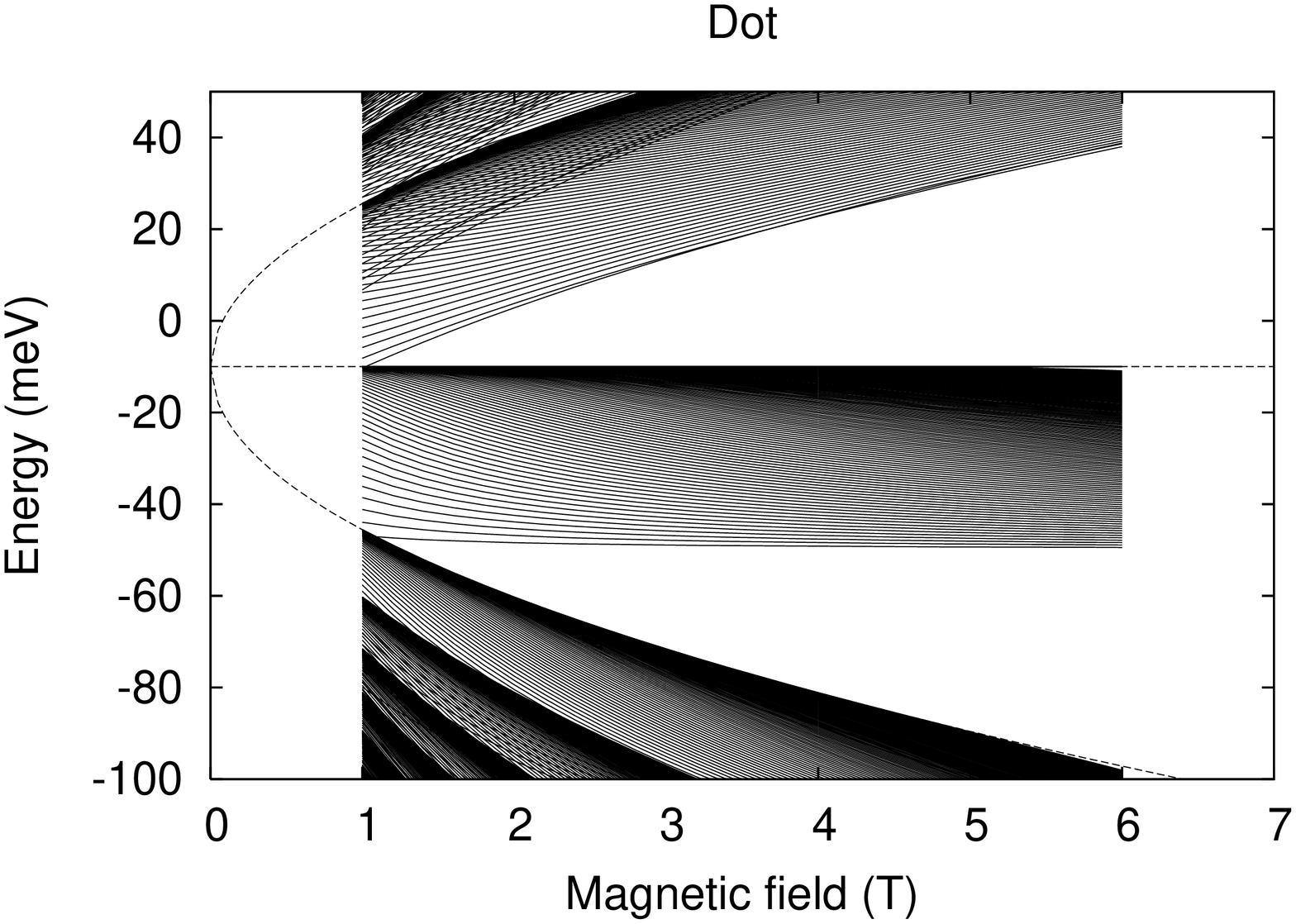}
\includegraphics[height=8.0cm,angle=270]{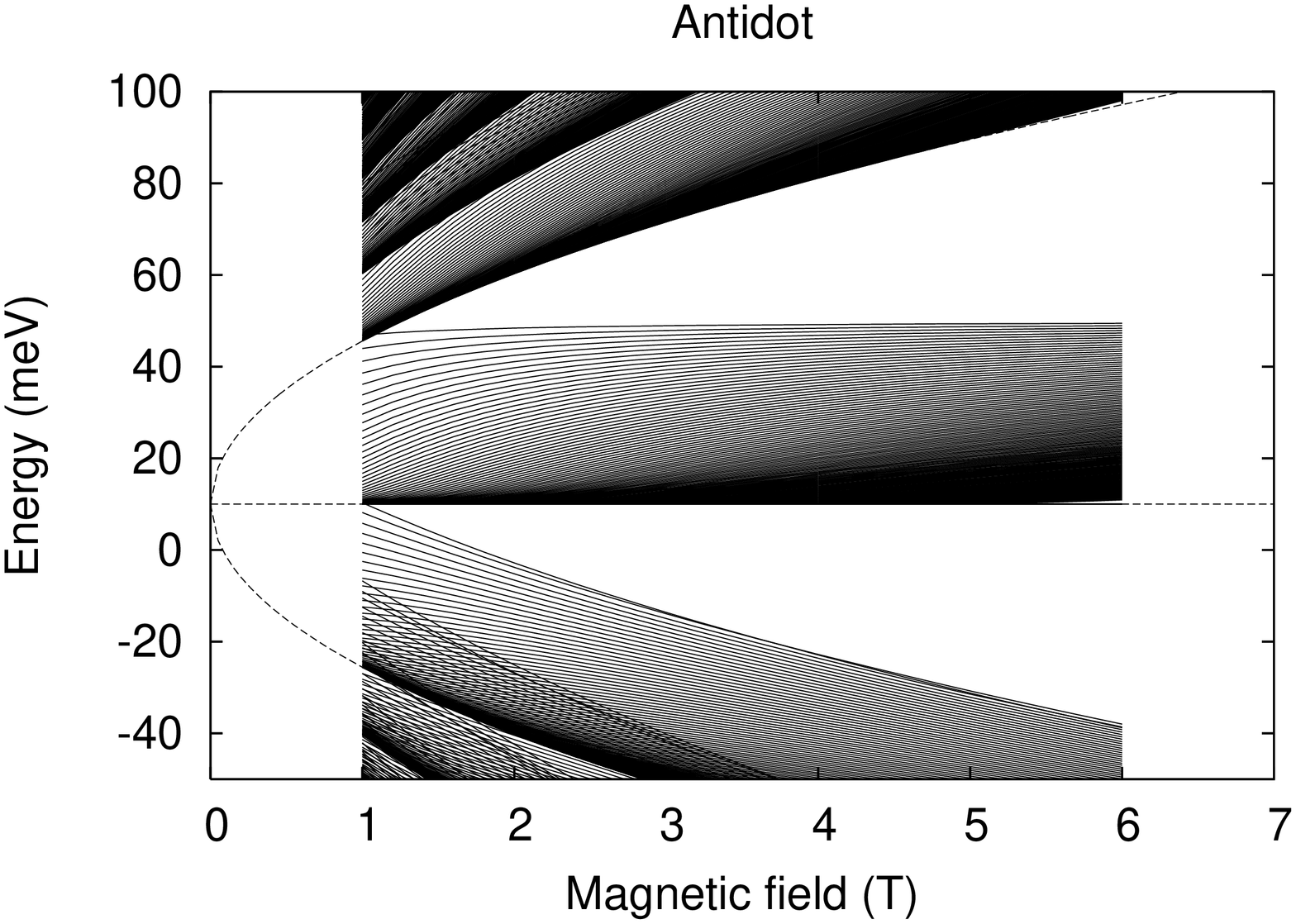}
\caption{A graphene dot or an antidot can be formed with the help
of a uniform magnetic field applied perpendicular to the graphene
sheet~\cite{Maksym2010}. (top) The electrostatic potential profile
of an antidot (solid line) and a dot (dashed line) along the
radial direction. Energy spectra as a function of the applied
magnetic field of dot (left) and antidot (right). The dashed lines
show the Landau levels of an ideal graphene sheet. The potential
is adjusted so that the confined dot (antidot) states lie in the
gap between Landau level 0 and Landau level $-$1 ($+$1). The
symmetry between the energy levels ($E\rightarrow-E$) of the two
systems is a direct consequence of the Dirac cone band-structure
in graphene.}\label{dotmaksym}
\end{center}
\end{figure}

The application of a magnetic field can completely suppress the
Klein tunneling, leading to bound
states~\cite{DeMartino2007,Masir2009,Giavaras2009,Giavaras2010a}.
Thus graphene quantum dots can be formed with the help of a
nonuniform magnetic field, whereby the field is zero within a disc
area defining the spatial region of the dot, and nonzero outside
the dot~\cite{DeMartino2007}. The combination of an electrostatic
potential and a vector magnetic potential allows the
confined-deconfined character of the dot states to be tuned at
will~\cite{Giavaras2009}. Most interestingly, it allows graphene
dots to be formed in a uniform magnetic field using standard gate
electrodes as in common semiconductors~\cite{Giavaras2009}. Then
the quantum states can be tuned with the strength of the magnetic
field and this property allows the Klein tunneling mechanism to be
probed experimentally in graphene dots. A dot design suitable for
this experiment was suggested in Ref.~\cite{Giavaras2009}. The
concept of defining a magnetic graphene dot was further developed
theoretically in Ref.~\cite{Maksym2010}. In particular, in a
strong magnetic field the electrostatic potential of the dot is
adjusted so that the confined dot states lie in the gap between
Landau level 0 and Landau level $-1$ (see Fig.~\ref{dotmaksym}).
This ensures that the dot states are energetically isolated in a
region of low density of states and thus they can be probed using
standard charge-sensing measurements as in a GaAs dot. Numerical
estimates showed that a typical spacing between the dot levels is
$\sim2$ meV at a magnetic field of 5 T. In addition,
Ref.~\cite{Maksym2010} considered how this basic idea can be
extended to a graphene antidot for which the levels of the
confined states lie in the region between Landau level 0 and
Landau level +1. For a confined state with energy $E$ in the dot,
there is a corresponding confined state with energy $-E$ in the
antidot. This is a unique property of graphene due to the symmetry
of the Dirac cone. The physics of graphene antidots in a magnetic
field was also examined in Ref.~\cite{Park2010}.


\begin{table}
  \centering
\begin{tabular}{||c|c|c||}
\hline\hline \multicolumn{3}{||c||}{\textbf{Quantum dot
states}}\cr \hline\hline
 \quad  Type  \quad & \quad Lifetime \quad \quad & \quad Spatial asymptotic behaviour of wavefunction \quad \cr
  (other used names)  & & \cr
    \hline\hline
  Bound   &   Infinite & \quad  Exponential decay \quad \cr
 (confined, stable)  & &  \cr
  Quasi-bound  &  Finite  & Oscillatory \cr
 \quad (deconfined, resonant) \quad &  &  \cr
   \hline
   \hline
\end{tabular}
\caption{The quantum states of a dot, formed within the ``bulk''
graphene sheet, can be either bound or quasi-bound. Because of the
Klein tunneling, both types of states can have a large amplitude
in the barrier region of the quantum dot, though their asymptotic
behaviour is different. Exponential decay is characteristic of
bound states, whereas oscillatory behaviour is characteristic of
quasi-bound states. As shown in Ref.~\cite{Giavaras2009} the type
of states can be tuned with an electrostatic potential and a
uniform magnetic field perpendicular to the graphene sheet (see
also Table~\ref{circular}).}\label{states}
\end{table}

Figure~\ref{dotklein} illustrates the magnetic field-induced
suppression of the Klein tunneling for an electron inside a dot,
and Table~\ref{circular} summarizes the general conditions for
confinement in a circular quantum dot, as derived in
Ref.~\cite{Giavaras2009}. The physics of electrostatic barriers in
the presence of uniform and nonuniform magnetic fields is analyzed
in Sec.~\ref{barrier}.

\begin{table}
  \centering
\begin{tabular}{||c|c|c||}
    \hline
    \hline
    \multicolumn{3}{||c||}{\textbf{Circular graphene dot: $V=V_{0}r^{s}$, $A_{\theta}=A_{0}r^{t}$}}\\
     \hline\hline
 \quad Asymptotically \quad & \quad $s$, $t$ \quad & Dot states \quad \\
    \hline
     \hline
$A_{\theta}<V$    & \quad $t<s$  \qquad & \quad Quasi-bound for all $V_{0}$, $A_{0}$ \quad\\
$A_{\theta}\sim V$& \quad $t=s$  \qquad & \quad Quasi-bound for $V_{0}> v_{F} eA_{0}$\quad \quad\\
$A_{\theta}\sim V$& \quad $t=s$  \qquad & \quad Bound for $V_{0}< v_{F} eA_{0}$ \quad \\
$A_{\theta}>V$    & \quad $t>s$  \qquad & \quad Bound for all  $V_{0}$, $A_{0}$ \quad \\
\hline \hline
\end{tabular}
\caption{Confinement of electrons in a circular graphene dot is
conditional because of the Klein tunneling~\cite{Giavaras2009}.
Consider a graphene dot defined by the electrostatic potential
$V=V_{0}r^{s}$ and the magnetic vector potential
$\mathbf{A}=(0,A_{\theta},0)$, with $A_{\theta}=A_{0}r^{t}$ and
$s$, $t>0$. In such a situation, if asymptotically $A_{\theta}<V$,
then the quantum states are quasi-bound; if $A_{\theta}>V$, they
are bound. In the special case $A_{\theta}\sim V$, the states are
bound only when $V_{0}<v_{F}eA_{0}$ ($e$ is the absolute value of
the electron charge). This suggests that the states can be tuned
at will with an electric or magnetic field. A simple limit occurs
for a constant potential ($V=$ const.) and uniform magnetic field
$B$ perpendicular to the graphene sheet ($t=1$,
$A_{0}=\frac{B}{2}$). Then the states are always bound and
correspond to the well-known Landau levels.}\label{circular}
\end{table}

\subsection{Nanoribbons of graphene and dots}

\begin{figure}
\begin{center}
\includegraphics[height=8.0cm,angle=0]{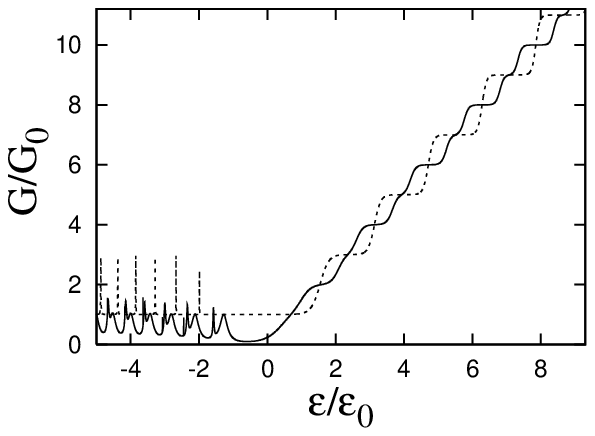}
\caption{Theoretical studies showed that a graphene quantum dot
can be formed by an external electrostatic potential in a
nanoribbon system~\cite{Silvestrov2007}. The conductance through
the dot as a function of the Fermi energy depends on whether the
system is semiconducting (solid line) or metallic (dashed line).
In both cases, the nanoribbons have armchair edges. For
$\varepsilon<$ 0 the device operates as a quantum dot and the
conductance exhibits resonances. For $\varepsilon>$ 0 the device
operates as a point contact and the conductance exhibits steps.
For $\varepsilon>$ 0 the conductance steps in the semiconducting
system are twice smaller than those in the metallic. Reprinted
figure with permission from P. G. Silvestrov and K. B. Efetov,
Physical Review Letters {\bf98}, 016802 (2007). Copyright (2007)
by the American Physical Society.}\label{ribbondot}
\end{center}
\end{figure}

As discussed in the previous section, the most commonly studied
graphene nanoribbons are either of armchair or zigzag type. A
quantum dot defined by an external electrostatic (parabolic)
potential along the nanoribbon was investigated theoretically in
Ref.~\cite{Silvestrov2007}. Quasi-bound states inside nanoribbons
with either zigzag or armchair edges, in both metallic and
semiconducting samples, were predicted. The lifetime of such
states can be made long enough by increasing the characteristic
length of the external potential, for example, with a gate
electrode. Further, the dependence of the conductance on the gate
voltage was found to be sensitive to the type of edges as
illustrated in Fig.~\ref{ribbondot}. The proposed device can
operate either as a quantum dot or as a point contact. For the
former, the conductance displays resonances at negative Fermi
energy, whereas for the latter it has a step-like behaviour at
positive Fermi energy. The conductance steps in the semiconducting
system are twice smaller than those in the metallic system.

Semiconducting nanoribbons with armchair edges were proposed for
the formation of spin qubits in graphene
dots~\cite{Trauzettel2007,Recher2010}. Confinement in one
direction is achieved naturally by the nanoribbon and in the
second direction electrically by gate voltages. In this set-up,
the valley degeneracy is lifted, thus allowing Heisenberg spin
exchange coupling in tunnel-coupled dots. Such graphene dots can
be coupled over long distances as a consequence of the
relativistic nature of electrons in graphene, exhibiting Klein
tunneling.

The electrostatic confinement of electrons in graphene nanoribbons
as well as the Coulomb-blockade effect were experimentally
demonstrated~\cite{Liu2009}. In particular, electrons are confined
between gate-induced $pn$-junctions acting as barriers. However,
even when no $pn$-junctions are formed, the electrons are still
confined, though in a larger area due to strong disorder.

\subsection{More dots}

Tunable quantum dots which take advantage of a gap in the energy
dispersion were also proposed in both monolayer and bilayer
graphene~\cite{Recher2009,Pereira2007,Giavaras2010b,Giavaras2011}.
Experimentally, the gap can be introduced via a chemical and/or
electrical technique~\cite{Zhang2009,Ohta2006} (see also
Table~\ref{energy_gap}), and it allows dots to be formed
electrostatically in a quite similar manner as in common
semiconductors. A finite gap introduces a mass term in the
Weyl-Dirac Hamiltonian. Then for a dot-confining potential with a
finite asymptotic value, the gap gives rises to an energy range
within which the Klein tunneling is suppressed, leading to the
formation of bound states. In this energy range, which is directly
proportional to the value of the gap, hole states do not exist and
therefore the electron states decay exponentially. Moreover, for
quantum dots formed in the gapped sample the valley degeneracy is
lifted by a uniform magnetic field. This property might be
attractive in order to define spin and valley
qubits~\cite{Recher2009}.

It was also shown theoretically that a spatially modulated Dirac
gap in the graphene sheet can lead to confined states with
discrete energy levels, thus giving rise to a dot. The basic
advantage of this proposal is that the dot is formed without
applying external electric and/or magnetic
fields~\cite{Giavaras2010b}. Thus magnetic fields can be used to
manipulate the spin states without affecting the confinement of
the corresponding orbital states. The properties of a Dirac
gap-induced graphene dot in the presence of an electrostatic
quantum well potential were studied in Ref.~\cite{Giavaras2011}.
It was shown that confined states which are induced thanks to the
spatially modulated Dirac gap couple to the states induced by the
potential. The resulting hybridised states are localised in a
region which can be tuned with the potential strength; an effect
which involves Klein tunneling. Numerical calculations of the
local density of states suggest that this effect could be
probed~\cite{Giavaras2011}.

Strain engineering is another proposal in order to generate
confinement in a sheet of graphene~\cite{Pereira2009}. Local
patterning of the substrate induces in-plane strain in the
graphene lattice, anisotropically changing the hopping amplitude
between the carbon atoms. As a result, in the continuum
approximation the quasi-particles are described by an effective
Weyl-Dirac equation in the presence of a gauge field. It turns out
that this field can act in a rather similar manner as a magnetic
field and therefore it can lead to confined quantum states. A
noteworthy advantage of this proposal~\cite{Pereira2009} is that
patterning can be made directly on the substrate, hence protecting
the graphene layer from possible damage.

Vacancy clusters in the graphane sheet were also suggested for dot
formation. In particular, DFT and tight binding calculations
showed that cluster of hydrogen vacancies can serve as quantum
dots. The stability as well as the shape and size of these dots
depend crucially on the graphene/graphane interface energy and the
degree of aromaticity~\cite{Singh2010}.

\subsection{Quantum rings}

Quantum rings in graphene have also attracted some interest,
mainly because these types of devices allow the investigation of
phase-coherence phenomena, as it is now well-known from studies in
usual semiconductor systems.

\begin{figure}
\begin{center}
\includegraphics[height=11.0cm,angle=0]{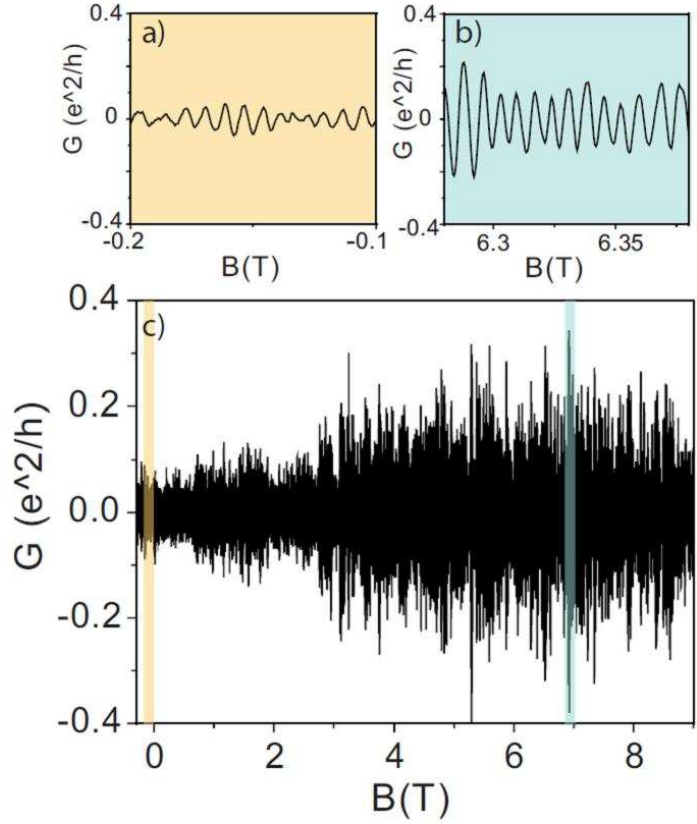}
\caption{ (Color online) The Aharonov-Bohm (AB) effect was
experimentally demonstrated in a graphene ring
device~\cite{Russo2008}. (a)-(c) AB conductance oscillations
versus applied magnetic field, for a back-gate voltage of 30 V and
temperature of 150 mK. For $B\sim$ 3 T, an increase of the AB
amplitude is observed. Panel (a) shows a magnified view of the
yellow region in (c), while panel (b) expands the blue part in
(c). Reprinted figure with permission from S. Russo, J B Oostinga,
D. Wehenkel, H. B. Heersche, S. S. Sobhani, L. M. K. Vandersypen,
and A. F. Morpurgo, Physical Review B {\bf77}, 085413 (2008).
Copyright (2008) by the American Physical Society.}
\end{center}
\end{figure}

In a graphene ring device, conductance oscillations versus
magnetic field were reported as a consequence of the Aharonov-Bohm
effect~\cite{Russo2008}. The amplitude of the oscillations
increases at high magnetic field in the regime where the cyclotron
diameter becomes comparable to the width of the arms of the ring.
For temperatures below 1~K the extracted phase-coherence length is
comparable to or larger than the diameter of the ring, which is
approximately 1 $\mu$m.

Theoretical investigations of graphene rings showed that the
valley-induced orbital degeneracy is lifted, as a result of the
ring confinement and the applied magnetic field~\cite{Recher2007}.
This lifting has observable consequences on the persistent current
and the ring conductance. An interesting finding is that the
degeneracy can be controlled with the induced Aharonov-Bohm flux,
and this can be achieved irrespective of the magnitude (weak or
strong) of the intervalley scattering.

Another theoretical work showed that both electrons and holes can
be confined in electrostatically formed quantum rings in bilayer
graphene~\cite{Zarenia2009}. There are two main advantages in this
proposal. First, bound states can be created owing to a
position-dependent energy gap that suppresses the Klein tunneling.
Second, the ring parameters can be tuned by external fields.

The role of Coulomb-induced electron-electron interactions and
their interplay with the valley polarization in a graphene quantum
ring were also examined~\cite{Abergel2008}.
In a few-electron ring, the interactions have a direct signature
on the fractional nature of the Aharonov-Bohm oscillations in the
persistent current and the absorption spectrum, and therefore they
could be observed.

\section{Graphene {\it pn}-junctions and {\it pnp}-structures}
\label{pnj}

Several graphene-based field-effect devices have been realized in
laboratories.
Reference~\cite{electronic_device}
reported the fabrication of a FET made of graphene which operates at a
record-breaking frequency of 100 GHz. In
Ref.~\cite{fed_paper}
a room-temperature-operated switch demonstrating an on/off ratio exceeding
10$^6$ was described. Also, the implementation of a digital integrated
circuit was reported in
Ref.~\cite{integr_circuit}.
The microcircuit consists of two transistors and performs the logical
inversion operation. A graphene FET used as a biosensor was described in
Ref.~\cite{fet_biosensor}.

All such devices are characterized by a spatially inhomogeneous Fermi level
inside the graphene sample. There are several basic types of such systems:
interfaces separating regions with different concentrations of the charge
carriers
({$pp^\prime$}-junctions,
{$nn^\prime$}-junctions),
or regions with carriers of opposite signs
({\it pn}-junctions),
or series of such interfaces
({\it pnp}-structure,
{$pp^\prime p$}-structure,
etc.). Transport properties of these systems is an important subject of
both theoretical and experimental investigations. These studies are
reviewed below.

\subsection{{\it pn}-junction}
\label{junction}

\begin{figure}[btp]
\centering
\leavevmode
\epsfxsize=8.5cm
\epsfbox{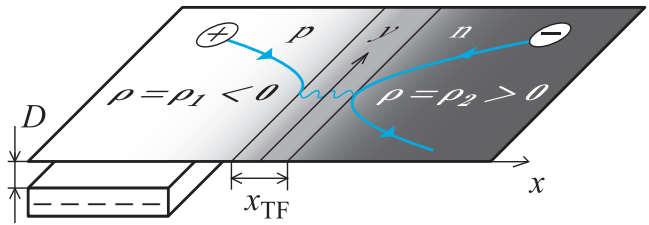} \caption[] {\label{pn-junction} (Color
online) The {\it pn}-junction studied in
Ref.~\cite{zhang_screening}.
There are two gate electrodes in this device. The first one is the
semi-infinite gate on the left side. It controls the density drop
$\rho_2 - \rho_1$
across the junction. The second electrode is an infinite back gate above
the sheet (not shown). It fixes the density $\rho_2$ at far right. Lines
with the arrows show trajectories of an electron ($-$) and a hole ($+$).
The electron current in $n$-region is converted into hole current in
$p$-region. Note that the direction of the incident electron current in
$n$-region and the direction of the hole current in $p$-region are
symmetric with respect to $y$-axis reflection (the same type of refraction
is shown in
Fig.~\ref{meta_focus}). 
Reprinted figure with permission from 
L.M.~Zhang and M.M.~Fogler, Phys. Rev. Lett. {\bf 100}, 116804 (2008).
Copyright (2008) by the American Physical Society.
}
\end{figure}

If two planar electrostatic gates separated from the graphene sample by an
insulating layer are charged in such a way that the chemical potential at
$x>0$ is shifted above the electroneutrality level and at $x<0$ below the
electroneutrality level, a
{\it pn}-junction
is formed at $x=0$, see
Fig.~\ref{pn-junction}.
The simplest model of the electrostatically-defined {\it pn}-junction was studied in
Refs.~\cite{cheianov_np_simple,cayssol_pnp}.

In Ref.~\cite{cheianov_np_simple} the current transmission through
a {\it pn}-junction was investigated. When the current approaches the {\it pn}-junction at the
right angle, it passes through with no reflection. This is a
manifestation of the Klein tunneling. Otherwise, the reflected
current appears. The primary source of reflection in such a system
is a classically-forbidden strip near the center of the junction,
which can be crossed only by quantum tunnelling. The
strip's width $l$ depends on the incidence angle $\theta$ [for
normal incidence $l(\theta=0)=0$, hence, Klein tunneling]. In the
model of Ref.~\cite{cheianov_np_simple}, parameters of the
potential barrier under which the particle has to tunnel depend on
the geometry of the {\it pn}-junction and the gates' potentials.

The main finding of Ref.~\cite{cheianov_np_simple} was that the
current transmission through such {\it pn}-junction is very sensitive to the
angle $\theta$: for normal incidence the transmission is perfect,
but it exponentially quickly deteriorates when $\theta$ grows. This allows
one to create very collimated beams of current. Additionally, the
selectivity to $\theta$ can be used to detect the magnetic field: since the
magnetic field bends the trajectory of a charged particle, then, in a
properly designed device, a particle hits the interface at a
magnetic-field-dependent angle. As a result, the transmission becomes
sensitive to the field. Different devices utilizing properties of the
graphene
{\it pn}-junction
were proposed in
Ref.~\cite{cheianov_np_simple}.

The treatment of
Ref.~\cite{cheianov_np_simple}
was re-examined in
Ref.~\cite{zhang_screening}.
It was noted there that the non-linear charge screening affects the
{\it pn}-junction
characteristics. When the non-linear screening is accounted, significant
deviations from the findings of
Ref.~\cite{cheianov_np_simple},
which neglects many-body effects, are discovered. The results in
Ref.~\cite{zhang_screening}
indicate that the interaction significantly reduces the {\it pn}-junction
resistance.

A more general study, including not only the electron-electron
interaction, but also the disorder, was presented in
Ref.~\cite{fogler_pn_disorder}. It was shown that, depending on
the junction's parameters, it may be in either of three regimes:
$(i)$ ballistic, where the {\it pn}-junction resistance is
dominated by the ballistic contribution, $(ii)$ diffusive, where
the resistance is dominated by the diffusive contribution, and
$(iii)$ the crossover regime, when both ballistic and diffusive
contributions are comparable. In Ref.~\cite{fogler_pn_disorder}
several experimental {\it pn}-junctions
\cite{Lemme2007,huard_2007,Oezyilmaz2007,Williams2007} were
analyzed trying to find junctions in the ballistic regime. It was
concluded that the considered experimental systems satisfy the
conditions for ballistic propagation only marginally at best. It
was suggested that higher mobility or a larger carrier
concentration gradient near the junction is required to create a
ballistic {\it pn}-junction.

\begin{figure}[btp]
\centering \leavevmode \epsfxsize=10.5cm
\epsfbox{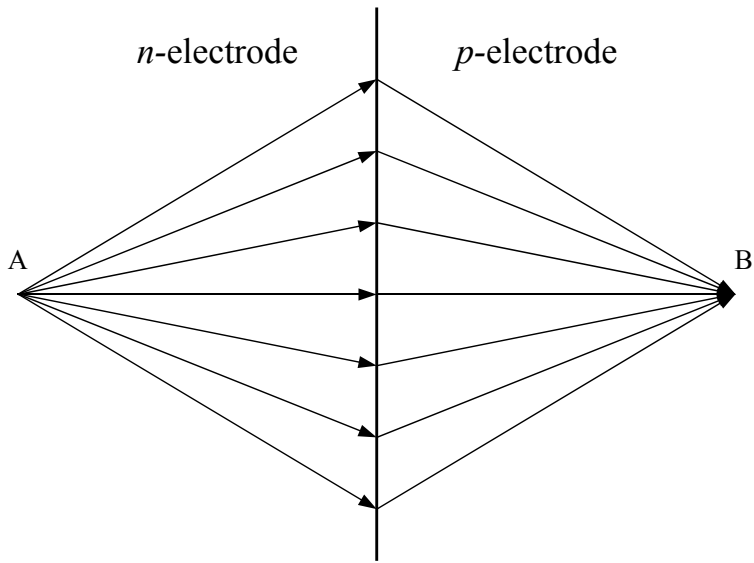}
\caption[]
{\label{meta_focus}
(Color online) {\it pn}-junction as an electron-focusing device, as
described 
in~\cite{graphene_veselago}.
If the gates' voltages are such that the concentration of electrons in the
$n$-region and the concentration of holes in the $p$-region are the same, a
particle hitting the {\it pn}-junction interface at the incidence angle
$\theta_c$ would be converted into a hole on the other side of
the junction propagating at the refraction angle
$\theta_v = - \theta_c$.
Under such conditions, a current emanating from a point source is focused
into a small spot, ``the image", on the other side of the
{\it pn}-junction
\cite{graphene_veselago}.
}
\end{figure}

The ballistic {\it pn}-junction has attracted considerable attention due to
its unusual electron-refraction properties. In
Ref.~\cite{graphene_veselago}
it was observed that, under certain conditions, the electron beam passing
through a graphene {\it pn}-junction experiences refraction in such a
manner that the refraction angle equals to {\it minus} the angle of
incidence, see
Fig.~\ref{meta_focus}.
If this is the case, then current emanating from a point source on one side
of the {\it pn}-junction is focused into ``a point image" on the other side
of the junction. This situation is similar to the refraction of light at
the interface with a metamaterial whose refraction index is minus unity.
The focusing properties of the {\it pn}-junction with circular geometry
were investigated in
Ref.~\cite{graphene_caustics}.

However, this ability to focus the electrical current is easy to spoil.
Ref.~\cite{graphene_veselago}
pointed out that at the level of the geometrical optics the focus is
perfect only if the density of holes in the
$p$-electrode
is the same as the electron's density in the
$n$-electrode.
In Ref.~\cite{fogler_pn_disorder}
it was shown that disorder destroys the focus as well. Finally, since the
transmission of a {\it pn}-junction decays quickly as the incidence angle
deviates from $\pi/2$, only a small fraction of the initial current is able
to pass through the junction to form ``the image".


When a graphene
{\it pn}-junction
is placed in a non-uniform magnetic field, it acquires new interesting
features. This type of devices is discussed in
section~\ref{barrier}.

\subsection{Doping graphene by contact with metals}
\label{metal_contact}

In addition to electrostatic doping, it is possible to change the charge
density in graphene by making contact with a metal electrode. In such a
case, depending on the electrode's material, the electrons either leave the
graphene sample to the electrode or flow into the graphene from the
electrode.

Junctions created with the help of this kind of doping were
investigated experimentally in Ref.~\cite{huard_contact_doping}.
Materials for the metallic electrodes were chosen in such a way as to dope
the graphene with holes. Then, depending on the voltage of the back gate,
either $pp^\prime$- or $pn$-junctions were formed.

The graphene-metal interface was investigated theoretically in
Refs.~\cite{blanter_martin,robinson_metal,golizadeh_metal}. Charge
transfer between the metal electrode and the graphene sample was
studied in
Refs.~\cite{PhysRevLett.101.026803,PhysRevB.79.195425}
with the help of DFT. According to
Ref.~\cite{PhysRevLett.101.026803}
the Fermi energy shift inside the graphene sample is a monotonous function
of the metal work-function, as one should expect. However, when the
work-function of the metal coincides with that of graphene, the graphene
sample is not neutral, as one naively might expect, but rather it is
predicted that the sample is $n$-doped. This happens because of the
chemical interaction between the metal and graphene.

\subsection{{\it pnp}-structure}
\label{pnp}

The theory of electronic transport in clean $pnp$-structures was presented
in Ref.~\cite{cayssol_pnp}.
There, they demonstrated that the conductance of ballistic
{\it pnp}-structure exhibits oscillations (`Fabry-P\'erot' resonances) as a
function of the carrier concentration in the middle ($n$) area of the
{\it pnp}-structure. These resonances are due to quasi-bound electron
states in the $n$-region of the {\it pnp}-structure.

A more general numerical study, which accounts for interaction and
impurities, was performed in
Ref.~\cite{rossi_pnp}.
It reported a crossover from ballistic to diffusive regime when the
mean-free-path becomes comparable to the length of the middle
region. The disorder wipes out the `Fabry-P\'erot' resonances.
However, it is conceivable that these survive under a small concentration of
impurities, and, thus, could be seen experimentally.

A phenomenon analogous to the `Fabry-P\'erot' resonances was
discussed in
Ref.~\cite{bliokh_superlatt,arovas_superlattice},
where the transmission
through several junctions connected in series was studied. Because
of electron wave function interference, the transport through such
structure demonstrates a non-monotonous dependence on the current
incidence angle and the distance between the junctions.

\begin{figure}[btp]
\centering \leavevmode \epsfxsize=10.5cm
\epsfbox{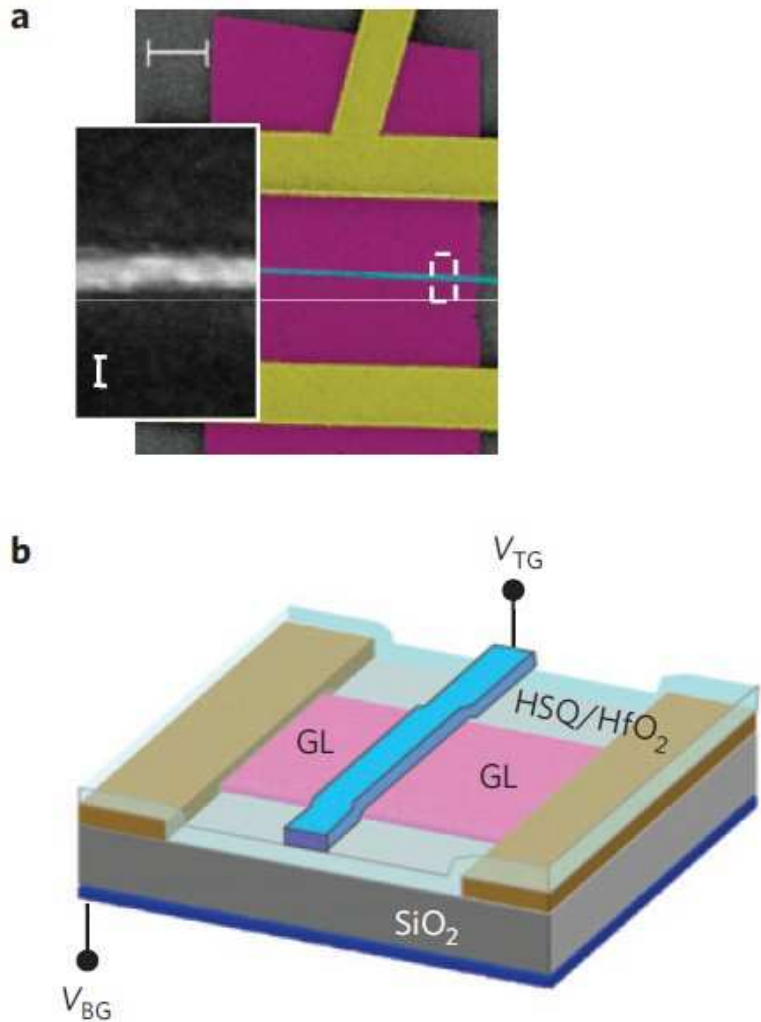}
\caption[]
{\label{pnp_experiment}
(Color online) Experimental realization of a graphene
$pnp$-structure,
from
Ref.~\cite{young_fabry_perot}.
Panel~(a) shows the scanning electron microscope image of the structure.
Large purple rectangle is the graphene sheet. Two bulk yellow electrodes
(source and drain) and one narrow blue electrode (top gate) are placed on
top of the graphene. The back gate beneath the structure is not visible.
The inset presents an enlarged view of the top gate. Panel~(b) shows a
schematic diagram of the same setup.
Reprinted by permission from Macmillan Publishers Ltd:  
\href{http://www.nature.com/nphys/index.html}{Nature Physics},
A. F. Young and P. Kim, Nat. Phys. {\bf 5}, 222 (2009),
copyright (2009).
}
\end{figure}

Electrostatically-defined $npn$- and $pnp$-structures were realized
experimentally
\cite{young_fabry_perot,velasco_fabry_perot,gorbachev_air_bridge}.
For example,
Fig.~\ref{pnp_experiment}
shows a scanning electron microscope image of a
$pnp$-structure
from
Ref.~\cite{young_fabry_perot}.
The experimental observation of `Fabry-P\'erot' oscillations in
$pnp$-structures
was reported in
Ref.~\cite{young_fabry_perot,velasco_fabry_perot}.
In
Ref.~\cite{gorbachev_air_bridge}
experimental data were analyzed within the theoretical framework of
Ref.~\cite{cheianov_np_simple}.
Reference~\cite{gorbachev_air_bridge}
concluded that, in the fabricated
{\it pnp}-structure,
the individual
$pn$-junctions
are ballistic, and that the fabrication of a ballistic graphene
{\it pnp}-structure
is feasible.

In the previous subsection we discussed the peculiar electron refraction at
the $pn$-interface.
In Ref.~\cite{graphene_veselago}
several possible applications of this effect were proposed, among which the
most known is the so-called ``electron Veselago lens".
The latter device is a ballistic graphene $npn$- or $pnp$-structure in
which both junctions are tuned to operate in such a manner that the
electronic current emitted from a point current source in the left
$n$-electrode
travels through two junctions and would be focused into a point (image of
the source) in the right
$n$-electrode.
A similar phenomenon was predicted by Veselago
\cite{veselago_original}
in the optics of materials with a negative-refractive index: the
electromagnetic rays emitted from a point source are focused upon passing
through a slab of such material. This slab is called the Veselago lens. It
is an analog of the $npn$-structure under discussion.

However, the above analogy is incomplete. The ``superresolution", the most
advantageous property of the optical Veselago lens
\cite{pendry,lagarkov_kissel}
(see also
Ref.~\cite{veselago_uspekhi,bliokh_meta,kats_2d_lensing,graphene_optics}),
is absent for the graphene device
\cite{Yampol'skii2008}.
Also, since the current transmission decays quickly for non-normal
incidence, the graphene lens is very opaque. This might make its
application problematic.


Our discussion in this section suggests that graphene
{\it pn}-junctions
and
{\it pnp}-structures,
due to their interesting properties, may, in principle, be used for current
control and magnetic field sensing applications, provided that a way to
attenuate the effects of the disorder is found.
If a graphene
{\it pnp}-structure
is placed into a non-homogeneous magnetic field, it may, under certain
condition, act as a electron waveguide. The relevant discussion can be
found in Sec.~\ref{wguide}.

%
%
%
%
%

\section{Quantum barriers, wires, and waveguides}
\label{barrier}

In this section, we address  charge transport  for designing tunable charge-conducting
elements. Unlike quantum dots (Section V), where electrons are bound in a closed space,
in graphene-based quantum wires and waveguides the charged particles should be confined
only in \textit{one} direction and be freely propagating in another one, as in
$pnp$-structures.

There are several methods of charge confinement in graphene (see
Sections IV-VI above). Typically, this is achieved either $(i)$
chemically (by binding graphene atoms to foreign atoms: e.g.,
oxygen \cite{Jia-AnYan2009}, fluoride
\cite{cheng_flouridation_2010, withers_fluoridation,
xiang_hydro_fluo}, hydrogen \cite{Novoselov2009,
xiang_hydro_fluo}, or aryl groups \cite{nitrophenyl}); $(ii)$
mechanically (either cutting or bending graphene sheets
\cite{Pereira2009, Guinea2010, Pereira2010}, or by creating
inhomogeneous spatial strain distributions \cite{Pereira2010,
Pereira2009a, Levy2010, low_gap_strain}); $(iii)$ thermally
\cite{du_high-mobility}, and $(iv)$ electronically (by applying
electromagnetic fields).

Each of these methods has its advantages and disadvantages. The first two are very
effective; however, they are rather difficult to control, in the sense that any tuning
(change of parameters) requires a reconstruction, either chemical or geometrical, of the
whole graphene sample, which usually cannot be done quickly.  Thus, these methods are not
very suitable in designing \textit{tunable} electronic devices.

Here we concentrate on confining electrons using  electromagnetic fields. This approach
is more flexible than chemical and mechanical approaches. Not only electromagnetic fields
are easy to control, but being tailored properly, they enable the creation of
graphene-based tunable elements, including quantum wires and waveguides with unique
properties, such as unidirectional conductivity, robustness to disorder, etc. The
diversity of methods and approaches makes it increasingly difficult to summarize of the
current state-of-the-art in this area, and calls for a systematic classification by both
methods and results. An attempt of such classification is a goal of this chapter.

Manipulating charge carriers by \textit{electric fields} (i.e., adding scalar potentials
of different shapes to the Dirac equation) is a very popular approach (see, e.g., the
reviews in Refs.~\cite{neto_etal, Peres2009}). This method has provided interesting and
surprising outcomes (i.e., \cite{Titov2007, chakraborty_review, Peres2009, Geim2009}).
Unfortunately, it turned out to be impossible to repel or localize electrons in all
directions and at all energies by only using an electric field \cite{beenakker_colloq}.
There is always a channel in any electric-field barrier where the charge can escape
through.

The charge-confining and current-guiding capabilities produced by \textit{magnetic
barriers} are well known and have already opened certain possibilities for practical
applications (e.g., \cite{Ghosh2009}). However, when it comes to designing fast-tunable
electronic devices (switches, filters, etc.) a difficulty emerges:  most of the existing
magnetic-barrier technologies usually involve the deposition, either on top or beneath
the graphene sheet, of a pattern of magnetic material, which reproduces the desired
magnetic field distribution in the sample. Any subsequent change of parameters would
require building a new setup, creating formidable (if surmountable) obstacles for
harnessing magnetic barriers as elements of fast-acting electronic devices. Grathene
structures based on the effective magnetic field created by applying inhomogeneous
pressure or strain to the graphene sheet \cite{Pereira2010, Pereira2009a, Levy2010}might
also be useful for applications.

An efficient way around this problem is the simultaneous use of inhomogeneous magnetic
and electric fields. The proper combination of these two not only preserves (or even
improves) the necessary transport properties of graphene samples, but makes them easy to
control by tuning the spatial distribution of  the electric potential, for a fixed
magnetic field.

Hereafter, we focus on low-energy excitations when the
inter-valley scattering \cite{Akhmerov2008} is negligible and
quasiparticles can be considered as massless Dirac fermions. To
simplify the presentation, all spin-related effects are also
neglected below.

\begin{figure}[tbh]
\centering \scalebox{0.8}{\includegraphics{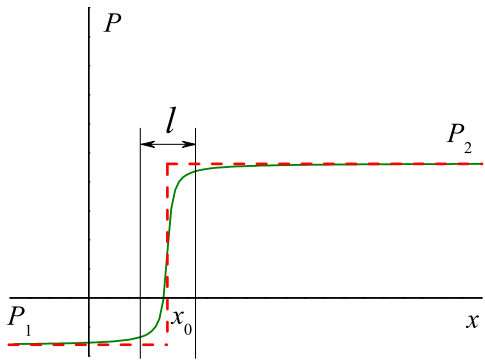}} \caption{ (Color online) Depending
on the problem being considered in this section, the function $P(x)$ is either one of the
potentials $V(x)$, $A(x)$, or the magnetic field $B(x)$. Here, $P(x)$ is shown as a
continuous green line, while the red dashed line repeats the step-like approximation.}
\label{Fig_0}
\end{figure}

The building block of all charge-confining elements is a field-induced barrier (i.e., a
reflecting wall). To introduce this in the most general way, consider two graphene
half-planes, $x<x_{0}-l/2$ and $x>x_{0}+l/2$, subject to different stationary electric
$V_{i}$ and magnetic $\bm{A}_{i}$
potentials ($i=1,2$), as shown in Fig. \ref{Fig_0}. Here we assume that $%
\bm{A}\equiv A(x)\widehat{\bm{y}}$, which means that the magnetic field $%
\bm{B}=B(x)\widehat{\bm{z}}$. When $|x-x_{0}|<l/2$, the fields and
potentials are $x$-dependent, providing the transition from
$\{V_{1},{A}_{1},\bm{B}_{1}\} $ to $\{V_{2},{A}_{2},\bm{B}_{2}\}$
(see Fig.~\ref{Fig_0}). If the width $l$ is large,  compared to
the graphene lattice spacing $a,$ and much smaller than the Fermi
wavelength $\lambda _{F}$, $a\ll l\ll \lambda _{F}$, then the
smooth profile can be replaced in the calculations by a step
function as shown by the dotted line in Fig.~\ref{Fig_0}. When
$B=0$ we have \textit{pn} junction, which is discussed in
Section~VI.

\subsection{Magnetic barriers: ${B}\neq 0,$ $V=0$}\label{NewA}

As it was demonstrated theoretically in Refs.~\cite{DeMartino2007, DeMartino2007a,
Dell'Anna2009}, charge carriers in graphene could be confined in space by applying a
properly tailored static inhomogeneous magnetic field perpendicular to the graphene plane
($x,y$). There are numerous unusual and intriguing features in this phenomenon, but at
its basic level, this happens
because the trajectory of the quasiparticle incoming from  free space $%
x<0 $ bends inside the magnetic barrier (Fig.~\ref{Fig_0}) and
eventually exits backward, independently of its cyclotron radius,
thus making the wall a perfect reflector for Dirac electrons of
any energy \cite{Masir2008}. Obviously, a barrier of finite width
$d$ reflects only charges with energies bellow a threshold, namely
those whose cyclotron radius is smaller than $d$. Below we
consider two types of these magnetic barriers.

\subsubsection{Step-wise vector potential $A$: $\protect\delta$-function-like magnetic field $B$}\label{NewA1}

This basic magnetic field profile, $B(x)=B_{0}\,\ell_{B_0}\,\delta (x)$ (here
$\ell_{B_0}=\sqrt{\hbar c/eB_0}$ is the magnetic length), already displays an important
advantage of magnetic barriers over electric ones: there is a range of energies for which
it is a perfect mirror for Dirac electrons for all angles of incidence including
perpendicular \cite{Ghosh2009}. Outside this energy range, the transparency is
angular-dependent, and the analog of the total internal reflection (TIR) of light takes
place when the angle of incidence exceeds some critical angle $\phi _{{\rm TIR}}$, which
depends on the applied magnetic field. The resulting charge refraction is not of the
classical Snell's law type, and is
asymmetric with respect to the transformation of the angle of incidence $%
\phi \rightarrow -\phi $ \cite{Ghosh2009}.

Rather counter-intuitive is also the fact that a ``point-like'' magnetic field,
$B(x)\propto\delta (x)$, creates a bound (in the $x$-direction) state at zero energy.
This state is a ``linear'' analog of two-dimensional surface waves: it is exponentially
localized in the $x$-direction around the line $x=0$, and propagates along the $y$-axis
with  wave numbers $\pm $ $k_{y}$. The longitudinal wave number $k_{y}$ is enclosed in
the finite interval $( -\ell_{B_0}^{-1}/2,\,\ell_{B_0}^{-1}/2) $, and the $x $-dependence
of the wave function is asymmetric, so that the rates of the exponential decay at $x<0$
and $x>0$ are different, depending on the sign of $k_{y}$ (direction of propagation)
\cite{Masir2008}. The possibilities to experimentally produce highly localized magnetic
fields have been discussed in Ref.~\cite{Ghosh2009}, where charge transport in the
presence of various arrangements of $\delta $-type magnetic barriers was studied in terms
of ``electron optics''.

\subsubsection{Step-wise magnetic field $B$: piece-wise linear vector
potential $A$}\label{NewA2}

The transport properties of Dirac electrons in a step-wise magnetic field [i.e.,
$B(x)=B_1\theta (-x)+$ $B_2\theta (x)$] are different and of greater variety than in the
step-wise potential considered in the previous subsection \ref{NewA1}. While in the later
case, the system is obviously invariant with respect to a shift of the $A$-step [namely,
$A(x)\rightarrow A(x)+\mathrm{const}$], for a step-wise $B$ this is not true: if $B_1$
and $B_2$ are  parallel ($\gamma=B_1/B_2>0)$ or antiparallel $(\gamma<0)$ makes a
significant difference.

To better understand  the behavior of Dirac electrons in nonhomogeneous magnetic fields,
it is illuminating to compare it with the conventional two-dimensional electron gas.
Fundamental differences  already exist in
the uniform field. In contrast with the classical Landau levels $\mathcal{%
E}_{n}$ in infinite space, $\mathcal{E}_{n}\propto (n+1/2)$,
$n\geq 0,$ the
quantization of the eigen-energy of the Dirac equation produces $\mathcal{E%
}_{n}\propto \mathrm{sign}(n)\sqrt{|n|}$, with $-\infty \leq n\leq \infty $. Positive
values of $n$, $n>0$, are associated with electron-like charge carriers, while $n<0$
corresponds to holes. The eigenstates with $n\neq 0$ are similar to the states of the
conventional two-dimensional electron gas, whereas the zero-energy state $n=0$ possesses
different properties. Bound states associated with this $n=0$ Landau level have different
features than the states associated with $n\neq 0$.

The energy in homogeneous magnetic fields does not depend on the
wave number $k$, therefore the group velocity is zero,
$v_{g}=d\mathcal{E}_{n}/dk=0$, and the states
carry no current. A dispersion, and therefore a non-zero group velocity $%
v_{y}=d\mathcal{E}_{n}/dk_{y}\neq 0$ can appear either for states localized near the
boundary of a finite sample (so-called edge states), or due to the spatial inhomogeneity
of the magnetic field, $\nabla_x B\neq 0$, which creates bound states localized in the
$x$-direction and propagating along the $y$-axis with its drift velocity proportional to
$\nabla B\times \mathbf{B}$ (denoted, by analogy with edge states, as \textit{magnetic}
edge states \cite{Muller1992}).

For the particular case $B_{1}=0$ (free half-space for $x<0$, constant magnetic field
$B_{2}>0$ at $x>0$), there is an infinite number of bound (in the $x$-direction)
dispersive states labeled by the Landau-level index $n$, whose energies are proportional
to $\mathrm{sign}(n)\sqrt{|n|}$. These states are localized as functions of $x$, centered
around points whose locations depend on the wave number $k_{y}$. Remarkably, these
localized states exist only with one sign of the wave number $k_{y}$, either positive or
negative, depending on the orientation of the magnetic field $B_{2}.$ Since the group
velocity has opposite signs for $+n$ and $-n,$ the direction of the charge flow created
by electrons and holes is the same, and therefore any bound $n$-state carries a finite
unidirectional current along the $y$-axis \cite{Ghosh2009}. Therefore, a particle in such
a state never undergoes backscattering; hence it is practically insensitive to disorder
and Anderson localization never takes place, no matter how strong the disorder. There is
also a bound state with $\mathcal{E}=0$, when $k_{x}=ik_{y}$. However, this state is
dispersionless and does not carry any current.

When $B_{1}\neq 0$ \cite{Park2008} and it is parallel to $B_{2}$ ($\gamma >0$), the
Landau levels of Dirac quasiparticles at large positive $k_{y}$ are localized in an
effective potential well around $x=-k_{y}\ell_B^{2}$. With $k_{y}$ decreasing to negative
values, the
dimensionless energy levels gradually change to $\mathrm{sign}(n)\sqrt{%
\gamma |n|}$ at large negative $n$ and shift in space to $%
x=-k_{y}\ell_B^{2}/\gamma $. It is important to note that the
directions of the drift (signs of $d\mathcal{E}_{n}/dk_{y}$) are
opposite for electron- and hole-like particles ($\pm n$), thus
providing a non-zero total current. In the vicinity of $k_{y}=0,$
the corresponding states become localized at $x=0$. Similarly to
$B_{1}=0$ case, the magnetic barrier with $\gamma
>0$ supports bound states, which create unidirectional conductivity at $%
n\neq 0$, while zero-energy solutions carry no currents.

When $\gamma <0$ (for antiparallel $B_{1}$ and $B_{2}$) \cite{Park2008}, the effective
potential at large positive $k_{y}$ has two minima (two connected harmonic wells),
located far away from the boundary $x=0$, so that the states are localized in each well.
As $k_{y}$ moves to negative values, the effective potential shifts toward $x=0$ and
transforms into a single non-harmonic potential well. The eigen-energies with $n\neq 0$
correspond to states which support non-zero unidirectional current following classical,
so-called snake, orbits \cite{Muller1992} confined to a narrow one-dimensional channel
centered at the line $x=0$ where the magnetic field changes its sign. It is shown in
\cite{Oroszlany2008, Ghosh2008} that in a symmetric graphene sample of a finite width $L$
($-L/2\leq x\leq L/2$) this current is compensated by real edge states localized close to
the sample boundaries. The states with $n=0$ exhibit both electron and hole features,
which is highly unusual and is unique for Dirac quasi-particles in graphene.

\subsection{Combined magneto-electric barriers: $B\neq 0$, $V\neq 0$}\label{NewB}

In principle, the charge-confining and guiding capabilities of magnetic walls presented
above open up certain possibilities for practical applications. However, as it was
mentioned before, the parameters of the magnetic barriers cannot be changed fast enough,
which makes it problematic to use them as elements of fast-acting electronic devices.

To overcome this problem, it is convenient to use a combination of inhomogeneous magnetic
and electric fields, which enables the efficient control of the transport properties of
graphene samples by tuning the electric potential without changing the parameters of the
magnetic field.

To introduce a basic setup combining magnetic and electric fields, we now consider a
single magneto-electric barrier produced by superimposing a scalar potential $V$ of the
same step-like shape on the magnetic structure with a $\delta$-like magnetic field [i.e.,
a step-wise vector potential $A(x)=A_1\theta (-x)+$ $A_2\theta (x)$]. This system
possesses unique properties that make it different from other types of barriers. Graphene
subject to mutually perpendicular electric and magnetic fields supports states which are
localized near the barrier. These current-carrying states (surface waves) correspond to
quasiparticles moving along the barrier only in one direction \cite{Bliokh2010}. This
direction, as well as the value of the quasiparticle velocity, are easily controlled by
the electrostatic potential. These states correspond to the classical drift of charged
particles in crossed electric and magnetic fields. They exist if and only if the drift
velocity $v_{d}=cE/B$ is smaller than the Fermi velocity $v_{F}$ [here $E\simeq
(V_{2}-V_{1})/l$ and $B\simeq (A_{1}-A_{2})/l$ are the electric and magnetic fields in a
barrier of  finite length $l$]. The absence of counter-propagating states prevents the
backscattering induced by either irregularities in graphene \cite{Titov2007,
bliokh_superlatt} or by the fluctuations of the magnetic field.

For potential applications, the important feature of a single magneto-electric barrier is
that the transport (electric current) across or along this structure can be controlled by
manipulating only the electric potentials $V_{1}$ and $\mathit{V}_{2}$. In particular:

--- The transmission and reflection coefficients across the junction
between two areas with different values $V_{1},A_{1}$ and $V_{2},A_{2}$ (Fig.
\ref{Fig_0}), and the angle of refraction (i.e., the direction of the transmitted
current) depend on the electric potentials. Specifically, tuning $V_{1}$ and/or $V_{2}$
can change the angle of incidence where the barrier is totally transparent, and thus the
Klein tunneling can be observed;

--- When the inequality
\begin{equation}
\left\vert V_{1}-\mathcal{E}\right\vert +\left\vert V_{2}-\mathcal{E}%
\right\vert <\left\vert A_{1}-A_{2}\right\vert  \label{1}
\end{equation}%
holds (here $\mathcal{E}$ is the energy of the quasiparticles; all units are
dimensionless), the step is a perfect reflector for electrons at all angles of incidence
and the junction is locked for the electric current.

--- If
\begin{equation}
\left\vert V_{1}-V_{2}\right\vert<\left\vert A_{1}-A_{2}
\right\vert,  \label{2}
\end{equation}%
a wave (current) exists which propagates unidirectionally along
the barrier (in the $y$-direction) with the dimensionless group
velocity $\nu=v_d/v_F<1$ and is exponentially localized in the
$x$-direction.

\subsection{Waveguide with electrically-tuned parameters}
\label{wguide}

While one barrier forms a wire, two such barriers constitute a
waveguide. This waveguide supports modes that are similar to the
electromagnetic eigenmodes of a dielectric waveguide and likewise
have quantized transverse wave numbers (The analogy between the
transport of Dirac electrons in graphene and light propagation in
dielectrics is described in Refs.~\cite{graphene_veselago,
bliokh_superlatt, Bliokh2010, graphene_optics}). However, along
with them there is another set of waves that is appropriate to
call ``extraordinary'' \cite{Bliokh2010}. They are formed by two
coupled surface waves propagating along the waveguide walls
(barriers). There is an energy gap where only extraordinary modes
exist. Decreasing the spacing between the barriers broadens this
gap. The extraordinary modes are also stable against
backscattering.

An important feature of field-induced waveguides in graphene, which is favorable for the
creation of tunable electronic devices is that the transport properties of these
structures are strongly dependent  on the parameters of the barriers. These parameters
are the potentials $A_{l,r}$ and $V_{l,r}$ of the left and right semiplanes,
respectively, surrounding the central region where the potentials are equal to zero,
$A_c=V_c=0$. In particular:

--- When
\[A_{l}=A_{r} \hspace{3mm}{\rm and}  \hspace{3mm} V_{l}=V_{r}\]
and   the inequalities~(\ref{1}) and (\ref{2}) are valid for both barriers (waveguide
walls), then the extraordinary modes are unidirectional. This makes them immune to
backscattering, and therefore robust against $y$-dependent disorder.\

 --- If
 \[A_{l}=-A_{r}  \hspace{3mm} {\rm and}  \hspace{3mm} V_{l}=V_{r},\]
then the surface waves ``attached'' to the barriers propagate along the $y$-axis in
opposite directions and the extraordinary modes are bidirectional. Nevertheless, the
backscattering also does not affect the total current, due to the spatial separation of
the charge fluxes with opposite directions.

 --- When \[A_{l}=-A_{r}  \hspace{3mm} {\rm and} \hspace{3mm} V_{l}=-V_{r},\]
the spectrum of the extraordinary modes is independent of the distance between the
barriers, and therefore there is no cutoff energy for them. This means that extraordinary
modes can penetrate through an \textit{arbitrary} narrow part of the waveguide.

In Tables~\ref{Tab1} and \ref{Tab2}, the barriers and waveguides of the above mentioned
types and combinations of fields are categorized according to the following features:

\begin{itemize}
\item Their ability to reflect all incident current (perfect wall)

\item Their ability to support bounded electron states

\item The type of spectrum of the propagating modes (either
continuous or discrete)

\item The directionality of the current (either uni- or in
bidirectional)

\item  A separate column lists some other distinctive features because graphene in
magnetic fields has unusual properties.
\end{itemize}

In summary, changing the electric potential (with the magnetic field unchanged) one can
switch on and off the current through the barrier and create/destroy a unidirectional
quantum wire along the barrier. Moreover, changing the electric potential one can create
waveguides with unique, exotic transport properties.



\begin{table}
\begin{tabular}{||p{4.7cm}||p{0.9cm}|p{1.1cm}|p{3.8cm}|p{5.5cm}||}
\hline \centering\textbf{Barrier type} & \textbf{Per\-fect wall} &
\textbf{Bound sta\-tes} & \centering\textbf{Directionality of the
bound state current} & {\textbf{Comments}} \\ \hline\hline

\,\scalebox{0.4}{\includegraphics{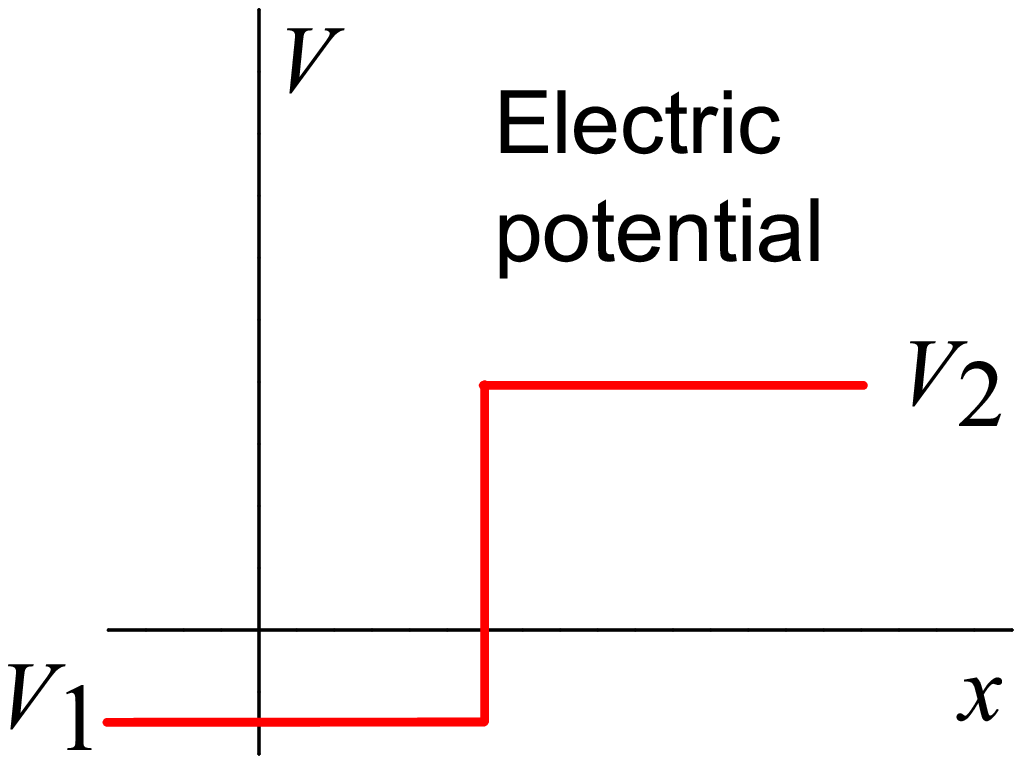}} & No & No & No bound state current.
 & When $\min\{V_i\}<\mathcal{E}<\max\{V_i\}$, the optics analogy of the
barrier is the interface between two dielectrics with opposite signs of the refraction
indexes. Otherwise, this is the interface between usual dielectric media. Total internal
refraction is
possible. Cannot be opaque for all angles of incidence (i.e., Klein tunneling). \\
\hline

\,\scalebox{0.4}{\includegraphics{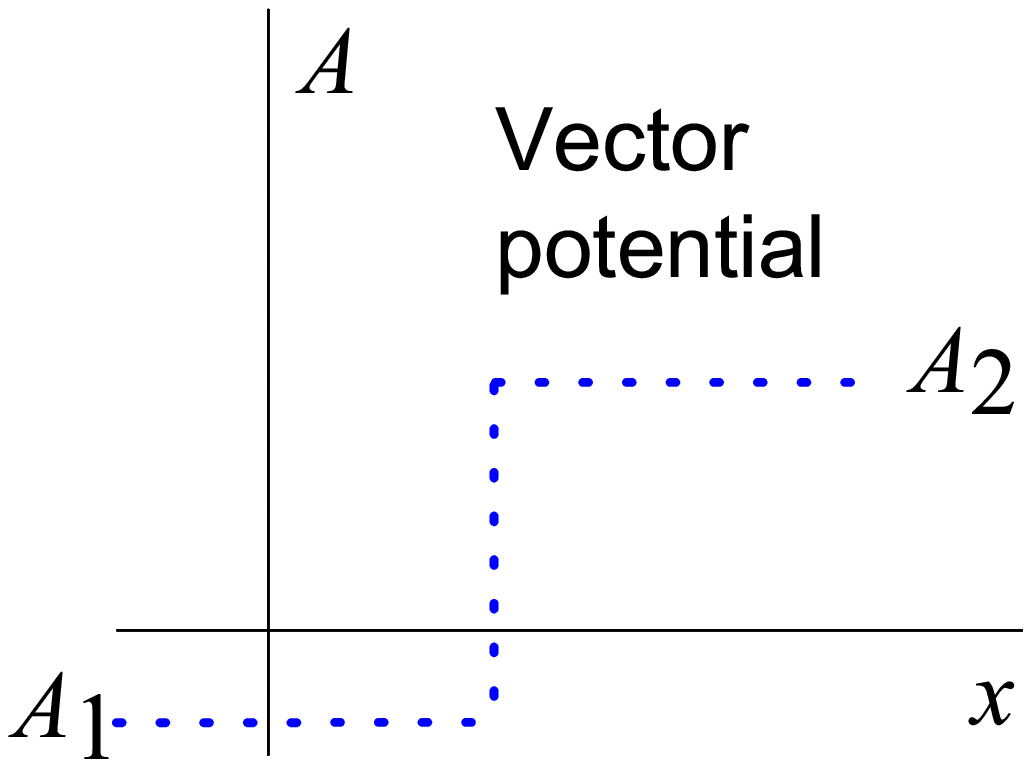}}\, & \,Yes\, &
\,Yes\, &
 One bound state with zero energy ($\mathcal{E}=0$) and zero group velocity ($v_g=0$)
along the barrier, and, therefore, carries no current. This bound state is associated
with the Landau level with $n=0$.\, & In a certain range of energies the barrier is
opaque for all angles of incidence. This barrier is similar to the barrier generated by a
graphene sheet strain. The difference is that the strain generates an effective vector
potentials jump (effective magnetic fields) with opposite (due to the time-reversal
symmetry) signs
in two valleys, whereas the real magnetic field has the same sign in both valleys. \, \\
\hline

\,\scalebox{0.4}{\includegraphics{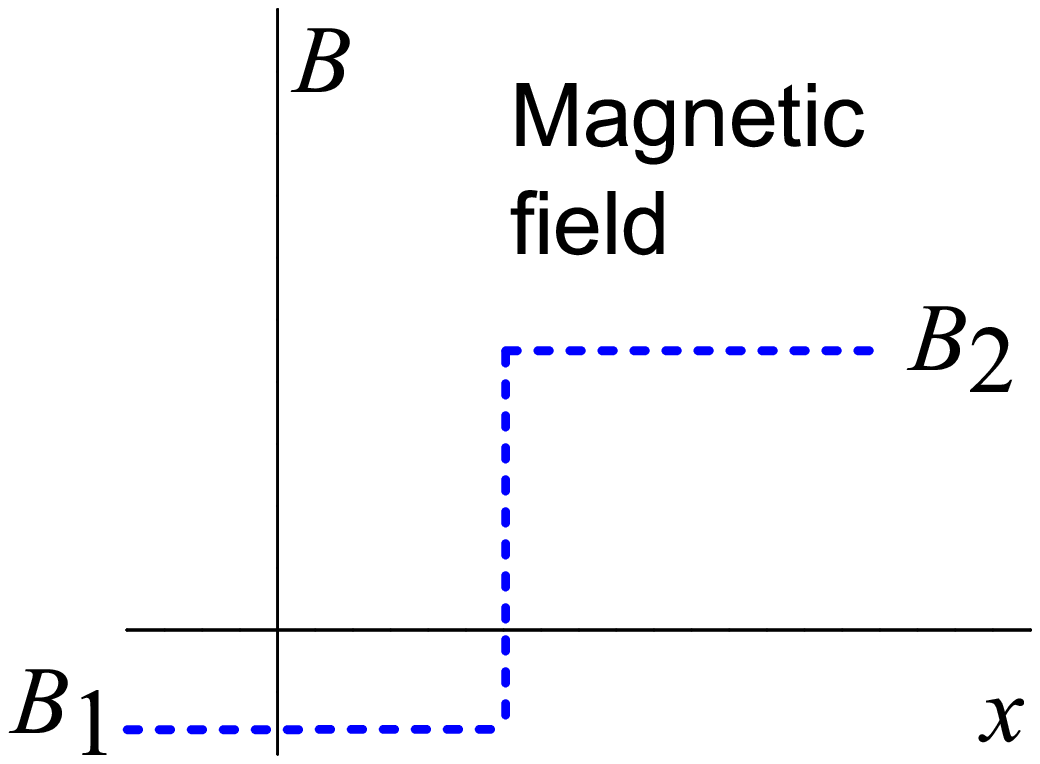}}\, & \,Yes\, & \,Yes\, & Bi-directional
conductivity when both $B_1$ and $B_2$ are non-zero; otherwise uni-directional
conductivity along the barrier. & When $\gamma=B_1/B_2>0$, the bound state is similar to
the classical electrodynamics  state with $\vec{\nabla}B\times\vec{B}$ drift. When
$\gamma<0$ and Landau-level index $n\neq 0$, the state is similar to a snake state
(charged particle motion along the $B=0$ line). Can be opaque for all angles of
incidence.
\\ \hline

\,\scalebox{0.4}{\includegraphics{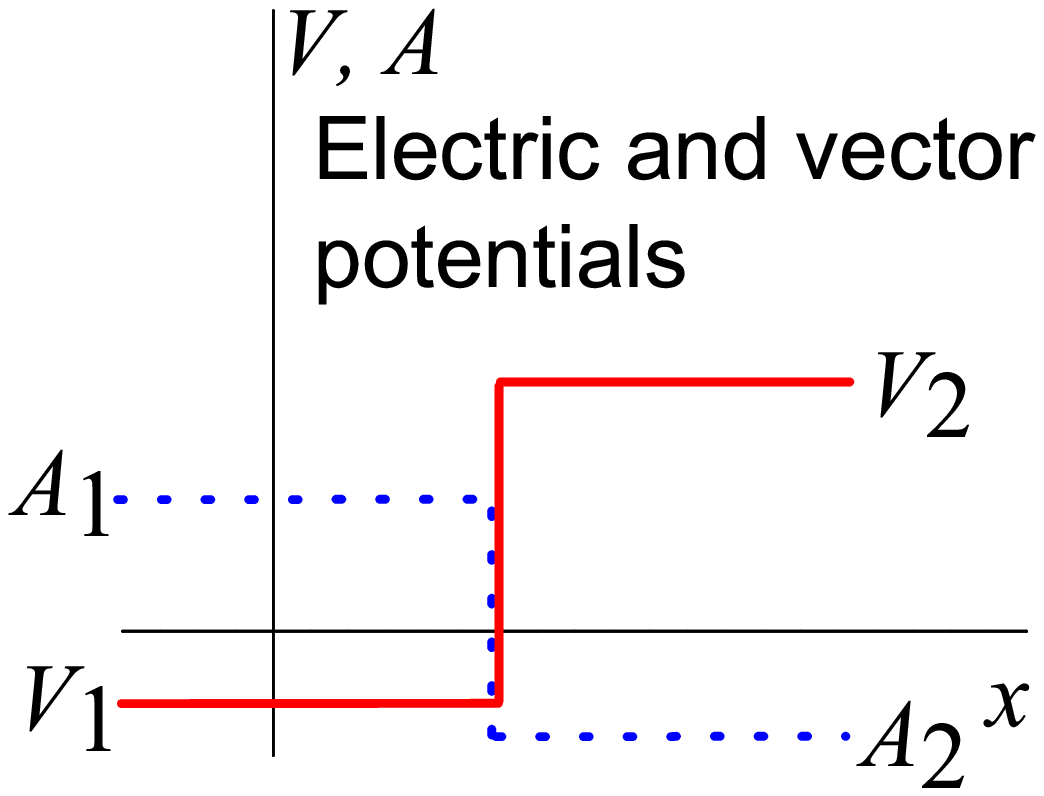}}\, & \,Yes\, & \,Yes\, & Confined
uni-directional state with linear spectrum $d\mathcal{E}/dk_y=v_d$ when
$|v_d|=c|(V_2-V_1)/(A_1-A_2)|<v_F $. In a certain range of energies, the barrier is
opaque for all angles of incidence. Easily controlled by the electric potential. & The
classical electrodynamics analogy of the bound state is the charged particle drift in
crossed
electric and magnetic fields. The drift velocity is $v_d=cE/B=c(V_2-V_1)/(A_1-A_2)$ \\
\hline \,\scalebox{0.4}{\includegraphics{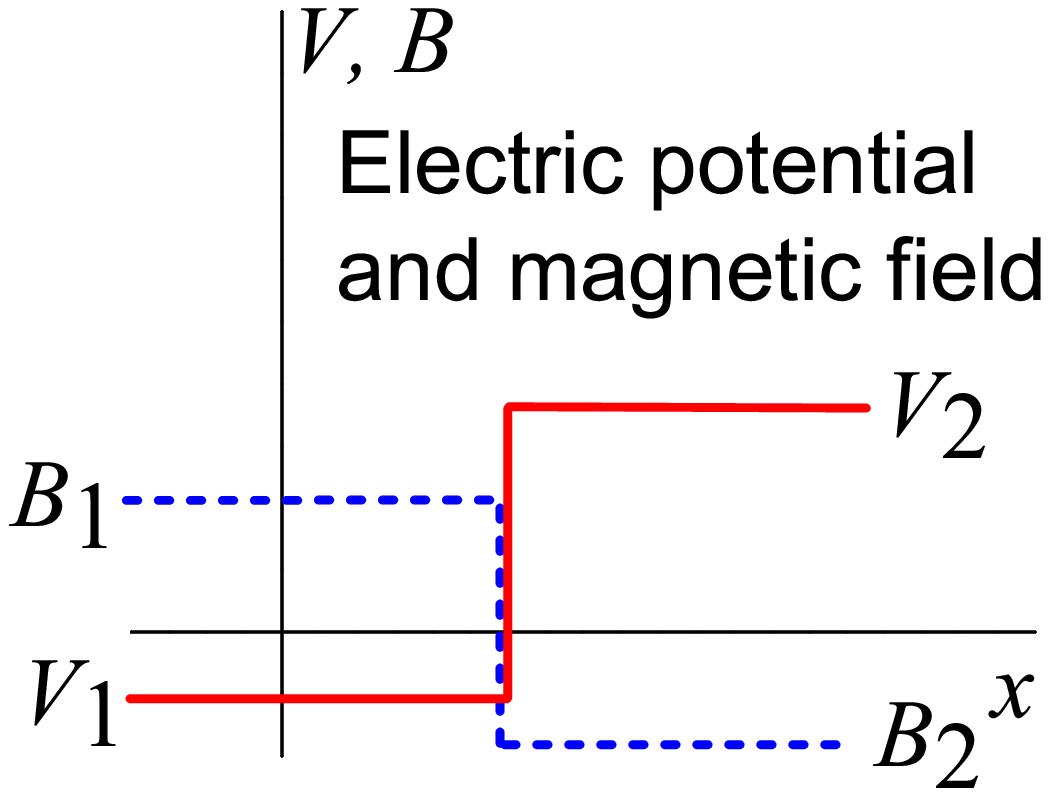}}\, & \,Yes\, & \,Yes\, & Same as the
magnetic field barrier shown above. & Same as the magnetic field barrier shown above. \\
\hline
\end{tabular}%
\caption{Basic properties of electric and/or magnetic barriers in
graphene. The ability to reflect all incident electric currents
(or the ability to be a perfect reflecting wall) is indicated in
the second column. The presence or absence of bound states is
listed in the third column. Additional physical properties are
summarized in the remaining columns.}
\label{Tab1}%
\end{table}

\newpage

\begin{table}
\begin{tabular}{||p{5.0cm}||p{4.0cm}|p{5.0cm}||}
\hline
\centering\textbf{Waveguide type} & \centering\textbf{Guided modes} & %
\textbf{Comments}  \\ \hline\hline

\,\scalebox{0.4}{\includegraphics{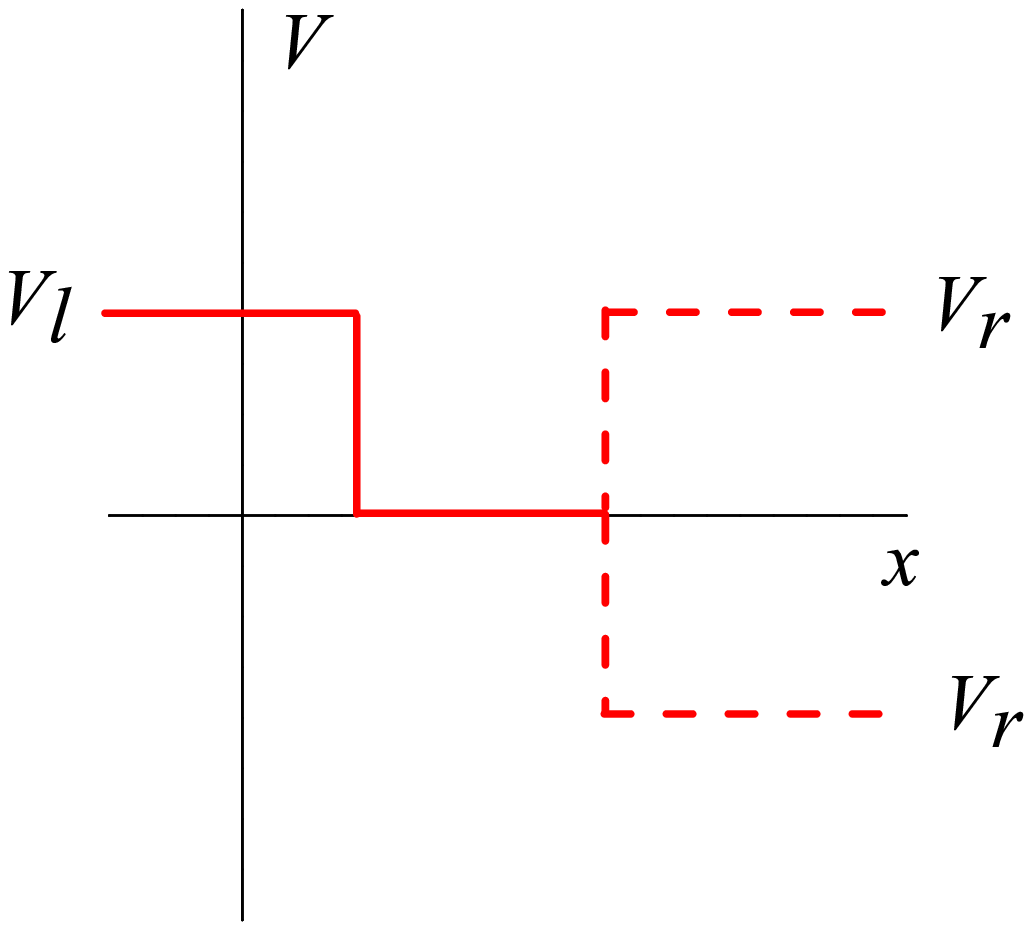}}\, & Discrete set of bulk waves. & Optics
analogy: this is a dielectric waveguide based on  total internal reflection.
 \\ \hline

\,\scalebox{0.4}{\includegraphics{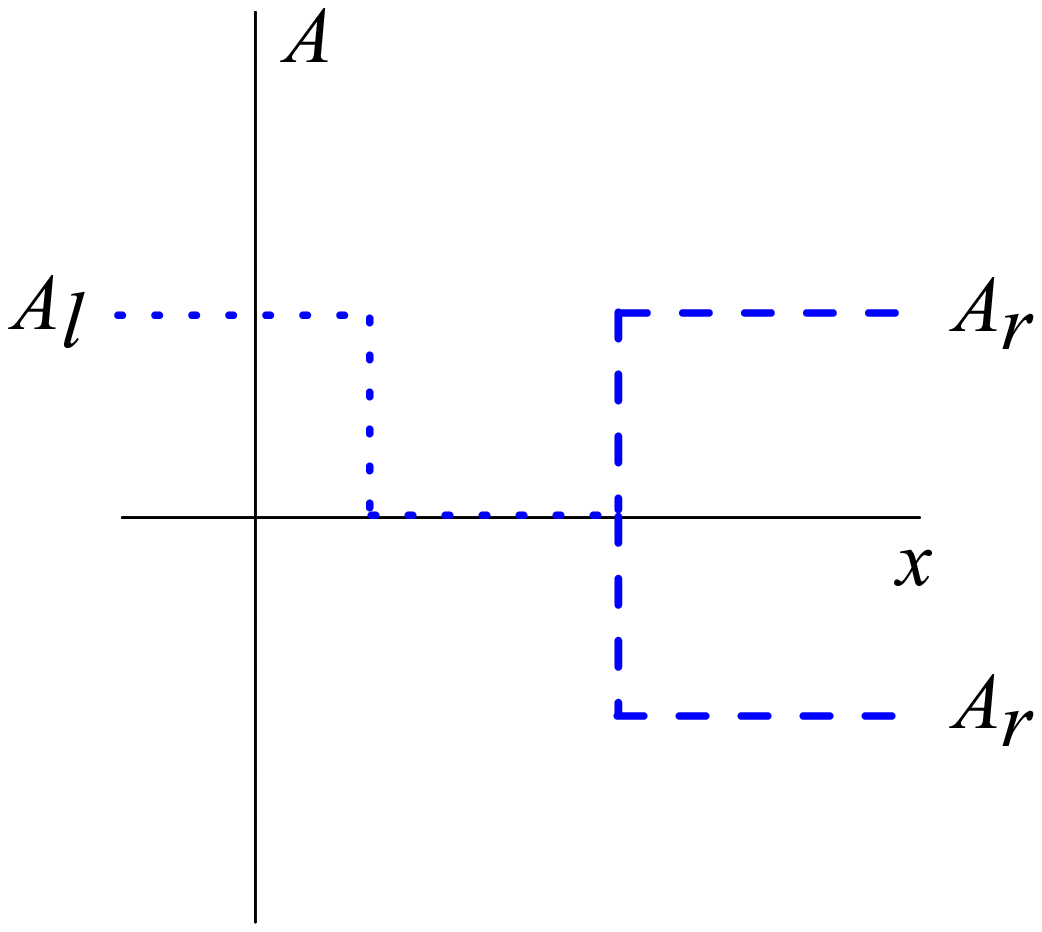}}\, & Discrete set of bulk waves and one
surface mode with zero group velocity along the waveguide. & The bulk waves are similar
to dielectric waveguide modes. The surface mode is localized near the waveguide walls
and consists of two coupled surface modes associated with the barriers (the waveguide walls).  \\
\hline

\,\scalebox{0.4}{\includegraphics{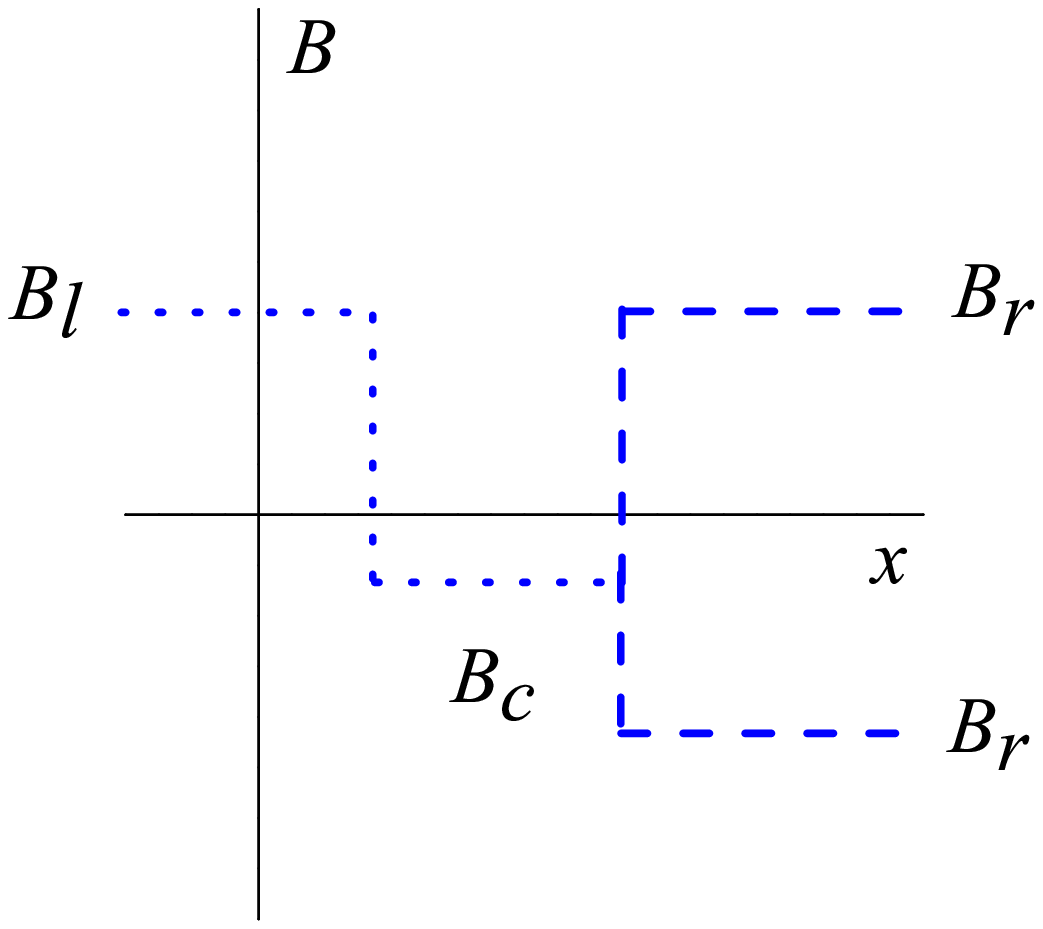}}\, & Discrete set of bulk waves when
$\sigma=B_l/B_r=\pm1$ and $B_c=0$. The mode is localized near the boundaries when
$\sigma=1$ and $B_c<0$. & The discrete set of bulk waves is similar to the eigenmodes of
a dielectric slab surrounded by a medium with negative permittivity/permeability. It
always has a  snake state with noncompensated current. When $\sigma=1$ and $B_c<0$, there
are counter-propagating currents near the walls (two snake states near the walls).  \\
\hline

\,\scalebox{0.4}{\includegraphics{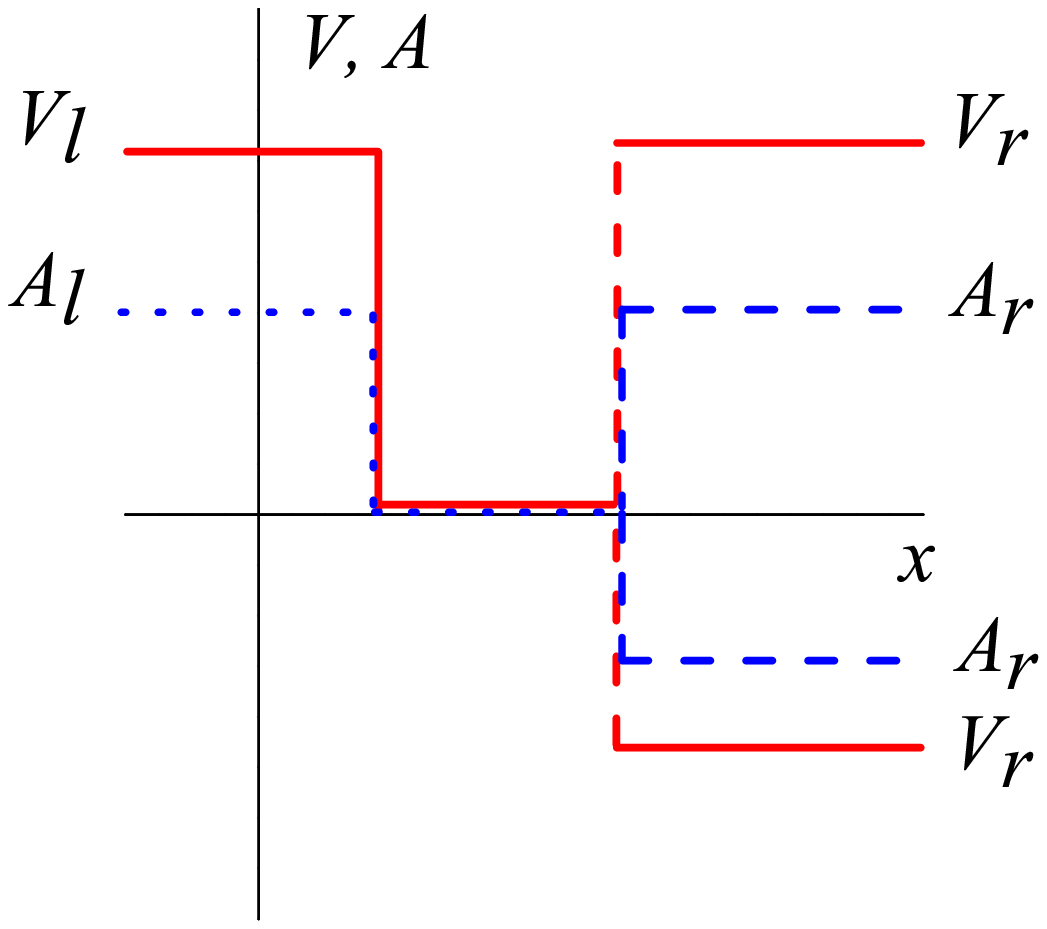}}\, & Discrete set of bulk waves and two
surface modes. & Bulk waves are similar to the eigenmodes of dielectric waveguides. When
the spacing between the barriers is rather small, there are only surface modes.
The propagation direction of surface modes is controlled by the electric potentials.  \\
\hline

\end{tabular}%
\caption{Basic properties of waveguides in graphene. These are
produced by various configurations of the applied electric and
magnetic fields. Note that the convention used here for dotted and
dashed lines is different from the one used in the previous
Table.} \label{Tab2}%
\end{table}






\hspace{4in}

\section{Conclusions}
\label{conclusions}

Graphene is a material with many interesting features which make
it an attractive candidate for microelectronic and micromechanical
applications. In the above pages we very briefly outlined several
ideas and notions driving current graphene mesoscopic research:
edge states, geometric quantization, quasi-bound states, Coulomb
blockade, Klein tunneling through {\it pn}-junctions, etc. Some of
them, like Klein tunneling, are unique to graphene. Others, e.g.,
Coulomb blockade and geometric quantization, have a much longer
history. Yet, even in the latter case, the peculiar properties of
graphene give rise to new features, for example, a much larger
energy scale for the confinement inside a quantum dot.

Although, many of the theoretical studies of graphene mesoscopic
systems are done in the single-electron approximation, the use of
many-body techniques are often warranted. Indeed, the
single-electron approximation could introduce qualitative errors,
which may be corrected only if proper many-body effects are
accounted.

Numerous theoretical proposals have not been explored
experimentally. For instance, the realization of nanoribbons with
atomically-sharp edges remains a distant possibility. Many
suggested devices impose stringent conditions on samples, in terms
of purity and regularity of the sample geometry. Some of these
proposals are stimulating various current experiments. An
important direction in this quest is to control and to understand
the disorder, which may enter through many routes: as foreign
atoms adsorbed on samples or chemically attached to the edges, as
imperfections of the sample edges, as random elastic deformations,
or as bulk defects of varied nature. On the other hand, disorder
is not always an enemy, since its use may be beneficial under
certain circumstances.

Graphene studies are still in their infancy, and it is too early
to guess which of its unusual features will be more useful for
applications. Yet, fabrication of several prototypic
microelectronic devices, like field-effect transistors, biosensor,
and integrated circuit, have been reported. In addition, graphene
presents an excellent playground for fundamental condensed matter
research, exciting enthusiasm of both experimentalists and
theorists in numerous subfields of condensed matter physics.

\section*{Acknowledgements}

We are grateful to L.A.~Openov who provided a high-resolution file
used to create Fig.~\ref{gg_interface}. We are grateful for the
support provided by the grant RFBR-JSPS 09-02-92114. A.V.R. is
partially supported by the grant RFBR 09-02-00248. G.G.
acknowledges support from the Japan Society for the Promotion of
Science (JSPS). F.N. acknowledges partial support from the
National Security Agency (NSA), Laboratory Physical Sciences
(LPS), Army Research Office (ARO), DARPA, Air Force Office of
Scientific Research (AFOSR), and National Science Foundation (NSF)
grant No.~0726909, Grant-in-Aid for Scientific Research (S), MEXT
Kakenhi on Quantum Cybernetics, and Funding Program for Innovative
R\&D on S\&T (FIRST).

\newpage
\appendix

\section{Tight-binding model of graphene lattice}
\label{appendix::basic}
In this Appendix we present basic notions necessary for the theoretical
description of the single-electron properties of a graphene sheet. It
provides some details omitted in Sec.~\ref{graphene}.

It is common to describe a graphene sample in terms of a tight-binding model
on the honeycomb lattice. Such lattice can be split into two sublattices,
denoted by
${\cal A}$
and
${\cal B}$.
The Hamiltonian of an electron hopping on the graphene sheet is given by
Eq.~(\ref{H}),
in which
${\bf R}$
runs over sublattice
${\cal A}$
\begin{eqnarray}
{\bf R} 
=
{\bm \delta}_1 + {\bf a}_1 n_1 + {\bf a}_2 n_2,
\label{subA}
\end{eqnarray}
where the primitive vectors of the honeycomb lattice are
\begin{eqnarray}
{\bf a}_1 &=& a_0 (3/2, \sqrt{3}/2),
\\
{\bf a}_2 &=& a_0 (3/2, -\sqrt{3}/2),
\end{eqnarray}
and
$n_{1,2}$
are integers.
The vectors 
${\bm \delta}_i$
($i=1,2,3$)
connect the nearest neighbors. They are
\begin{eqnarray}
{\bm \delta}_1 
&=&
a_0 (-1, 0),
\\
{\bm \delta}_2 
&=& 
a_0 (1/2, \sqrt{3}/2),
\\
{\bm \delta}_3 
&=& 
a_0 (1/2, -\sqrt{3}/2).
\end{eqnarray}
The geometry of the graphene lattice is presented in 
Fig.~\ref{graphene_lattice}.

The Schr\"odinger equation can be written as
\begin{eqnarray}
\varepsilon \, \psi^{\cal A}_{\bf R} 
&=&
- t \,
\psi^{\cal B}_{{\bf R} + {\bm \delta}_1}
-
t \sum_{i=1,2}
	\psi^{\cal B}_{{\bf R} + {\bm \delta}_1 + {\bf a}_i},
\label{sch_a}
\\
\varepsilon \, \psi^{\cal B}_{{\bf R} + {\bm \delta}_1} 
&=& 
- t \, \psi^{\cal A}_{\bf R} 
- 
t \sum_{i=1,2}
	\psi^{\cal A}_{{\bf R} - {\bf a}_i},
\label{sch_b}
\end{eqnarray}
where 
$\psi_{\bf R}^{\cal A}$ 
($\psi_{{\bf R} + {\bm \delta}_1}^{\cal B}$) 
denotes the wave function value at the site 
${\bf R}$ 
(at the site ${\bf R} + {\bm \delta}_1$)
of sublattice ${\cal A}$ (sublattice ${\cal B}$).

%
The primitive cell of graphene contains two atoms, one at
${\bf R}$,
another at
${\bf R} + {\bm \delta}_1$.
Therefore, it is convenient to define the two-component (spinor) wave
function
\begin{eqnarray}
\Psi_{\bf R} 
= 
\left(
	\matrix{ 
			\psi_{\bf R}^{{\cal A} \hphantom{\delta_1}}\cr
			\psi_{{\bf R} + {\bm \delta}_1}^{\cal B}\cr
		}
\right).
\label{spinor}
\end{eqnarray} 
By construction, the function
$\Psi_{\bf R}$ 
is defined on sublattice 
${\cal A}$,
Eq.~(\ref{subA}). 

The action of $H$ on a plane wave
\begin{eqnarray} 
\Psi_{\bf R} = \Psi_{\bf k} \exp ( -i {\bf k} \cdot {\bf R} )
\end{eqnarray} 
can be expressed as
\begin{eqnarray}
H \Psi_{\bf k} 
=
\left(
	\matrix{ 
			0            &  -t_{\bf k}  \cr
       		       -t_{\bf k}^*  &        0     \cr
       		}
\right)
\Psi_{\bf k},
\label{matrix_ham}
\\
t_{\bf k} 
=
t
\left[
	1 
	+ 
	2 {\exp}	\left(
				-i\frac{3  k_x a_0}{2} 
			\right)
	\cos \left(
			\frac{\sqrt{3}}{2} k_y a_0
	      \right)
\right].
\label{tk}
\end{eqnarray}
For every ${\bf k}$ there are two eigenstates
\begin{eqnarray}
\Psi_{{\bf k} \pm}
=
\left(
	\matrix{
				1			\cr
			\mp {\rm e}^{-i \theta_{\bf k}} \cr
		}
\right),
\\
\exp 	\left(
	{i \theta_{\bf k}} 
	\right)
=
\frac{t_{\bf k}}{|t_{\bf k}|},
\label{theta_def}
\end{eqnarray} 
with eigenvalues:
\begin{eqnarray}
\varepsilon_{{\bf k} \pm}
&=&
\pm |t_{\bf k}|
\label{graphene_energy}
=
\pm t
\sqrt{
	3 
	+ 
	F({\bf k})
     },
\\
F({\bf k})
&=&
4\cos \left( 
		\frac{3}{2} k_x a_0 
	\right)
\cos \left( 
			\frac{\sqrt{3}}{2} k_y a_0 
     \right)
+
2 \cos \left( 
		\sqrt{3} k_y a_0 
	\right).
\label{F}
\end{eqnarray}
The states with negative (positive) energy are filled (empty) at $T=0$.
The allowed values of ${\bf k}$ lie within the Brillouin zone presented in
Fig.~\ref{bz}.

The reciprocal lattice is characterized by the following lattice vectors
\begin{eqnarray}
{\bf d}_1 = (4\pi/3 a_0, 0),
\label{d1}
\\
{\bf d}_2 = ( - 2\pi/3 a_0, 2\pi/\sqrt{3} a_0 ).
\label{d2}
\end{eqnarray} 
The amplitude
$t_{\bf k}$
and energy 
$\varepsilon_{{\bf k} \pm}$
are invariant under shifts over 
${\bf d}_{1,2}$.

The quantity 
$\varepsilon_{{\bf k} \pm}$
vanishes at the six corners of the Brillouin zone:
${\bf K}_{1,2} = (0, \pm 4\pi/(3\sqrt{3}a_0))$ 
and
${\bf K}_{3,4,5,6} = (\pm 2\pi / (3 a_0), \pm 2\pi/(3\sqrt{3}a_0))$.
These are the locations of the famous Dirac cones of graphene. These six
cones can be split into two equivalence classes: all cones inside a given
equivalence class are connected by a reciprocal vectors. We choose cones
located at 
${\bf K}_{1,2}$
to be representatives of these classes.

If one is interested in the low-energy description only, then the
tight-binding Hamiltonian may be replaced by the Dirac Hamiltonian with the
help of the following derivation. Near the cones, the Hamiltonian 
Eq.~(\ref{matrix_ham})
may be expanded in orders of
${\bf k} - {\bf K}_{1,2}$. 
The spinor itself may be represented as
\begin{eqnarray}
\Psi ({\bf R})
=
\Psi_1 ({\bf R}) \exp ( - i {\bf K}_1 \cdot {\bf R})
+
\Psi_2 ({\bf R}) \exp ( - i {\bf K}_2 \cdot {\bf R}),
\label{spinor_decomp}
\end{eqnarray}
and the Schr\"odinger equation splits into two copies of the Weyl-Dirac
equation~(\ref{dirac}),
where we treat 
$\Psi_{1,2} ({\bf R})$
as slowly varying functions of the continuous variable ${\bf R}$, and the plus
(minus) sign is chosen for 
$\Psi_1$ ($\Psi_2$). 

Describing the electronic properties of the graphene in terms of
Eq.~(\ref{dirac})
is very popular for two reasons. First, it is much simpler than the full
tight-binding Hamiltonian (\ref{H}). Second, unlike the tight-binding
Hamiltonian, which may require knowledge of several hopping amplitudes,
only one parameter
$v_{\rm F}$
has to be specified. Of course, the tight-binding model is needed when a
more detailed description is required.

\section{Edge states}
\label{appendix::es}
Usually, two types of edges are discussed in the literature: zigzag and
armchair. A variant of the zigzag is the Klein edge 
\cite{klein_edge}.
All three types are shown in
Fig.~\ref{graphene_lattice}.

\subsection{Armchair edge}
\label{appendix::armchair}

The physics of the armchair edge is simpler than that of the zigzag edge
because the latter supports zero-energy localized states, while the former
does not. The easiest way to describe an electron near an armchair edge is
to use the Weyl-Dirac equation
(\ref{dirac})
with the appropriate boundary condition. The general problem of the
boundary condition for the Weyl-Dirac equation is investigated in
Refs.~\cite{dirac_boundary_cond_nanotube,dirac_boundary_cond_graphene,
complicated_edges,volkov_zagorodnev_bc,volkov_zagorodnev_bc2}.
Here we use a simple explicit form of the boundary condition suitable for
armchair edge
\cite{brey_fertig_dirac_eq_nb}.
Namely, we demand that our spinor wave function 
Eq.~(\ref{spinor_decomp})
vanishes at the edge, which we assume to be located at
$y = 0$
\begin{eqnarray}
\Psi_1 ({\bf R})|_{y = 0}
=
- \Psi_2 ({\bf R})|_{y = 0}.
\end{eqnarray}
For an infinite half-plane, the solution of the Weyl-Dirac equations with
this boundary condition is equal to
\begin{eqnarray}
\Psi_{1\pm} ({\bf R}) 
=
\frac{1}{\sqrt{2}} 
\left(
	\matrix{
		1 
		\cr
		\mp(i k_x + k_y)/{k}
	}
\right)
\exp(- i k_x x - i k_y y ),
\label{wd_solution1}
\\
\Psi_{2\pm} ({\bf R}) 
=
-\frac{1}{\sqrt{2}} 
\left(
	\matrix{
		1 
		\cr
		\mp(i k_x + k_y)/{k}
	}
\right)
\exp(- i k_x x + i k_y y ).
\label{wd_solution2}
\end{eqnarray} 
The eigenfunctions 
$\Psi_{1,2+}$
($\Psi_{1,2-}$)
correspond to positive (negative) eigenvalues. The total wave function is
to be constructed according to
Eq.~(\ref{spinor_decomp}).

\begin{figure}[btp]
\centering
\leavevmode
\epsfxsize=8.5cm
\epsfbox{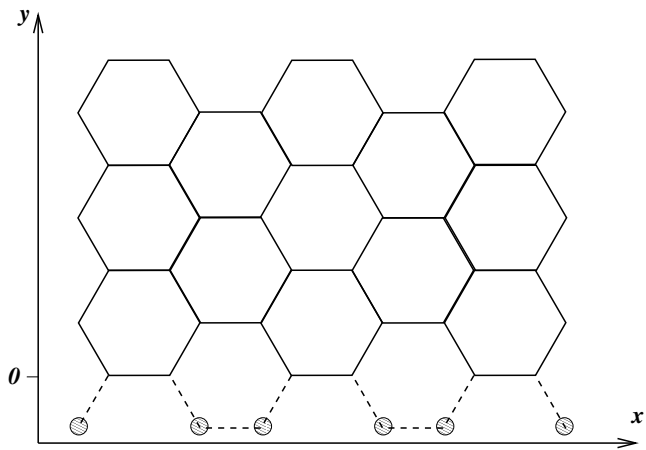}
\caption[]
{\label{aux}
Introduction of auxiliary atoms (hatched circles) at an armchair edge.
}
\end{figure}
However, it is not necessary to use the approximate description in terms of
the Weyl-Dirac equation. The tight-binding model may be solved near the
armchair edge as well. To construct such a solution, note first that the
atoms at the very edge are special: they have only two nearest neighbors,
while the atoms in the bulk have three. This means that
Eqs.~(\ref{sch_a},\ref{sch_b}) 
must be modified to account for this fact. It is more convenient, however,
to introduce auxiliary rows of carbon atoms (Fig.~\ref{aux}) and demand
that the wave function vanishes on these additional atoms. Since there is
no wave function density on the auxiliary atoms, they do not contribute to
the Schr\"odinger equation for the physical atoms at the edge. This
construction is used in several papers (e.g.,
Ref.~\cite{gunlycke_nanoribbon_gap,our_nanoribbon_paper_2009}).
The solution of Eqs.~(\ref{sch_a}) and (\ref{sch_b}) is
\begin{eqnarray}
\Psi_{\bf R} 
=
\Psi_{{\bf k} \pm} \exp( -i k_x x ) \sin [k_y(y+\sqrt{3}a_0/2)].
\label{tb_solution}
\end{eqnarray}
This wave function vanishes at 
$y = - \sqrt{3}a_0/2$,
which is where the auxiliary atoms are located. At the physical edge of
the nanoribbon 
($y = 0$),
however, the electron density remains non-zero. Note that momentum
components
$k_{x,y}$
in 
Eq.~(\ref{tb_solution})
is measured from the center of the Brillouin zone. However, in
Eq.~(\ref{wd_solution1})
and
Eq.~(\ref{wd_solution2})
they are measured from the Dirac cones. When this subtlety is accounted it
is easy to demonstrate that both solutions have the identical
dependence on
${\bf R}$.
They differ only in the value of the spinor part 
$\Psi_{\bf k}$.
This is because 
Eq.~(\ref{wd_solution1})
and
Eq.~(\ref{wd_solution2})
are only approximations which are accurate at small energy only.

The discussion presented above assumes that the properties of the
carbon-carbon bonds near the edge remain the same. In a real sample this
assumption is only an approximation. A variety of perturbations may be
present near the edge. For example, 
Ref.~\cite{density-func_zig_armch}
investigates the edge structure with the help of density-functional
methods. It is found that the carbon-carbon bonds at armchair edges are
shorter than in the bulk (1.26\,\AA\ versus 1.4\,\AA). This means that
the effective electron hopping at the edge 
$t_{\rm e}$
differs from its value $t$ in the bulk.

Also, non-carbon radicals may be attached to the unpaired chemical bond of
carbon atoms at the armchair edge. This means that, depending on the
description, either the Hamiltonian near the edge, or the boundary
conditions have to be modified. Fortunately, these alterations are small,
and may be treated as weak corrections.

\subsection{Zigzag edge}
\label{appendix::zigzag}

The physics of the zigzag edge is richer than the physics of the armchair
edge due to the zero-energy states localized at the edge. When graphene is
described by the Hamiltonian (\ref{H}), which only contains
nearest-neighbor hopping, these states are dispersionless and
macroscopically degenerate. 

The simplest way to detect the presence of edge states is to use the
Weyl-Dirac equation with the boundary condition appropriate for the zigzag
edge
\cite{dirac_boundary_cond_nanotube,dirac_boundary_cond_graphene,
complicated_edges}.
For the purpose of demonstrating the presence of edge states the simplest
version of the boundary condition is used here
\cite{brey_fertig_dirac_eq_nb}.
The zigzag edge, unlike the armchair edge, consists of atoms belonging to
the same sublattice. This is clearly seen in 
Fig.~\ref{graphene_lattice}:
all atoms at the left edge belong to 
${\cal A}$
sublattice (red). The right zigzag edge has all its atoms on the
${\cal B}$
sublattice. Extending the left edge by a column of the auxiliary atoms, we
demand that the wave function vanishes on them (assume that the auxiliary
atoms are located at $x=0$):
\begin{eqnarray}
\psi^{\cal B}|_{x=0} = 0.
\label{bc_zigzag}
\end{eqnarray}
Then the following spinor functions are the solutions of the Weyl-Dirac
equation:
when $k_y > 0$
\begin{eqnarray}
\Psi_1 ({\bf R})
=
\left(
	\matrix{
			1 \cr
			0
		}
\right)
\exp(-k_y x + i k_y y),
\quad 
\Psi_2 = 0;
\label{es_dirac1}
\end{eqnarray}
and when
$k_y < 0$
\begin{eqnarray}  
\Psi_1 = 0,
\quad
\Psi_2 ({\bf R})
=
\left(
	\matrix{
			1 \cr
			0
		}
\right)
\exp(-|k_y| x + i k_y y).
\label{es_dirac2}
\end{eqnarray}
Both solutions decay for large values of $x$ and correspond to the zero
eigenvalue. The difference between them is that
Eq.~(\ref{es_dirac1})
describes the solution near 
${\bf K}_1$,
while 
Eq.~(\ref{es_dirac2})
near 
${\bf K}_2$.
The wave function 
Eq.~(\ref{spinor_decomp})
constructed from
Eqs.~(\ref{es_dirac1}) and (\ref{es_dirac2})
is equal to
\begin{eqnarray}
\Psi ({\bf R})
=
\left(
	\matrix{
			1\cr
			0
		}
\right)
\exp ( - i {\bf K}_1 \cdot {\bf R} - k_y x + i k_y y)
+
\left(
	\matrix{
			1\cr
			0
		}
\right)
\exp ( - i {\bf K}_2 \cdot {\bf R} - k'_y x - i k'_y y),
\label{es_dirac_full}
\end{eqnarray}
where both
$k_y$ and $k'_y$
are positive and small compared to 
$1/a_0$. 
The quantity $k_y$ sets a parameter with dimension of length
\begin{eqnarray}
\lambda_{\rm edge} = \frac{1}{|k_y|}
\end{eqnarray}
which characterizes how deeply the edge states extend into the bulk of
graphene.

Since the Weyl-Dirac equation is only an approximation of the tight-binding
Hamiltonian (\ref{H}), we cannot reliably use 
Eq.~(\ref{es_dirac_full}) 
for large values of 
$k_y$ and $k'_y$.
However, the tight-binding problem may be solved exactly
\cite{nakada-fujita_nribb_edge_st}
to discover that there is a degenerate manifold of states labeled by the
momentum $k_y$, which stretches from 
${\bf K}_1$
to
${\bf K}_2$.

\subsubsection{Effect of the longer-range hopping}
\label{tb_es}

The degeneracy of edge states is purely accidental property of Hamiltonian
(\ref{H}).
It disappears when other terms are added to $H$. For example, in
Refs.~\cite{el_prop_disordered_graphene,sasaki_nnn,sasaki_gague_nnn,
decomposition}
the effect of next-to-nearest-neighbor hopping on edge states is studied both
analytically and numerically. In
Ref.~\cite{decomposition}
the authors made three keen observations: $(i)$ the next-to-nearest-neighbor
hopping effectively induces a shift in the local potential; $(ii)$ the
shift depends on site's position, namely, near the edge it is not the same
as in the bulk; and $(iii)$ this spacial variation of the potential induces
finite dispersion for an otherwise dispersionless edge states. To prove
that $(iii)$ holds true, consider the following argument. The states with
large 
$\lambda_{\rm edge}$
are insensitive to the potential variation (they ``feel" the potential
averaged over large 
$\lambda_{\rm edge}$),
however, those with small 
$\lambda_{\rm edge}$
are affected strongly; since 
$\lambda_{\rm edge}$
depends on 
$|k_y|$,
the edge states acquire the dispersion.

This discussion suggests that, if we were to describe this phenomena with
the help of Weyl-Dirac equation, the boundary conditions,
Eq.~(\ref{bc_zigzag}),
must be modified to account for the effect of the potential modulation
near the boundary. Indeed, it is demonstrated in
Ref.~\cite{volkov_zagorodnev_bc2}
that one can generalize
Eq.~(\ref{bc_zigzag})
and reproduce the findings of
Ref.~\cite{decomposition},
at least near the apexes of the Dirac cones.

\newpage
{\bf References}
\bibliographystyle{apsrevlong_no_issn_url}

\bibliography{all_bib}

\newpage

\end{document}